\documentclass{JHEP3}
\usepackage{amsfonts}
\usepackage{amsmath}
\usepackage{amssymb}
\usepackage{latexsym}
\usepackage{graphicx}
\usepackage{cite}

\numberwithin{equation}{section}

\def\fr#1{\mathfrak{#1}}

\def\LP{\ell_{11}}

\def\ZZ{{\mathbb Z}}
\def\IC{{\mathbb C}}
\def\IR{{\mathbb R}}

\def\EE{ {\cal E}_{({3\over 2},{3\over 2})} }

\def\cE{{\cal E}}

\def\cE{\mathcal{E}}
\def\cF{\mathcal{F}}
\def\cV{\mathcal{V}}
\def\hcV{{\hat{\mathcal{V}}}}
\def\cR{\mathcal{R}}
\def\cT{\mathcal{T}}

\newcommand{\be}{\begin{equation}}
\newcommand{\ee}{\end{equation}}
\newcommand{\bea}{\begin{eqnarray}}
\newcommand{\eea}{\end{eqnarray}}
\def\p{\partial }

\def\threeh{{\scriptstyle {3 \over 2}}}
\def\fiveh{{\scriptstyle {5 \over 2}}}

\def\bE{{\bf E}}

\def\hbE{{\hat{\bf E}}}

\def\calE{{\cal E}}

\def\calT{{\mathcal T}}

\def\bs{\backslash}

\def\calF{{\mathcal F}}

\def\calM{{\mathcal M}}
\def\calV{{\mathcal V}}

\def\nn{\nonumber} 
\def\half{{\scriptstyle {1 \over 2}}}
\def\quart{{\scriptstyle {1 \over 4}}}
\def\sevenh{{\scriptstyle {7 \over 2}}}

\def\EE{{\cal E}_{(0,1)}}

\preprint{ DAMTP-2010-1, IPHT-T-10/001, IHES/P/10/01, ICCUB-10-002}

\title{Automorphic properties of low energy string amplitudes in various dimensions}
\author{Michael B. Green\\
 Department of Applied Mathematics and
Theoretical Physics\\
Wilberforce Road, Cambridge CB3 0WA, UK\\
\email{\tt M.B.Green@damtp.cam.ac.uk}}
\author{Jorge G. Russo\\
Instituci{\'o} Catalana de Recerca i Estudis Avan\c cats (ICREA)\\
Departament ECM and Institut de Ciencies del Cosmos, \\
University de Barcelona,  Facultat de Fisica\\
 Av. Diagonal, 647,  Barcelona 08028 SPAIN\\
\email{\tt jrusso@ub.edu}}
\author{Pierre Vanhove\\
Institut des Hautes Etudes Scientifiques\\
Le Bois-Marie, 35 route de Chartres\\
F-91440 Bures-sur-Yvette, France\\
\textrm{and}\\
Institut de Physique Th{\'e}orique,\\
CEA, IPhT, F-91191 Gif-sur-Yvette, France\\
CNRS, URA 2306, F-91191 Gif-sur-Yvette, France\\
\email{pierre.vanhove@cea.fr}}

\abstract{ This paper explores the moduli-dependent coefficients of higher derivative interactions 
  that  appear  in  the low-energy  expansion of the four-supergraviton amplitude of maximally 
  supersymmetric string theory compactified on a $d$-torus. 
  These automorphic functions are determined for terms up to order $\partial^6\cR^4$ and various values of $d$ by imposing a variety of consistency conditions. They satisfy Laplace  eigenvalue equations  with or  without source  terms, whose
  solutions are  given in terms  of Eisenstein series, or more general
  automorphic  functions, for  certain parabolic  subgroups of  the relevant
  U-duality groups.  The ultraviolet divergences of the corresponding supergravity
  field  theory  limits are  encoded in various logarithms, although the string theory expressions are finite.  This analysis includes intriguing representations of $SL(d)$ 
and $SO(d,d)$ Eisenstein series in terms of toroidally compactified one and two-loop string and supergravity amplitudes.

}


\begin{document}


\newpage
\section{Introduction}
\label{sec:intro}

In  this  paper  we  will  pursue a  programme  of  elucidating  exact
properties of the four-supergraviton scattering amplitude\footnote{The term ``supergraviton'' refers to the supermultiplet of 256 massless states. The dependence on the helicities of these states arises in the amplitude through a generalised curvature,  $\cR$~\cite{Green:2008bf}.} in the low energy
expansion  of string  theory  compactified from  $10$  to  $D=10-d$
dimensions on a  $d$-torus, $\calT^d$.  Although this is  a very small
corner of M-theory it is one  in which precise statements can be made.
In particular, the combination  of maximal supersymmetry and U-duality
is very constraining~\cite{Hull:1995mz}.   The low energy expansion of
the scattering amplitude in $D$-dimensional space-time has the general
form 
 \be
 A_D(s,t,u)= A_D^{analytic}(s,t,u) + A_D^{nonan}(s,t,u)\,,
 \label{ampsplit}
 \ee
 where we have separated analytic and nonanalytic functions of the Mandelstam invariants, $s$, $t$ and $u$
 ($s=-(k_1+k_2)^2$, $t=-(k_1+k_4)^2$, $u=-(k_1+k_3)^2$ and $s+t+u=0$). Although it is not obvious that such a separation can be made in a useful manner to all orders in the low energy expansion, it is sensible and useful at the orders to be considered in this paper.  The analytic part of the amplitude has the expansion (in the Einstein frame) 
 \be
 A_D^{analytic}  =  \cE^{(D)}_{(0,-1)}(\calM_{K\bs  G})\, {\cR^4\over \sigma_3}+ \sum_{p=0}^\infty\sum_{q=0}^\infty \cE^{(D)}_{(p,q)}(\calM_{K\bs G})\, \sigma_2^p\, \sigma_3^q\, \cR^4 \,,
 \label{analytic}
 \ee
which  is   the  general  symmetric  polynomial in the Mandelstam invariants, which enter in the dimensionless combinations
 \be
 \label{sigdef}
 \sigma_n =  (s^n + t^n + u^n)\, {\ell_D^{2n}\over4^n}\,,
 \ee
 where $\ell_D$ is the Planck length in $D$ dimensions.  The factor of $\cR^4$ in~(\ref{analytic}) indicates the contraction of four powers of the Riemann curvature tensors linearised around flat space and contracted with a standard sixteen-index tensor, $t_8\, t_8$ \cite{Green:1987sp}.  
 The
 coefficient functions are  necessarily automorphic functions that are
 invariant  under the $D$-dimensional  duality group,  $G_d(\mathbb Z)$,
 appropriate to compactification  on a $d=(10-D)$-torus.  These groups
 are    listed     in    table~\ref{tab:Udual}.  They are functions of the
 symmetric  space, $\calM_{K\bs  G}$, defined  by the  moduli,  or the
 scalar fields,  of the coset space  $K\bs G$.   It is often convenient to express the analytic part of the amplitude in terms of a local one-particle irreducible effective action. 
  \begin{table}
   \centering
   \begin{tabular}{||c|c|c|c||}
   \hline
 $D$ & $G_{d}(\IR)=E_{d+1(d+1)}(\IR)$ & $K$&$G_d(\ZZ)$\\
 \hline
 10A&$ GL(1,\IR)$&1&1\\
 10B&$SL(2,\IR)$&$SO(2)$& $SL(2,\ZZ)$\\
 9&$GL(2,\IR)$& $SO(2)$& $SL(2,\ZZ)$\\
 8&$SL(3,\IR)\times SL(2,\IR)$& $SO(3)\times SO(2)$& $SL(3,\ZZ)\times SL(2,\ZZ)$ \\
 7&$SL(5,\IR)$ & $SO(5)$& $SL(5,\ZZ)$\\
 6 &$ SO(5,5,\IR)$ & $SO(5)\times SO(5)$& $SO(5,5,\ZZ)$\\
 5&$ E_{6(6)}(\IR)$& $USp(8)$& $ E_{6(6)}(\ZZ)$\\
 4 &$ E_{7(7)}(\IR)$& $SU(8)/\ZZ_2$& $E_{7(7)}(\ZZ)$\\
 3 &$ E_{8(8)}(\IR)$& $SO(16)$& $E_{8(8)}(\ZZ)$\\
 \hline
   \end{tabular}
   \caption{The   duality  groups  of   maximal  supergravity   in  $D=10-d\le 10$
     dimensions.  The groups $G_{d}(\IR)=E_{d(d)}(\IR)$ are the real split forms of rank $d+1$
     and $K$ are the maximal compact subgroups.
     In  string  theory  these  groups  are  broken  to  the  discrete
     subgroups, $G_d(\ZZ)$ as indicated in the last column.}
   \label{tab:Udual} 
 \end{table}

Although  this  paper  will  be  concerned almost  entirely  with  the
analytic part of~(\ref{ampsplit}), $A^{analytic}$, it is important to
consider its relationship to  the nonanalytic part, $A^{nonan}$.  This
part  of the  amplitude contains  the information  about  the massless
thresholds  that arise in  perturbation theory  and contribute  to the
nonlocal part of the effective action.  Such contributions include the
threshold    structure   of   supergravity    scattering   amplitudes,
and  depend  on the space-time dimension, $D$,  in a sensitive
manner.  At sufficiently high  values of $D$, a $L$-loop perturbative
contribution  in  supergravity has  ultraviolet  divergences that  are
power-behaved in a momentum  cut-off, $\Lambda$.  Such divergences are
absent in string theory and the  dependence on a power of $\Lambda$ is
replaced  by a  finite analytic  term  with a  corresponding power  of
$\ell_s^{-1}$, where  $\ell_s$ is the  string length scale. As  $D$ is
decreased it reaches  a critical value at  which supergravity develops a
logarithmic ultraviolet divergence.  Introducing a momentum cutoff now
produces a  nonanalytic factor of the  schematic form $A_{D}^{nonan}
\sim  \cR^4\,  s^k\,\log (-s/\Lambda^2)$,  which  is  replaced in  string
theory by 
\be
\label{nonanform}
A_{D}^{nonan} \sim \cR^4\, s^k\, \log (-\ell_s^2\, \mu\, s)\, ,
\ee
where  $\mu$  is a  dimensionless  scale,  which  is independent of the moduli and may be determined  by  a
detailed string loop calculation.    This  expression   is  merely
illustrative --  the detailed  dependence on the  Mandelstam variables
and  pattern  of  logarithms  is more  complicated.  For  a
discussion  of  such  effects  in  the  expansion  of  the  genus-one
contribution see~\cite{Green:2008uj}.  Of course, there is some ambiguity in how such constant terms are assigned to the analytic and non-analytic pieces since $\mu$ may
be changed to $\mu/\tilde \mu$  by adding  $\cR^4\,
s^k\  \log \tilde  \mu$ to the analytic term.
 In the subsequent discussions in this paper  our convention will be to associate all such moduli-independent logarithms with  the scale of non-analytic  $s^k\, \log(-\ell_s^2\,\mu/\tilde\mu\, s)$ contributions to the amplitude.
Furthermore, we will not  discuss the precise values of  the constant scales such as $\mu$, which can
be determined   by   explicit  string   perturbation   theory
computations,    such    as   that    carried    out   at    genus-one
in~\cite{Green:2008uj}.  As  $D$ is  decreased to values  $D<D_c$, the
nonanalytic terms are  proportional to inverse powers of  $s$, $t$ and
$u$. For $D\le 4$ the four-supergraviton amplitude possesses the standard infrared divergences of a perturbative  gravitational theory, which will not be discussed here.

  The first  term in the
 expansion~(\ref{analytic}) ($p=0,q=-1$) has coefficient $\cE^{(D)}_{(0,-1)}=3$ and is the classical
 supergravity tree-level  term, with  poles in $s$,  $t$, $u$,  and is
 determined  by  the   Einstein--Hilbert  action.   This  has  trivial
 dependence  on  the  moduli.    The  subsequent  terms  have  a  rich
 dependence   on   $\calM$   that   encodes  both   perturbative   and
 non-perturbative information.
This contrasts with supergravity, in which the continuous $G_d(\IR)$ duality symmetry is unbroken, and amplitudes are independent of the moduli. 
 The  simplest  non-trivial examples  of  automorphic  functions
 arise in the ten-dimensional IIB  theory, where the coset is $SO(2)\bs
 SL(2)$, so there is a single complex modulus, $\Omega = \Omega_1 +i
 \Omega_2$, and the duality group  is $SL(2,\mathbb Z)$.  In this case
 the first  two terms in the  expansion beyond the  classical term are
 given  by particular examples  of non-holomorphic  Eisenstein series
 for $SL(2,\mathbb Z)$ 
 \be
 \bE_{s}(\Omega) = \sum_{(m,n)\ne (0,0)} \frac {\Omega_2^s}{|m + n \Omega|^{2s}}\,,
 \label{firstterm}
 \ee
 which satisfies the Laplace equation
 \be
 \Delta_{\Omega}\, \bE_s(\Omega) \equiv \Omega_2^2\, (\partial^2_{\Omega_1} + \partial^2_{\Omega_2})\, \bE_s(\Omega) = 
 s(s-1)\, \bE_s(\Omega)\,,
 \label{laplacesl2}
 \ee
 and where $s$ is a (generally complex) index.
 Some important properties of these functions are reviewed in appendix~\ref{sec:Sl2Series}.  The Fourier expansion of $\bE_s$ in~(\ref{fourieres}) has a zero mode or ``constant term'' that consists of the sum of two powers, 
 \be
 \label{constterm}
 \int_{-\frac12}^{\frac12} d\Omega_1 \, \bE_s = 2\zeta(2s)\Omega_2^s + 2\sqrt \pi \, \frac{\Gamma(s-\half)}{\Gamma(s)}\,\zeta(2s-1) \Omega_2^{1-s}\,,
\ee 
which  correspond to a tree-level and genus-$(s-1/2)$ contribution to the interaction in string perturbation theory.  The non-zero modes correspond to exponentially suppressed $D$-instanton contributions to the interaction. 
 The  first term  of this  type  is the  lowest order  term beyond  the
 Einstein--Hilbert  term,  which is  the  $\cR^4$  interaction for  which
 $p=q=0$     and    the    coefficient   is    $\cE^{(10)}_{(0,0)}(\Omega)
 =\bE_{\threeh}(\Omega)$ that has  tree-level and one-loop perturbative
 contributions~\cite{Green:1997tv,Green:1997as}. The  next term in~\eqref{analytic}, with $p=1,
 q=0$,   corresponds to  a  $\partial^4 \cR^4$  interaction  in the  effective
 action, with a coefficient 
 $\cE^{(10)}_{(1,0)}(\Omega) =1/2\, \bE_{\fiveh}(\Omega)$ that has tree-level and two-loop contributions~\cite{Green:1999pu}.   Both the $\cR^4$ and $\partial^4\, \cR^4$ interaction coefficients can be determined by imposing constraints implied by modified supersymmetry 
transformations that incorporate higher-derivative contributions~\cite{Green:1998by,Sinha:2002zr}.

 The  next  term  has  $p=0,q=1$  and corresponds  to  the  $\partial^6  \cR^4$
 interaction.   Its coefficient $\cE^{(10)}_{(0,1)}(\Omega)$  is not
 an  Eisenstein  series~\cite{Green:2005ba},  but  satisfies  the 
 interesting  inhomogeneous  Laplace eigenvalue  equation,\footnote{We
   have  rescaled  this  interaction by  a factor  of 6
   compared to~\cite{Green:2005ba}.}
  \be
 (\Delta_{\Omega}- 12)\, \cE_{(0,1)}^{(10)}(\Omega) = - \left(\cE^{(10)}_{(0,0)}(\Omega)\right)^2  \,,
 \label{inhom}
 \ee
where the right-hand side is a source term proportional to the square of the coefficient of the $\cR^4$ interaction.
 In this case the constant term has power-behaved terms corresponding to perturbative string theory contributions at genus $0,1,2,3$, as well as exponentially suppressed contributions corresponding to an infinite set of $D$-instanton -- anti $D$-instanton pairs.

There is  a certain  amount  of information  about terms  of order $\partial^8\, \cR^4$ and higher, but these terms raise issues that go beyond the scope of this paper and  will not  be discussed
here   (see~\cite{Green:2008bf}  for
particular examples).  Our main aim will be to extend the results up to
order  $\partial^6\cR^4$ to the  higher-rank duality  groups that  arise upon
compactification to  $D$ dimensions on a  $d=(10-D)$-torus.  There has
been   some   work   in   this   direction  for   the   $\cR^4$   term
in~\cite{Green:1997as,Green:2005ba,   Kiritsis:1997em}  and   for  the
$\partial^4\cR^4$  and  $\partial^6\cR^4$  terms  in~\cite{Basu:2007ru,Basu:2007ck}.
Here we  will not only  amend these and  extend their scope,  but more
importantly, set it in  the general framework of automorphic functions
for higher-rank  groups.  Some of  our ideas overlap  with suggestions
in~\cite{Kiritsis:1997em,Pioline:1997pu,Pioline:Automorphic}        and
related  papers~\cite{Lambert:2006ny,Bao:2007er}, but  they  differ in
important respects. 

 Our  procedure, outlined in section~\ref{sec:limits}, will  be to
 constrain the  expressions for the  automorphic coefficient functions
 by  requiring them  to  reproduce the  correct  expressions in  three
 distinct degeneration limits:

 {\bf (i) The decompactification  limit from $D$ to $D+1$ dimensions.}
 When the  radius, $r_d$, of  one compact dimension becomes  large the
 part    of   the    $D=(10-d)$-dimensional    coefficient   function,
 $\cE^{(D)}_{(p,q)}$,  that leads  to a  finite term  in the  $r_d \to
 \infty$  limit  is  required  to  reproduce  the  $(D+1)$-dimensional
 coefficient function,  $\cE^{(D+1)}_{(p,q)}$.  In addition  there are
 suppressed  terms with powers  of $r_d^{-n_i}$  (where the  values of
 $n_i>0$  depend on $D$)  multiplying $\cE^{(D+1)}_{  (p',q')}$, where
 $2p'+3q'  < 2p+3q$.   There  are also  specific  terms with  positive
 powers  of  $s\,r^2_d$  that  are  necessary  to  account  for  the
 non-analytic  thresholds in $D+1$  dimensions  (see the
 discussion in~\cite{Green:2006gt}
 for more details).  The remaining terms are
 exponentially suppressed in $r_d$ and  will not be constrained in any
 direct fashion.  

 {\bf (ii) Perturbative  string theory limit.}  In the  limit in which
 the  $D$-dimensional  string  coupling  constant  becomes  small  the
 expansion of  $\cE^{(D)}_{(p,q)}$ in powers of  the $D$-dimensional  string coupling, $y_D$, 
 is  required  to  reproduce  the  known  perturbative  string  theory
 results.   In order to  make this  comparison the  contributions from
 genus-one  string theory  are derived  in appendix~\ref{sec:genusone}
 using the  methods of~\cite{Green:2008uj}.  Furthermore,  the leading
 low  energy  contribution to  $\partial^4\cR^4$  from  the genus-two  string
 theory   amplitude   compactified   on   $\calT^2$  is   derived   in
 appendix~\ref{sec:genus2}. 

 {\bf  (iii)  The semi-classical  M-theory  limit.}  In  the limit  of
 decompactification    to     eleven-dimensional    supergravity    on
 $\calT^{d+1}$ the  part of the  modular function that depends  on the
 geometric  moduli of the  torus, which  parameterise the  coset space
 $SO(d+1)\bs SL(d+1)$, should be  reproduced.  This will give the part
 of  the  coefficient function  that  transforms under  $SL(d+1,\ZZ)$.  This is the limit in which the effects of wrapped $p$-branes are suppressed and the Feynman  diagrams of  compactified
 eleven-dimensional quantum supergravity  should give a valid expansion
 in      powers     of     the      inverse     volume      of     the
 torus, $\calV_{d+1}$ ~\cite{Green:1997as,Green:1999pu,Green:2005ba,Green:2008bf}.
The  analysis of  one-loop  and two-loop  expressions  is reviewed  in
appendix~\ref{sec:sugra}. 

  As we will emphasize, our analysis of these three limits makes contact with properties of the ``constant terms''  of the  generalised Eisenstein series associated with various  parabolic subgroups of the U-duality groups~\cite{Langlands}.    This viewpoint indicates the extent of the very powerful symmetries that relate these three limits for any value of $n$.  Furthermore it gives a unified view of the relation between the theory in different dimensions by considering a
 nested set of (maximal) parabolic subgroups~\footnote{We here restrict our
   attention  to  the classical  Lie  groups  relevant to  supergravity
   theories in  $D\geq3$, although there are likely  to be interesting
   extensions to affine and hyperbolic cases~\cite{Damour:2002cu,Riccioni:2007au}.}
 \be
 E_{8(8)}\supset   E_{7(7)}\dots   \supset
 E_{1(1)}=SL(2)\,,
 \label{nest}
 \ee
 where the sequence corresponds to successive decompactifications, as outlined in point (i) above.
 We are here using the usual  economic notation for the duality groups in
 Table~\ref{tab:Udual} in which $G_d = E_{d+1(d+1)}$ refers to the real split form of the
 classical group of rank $d+1$ (and so is related to the coset for string theory compactified on a $d$-torus).
 
 In other words, we will use the explicit properties of string/M-theory in higher dimensions to constrain the particular automorphic functions that arise as coefficients in lower dimensions.  We will therefore be focussing on very special cases of the general Eisenstein series.  We will see that these particular cases have many interesting properties.

This analysis of the coefficients  in various dimensions is somewhat complicated, as well as repetitive, so the casual reader could choose to skip the details  in the bulk of the paper and read the brief summary in section~\ref{sec:briefsummary}.

The main arguments will  begin in section~\ref{rfour}, where we will describe
the   results    for   the   $\cR^4$    interaction.    The   explicit
$\cE^{(D)}_{(0,0)}$  coefficients  in  dimensions  $D\ge  6$  will  be
obtained in terms of Eisenstein series that satisfy Laplace eigenvalue
equations   on   moduli   space    space,   building   on   the   work
of~\cite{Green:1997as,Green:2005ba, Kiritsis:1997em,Pioline:Automorphic} .  The $D=8$ case
is  of interest because  it contains  the logarithmic  dependence that
encodes  the one-loop  logarithmic ultraviolet  divergence  of maximal
supergravity.  The fact that string  theory is finite is manifested by
the  cancellation  of  an  apparent divergence,  subject  to  suitable
regularisation.  This arises because $\cE^{(8)}_{(0,0)}$ is the sum of
two Eisenstein  series that each have  poles in the  parameter $s$ at
appropriate values of $s$.   A suitable analytic continuation leads to
a cancellation of the poles  in these two terms, leaving a logarithmic
dependence on a modulus that can be identified with the logarithm that
arises  in   the  low  energy  supergravity   limit.   Formally  these
considerations  extend to  lower  dimensions $D\ge  3$,  in which  the
duality groups are those in  the $E_{d+1(d+1)}$ sequence, where $d= 10-D$.
In all  cases these series  are finite, despite apparent  poles, which
cancel leaving crucial logarithmic  dependence on moduli that are also
expected for a consistent string theory interpretation.
   
In section~\ref{sec:dfourrfour} this analysis  will be extended to the
$\partial^4   \cR^4$   interaction,    for   which   the   coefficients   are
$\cE^{(D)}_{(1,0)}$.        Building       on      the       analyses
in~\cite{Green:2005ba,Basu:2007ru} we  will first discuss  the $D=9,8$
cases.   The  $D=7$  expression   will  then  be  analyzed.   This  is
particularly  interesting since it  reproduces the  two-loop logarithm
characteristic of the  ultraviolet divergence of maximal supergravity~\cite{Bern:1998ug}.
In order to satisfy the conditions  (i)-(iii) we are led to a specific
combination  of two  Eisenstein series  for $SL(5)$.   As  before, the
precise  combination  of  Eisenstein  series  is  one  for  which  the
divergent  pole terms  cancel, reflecting the absence of ultraviolet divergences in string theory.  
The  analysis of  the $D=6$  case with
duality group $SO(5,5)$ will be left for the discussion in section~\ref{sec:briefsummary}, since our analysis is incomplete. In this case we make strong use of results for constant terms of
Eisenstein series  by Stephen  Miller\footnote{We are very  indebted to
  Stephen  Miller  for many  illuminating  discussions concerning  the
  general structure  of Eisenstein series and their  specific form for
  the cases of interest to us.}   and is not as complete.  There is no
obvious  obstacle to  the extension  to  $D <  6$ higher-rank  duality
groups, although this will not be discussed in this paper. 
 
  Section~\ref{sec:dsixrfour} concerns the  $\partial^6 \cR^4$ interaction in
  $D=9$, $8$  and $7$  dimensions.  To some  extent the  $D=8,9$ cases
  overlap with  the analysis in~\cite{Basu:2007ck},  demonstrating how
  the Laplace equation  with a source term generalizes  for the larger
  duality groups.  In  each case the source term is  the square of the
  $\cR^4$ coefficient,  $\cE^{(D)}_{(0,0)}$.  In $D=8$  this source possesses
  both $\log$ and $(\log)^2$ terms  that are required for the solution
  to have requisite  interpretation in the low energy  limit of string
  theory.    For  example,   maximal  supergravity   has   a  two-loop
  logarithmic ultraviolet  divergence multiplying $\partial^6\cR^4$,  as well
  as a  logarithmic contribution from the  one-loop $D=8$ counterterm,
  which are reproduced by our modular coefficients.

  Section~\ref{sec:briefsummary}  will  summarize  our  results and describe some issues relating to the extension to
  higher-rank groups and higher derivative interactions.  In particular, we will summarise in a compact manner the set of homogeneous and inhomogeneous  Laplace eigenvalue equations satisfied by the coefficient functions for values of $D$ discussed in this paper, but which we argue should be valid   in any dimension in the range $3\le D\le 10$.
  We will also make comments about the form of certain coefficients in $D\le 6$ dimensions.
 
   Technical details are given in several appendices.

 \section{ Degeneration limits and Eisenstein series for parabolic subgroups}
 \label{sec:limits}
 \begin{figure}[ht]
 \centering\includegraphics[width=14cm]{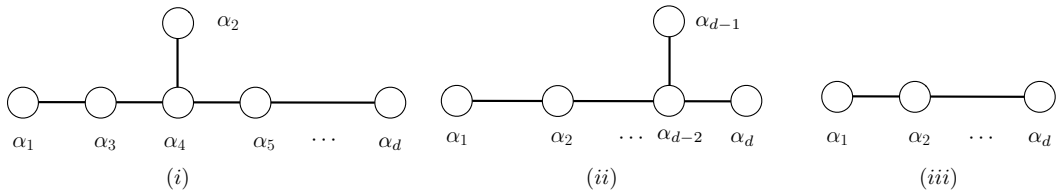}
 \caption{\label{fig:dynkin} The  Dynkin diagrams relevant  to: (i) the $E_{d(d)}$ ($d\le 8$)
   type II duality groups of type II string theory compactified to $D=11-d$ dimensions 
   on a $(d-1)$-torus.  Successive decompactifications to higher dimensions
   are obtained by deleting the nodes $\alpha_d, \alpha_{d-1}\ \dots$ in~(i);
   (ii) The  T-duality groups $SO(10-D,10-D)$ obtained  by deleting the
   left node $\alpha_1$ of~(i) are the symmetries of string perturbation theory in $D$ dimensions; 
   (iii) The $SL(11-D)$ groups obtained by deleting node $\alpha_2$ in~(i) are  associated  with    the   geometric
 compactification    of    eleven-dimensional    supergravity   on    a
 $(11-D)$-torus.}  
 \end{figure}

 The duality  groups of maximally  supersymmetric closed-string theory
 are   associated   with   the    series   of   Dynkin   diagrams   in
 figure~\ref{fig:dynkin}(i) that  may be obtained  from the $E_{8(8)}$
 diagram by  deleting the  right nodes in  a sequential  manner.  This
 generates the diagrams for the $E_{d(d)}$ series.  In terms of string
 theory  compactified on  a $d$-torus,  $\calT^d$, the  deletion  of a
 right    node   labelled    $\alpha_{d+1}$    corresponds   to    the
 decompactification of a radius,  $r_d\to \infty$ ($d\ge 2$).  This is
 the degeneration  limit (i)  of the previous  section.  The  limit of
 small string coupling, or  string perturbation theory, corresponds to
 deleting the left node labelled $\alpha_1$.  This is the degeneration
 limit (ii) and gives a series of terms with symmetry $SO(d,d)$ (where
 the   right   node   is   again   $\alpha_{d+1}$).    The   $\calT^d$
 compactification of string theory  may be viewed as the $\calT^{d+1}$
 compactification of eleven-dimensional  M-theory.  The limit (iii) is
 one  in which  the M-theory  volume of  $\calT^{d+1}$  becomes large,
 $\calV_{d+1} \to  \infty$, in which  semiclassical eleven-dimensional
 geometry is a good approximation  and the duality symmetry reduces to
 $SL(d)$.  This is the degeneration limit in which the node $\alpha_2$ in  figure~\ref{fig:dynkin}(i) 
 is deleted.

\subsection{Parabolic subgroups}
\label{sec:minim-maxim-parab}

 Parabolic   sub-algebras   of   a   semisimple   Lie   Algebra   $\fr
 g=\textrm{Lie}(G)$ with $\fr h$ a Cartan sub-algebra are defined as follows~\cite{Bourbaki,TauvelYu}. 
 If $\Delta$ is  the set of simple roots (a basis
 of roots) and $R^+$
 the set of positive roots spanned by $\Delta$. Then $\fr b=\fr
h+\oplus_{\alpha\in R^+}\,\fr g_\alpha$,  where  $\fr  g_\alpha$  is   the  root  space
 associated with the root $\alpha$, is the associated Borel sub-algebra.
 Consider a partition of the positive root space $\Delta$ into disjoint  sets $\Delta_1$
 and $\Delta_2$ so  $\Delta=\Delta_1\sqcup \Delta_2$. We define, $R_1$  the set  of
 positive roots  spanned by $\Delta_1$  and $R_2$ the set  of positive
 roots spanned by $\Delta_2$.  Define

\begin{equation}
\fr p_{\Delta_2}=\fr h+\bigoplus_{\alpha\in R^+\cup (-R_1)}\fr g_\alpha,\qquad
\fr l_{\Delta_2}=\fr h+\bigoplus_{\alpha\in R_1\cup (-R_1)} \fr g_\alpha,\qquad
\fr n_{\Delta_2}=\bigoplus_{\alpha\in R_2} \fr g_\alpha,
\end{equation}
This defines the parabolic  sub-algebra $\fr p_{\Delta_2}$ associated with
the set of positive roots $R_1$, $\fr l_{\Delta_2}$ is its Levi factor and
$\fr n_{\Delta_2}$ the unipotent radical.
Clearly if $ \Delta_2\subset \hat \Delta_2$ then $\fr p_{\hat\Delta_2}\subset \fr p_{  \Delta_2}$.

\medskip
\noindent{$\bullet$}
When $\fr p_{\Delta}=\fr b$,  $R_2$ is the set
of all the positive roots (and $R_1=\emptyset$)  the associated parabolic is {\bf the minimal parabolic sub-algebra.}

\medskip
\noindent{$\bullet$}
When $\fr p_{\emptyset}=\fr
g$  (equivalently when $R_2=\emptyset$), $R_1$ is the set
of all  the positive roots the  associated parabolic sub-algebra
is the Lie Algebra $\fr g$.

\medskip
\noindent{$\bullet$} {\bf
 Maximal  parabolic sub-algebras}  different from  $\fr g$  are  defined by
 singling out one simple root $\alpha_i$ and taking $\Delta_2=\{\alpha_i\}$.
 We  denote  the  maximal  parabolic   sub-group  by
   $P_{\alpha_i}$, with ${\rm rank}\ P_{\alpha_i} = {\rm rank}(G)-1$.

\medskip \noindent{$\bullet$} The (standard) parabolic subgroup  of
$GL(n)$ is defined as the group of matrices of the form, for $n=n_1+\cdots+n_q$,
\be
P(n_1,\dots,n_q)  = \begin{pmatrix}
  U_1 & *&* \\
  0  &\ddots&*\\
  0&0 &U_q
\end{pmatrix} , \qquad \textrm{where~} U_i\in GL(n_i)\,,
\ee
which can be factored in the form
\be
P(n_1, \dots, n_q) = L(n_1, \dots, n_q)\, N(n_1, \dots ,n_q)\,.
\label{e:pmndef}
\ee
Here 
\begin{equation}
\label{e:ParaDef} 
N(n_1, \dots, n_q) = \begin{pmatrix}
  I_{n_1} & *&* \\
  0  &\ddots&*\\
  0&0 &I_{n_q}
\end{pmatrix}\,   \qquad \textrm{where~}  I_n=\textrm{diag}(1,\dots,1)
\ee
is the unipotent radical and 
\be
L(n_1, \dots, n_q) = 
\begin{pmatrix}
  U_1 & 0&0 \\
  0  &\ddots&0\\
  0&0 &U_q
\end{pmatrix} \,,
\end{equation}
is the Levi component.  
The minimal parabolic subgroup is given by $P(1,\dots,1)$.
A given maximal parabolic subgroup has a characteristic pattern of zeroes in the upper off-diagonal elements of $N$.  For example,  the $SL(3,\IR)$ maximal parabolic subgroup~\cite{MillerThesis},
\be
P(1,2) =  \begin{pmatrix}
  * & *&* \\
  0  &*&*\\
  0&* &*
\end{pmatrix}
 \label{e:sl3p12}
 \ee
 has a unipotent radical of the form
 \be
N(1,2) =  \begin{pmatrix}
  1 & \nu_1&\nu_2 \\
    0&1&0\\
  0&0 &1
\end{pmatrix}\,,
 \label{e:sl3n12}
 \ee
 where  $\nu_1$  and $\nu_2$  are  real  angular  variables.

 Three cases will be of particular interest in this paper.  These concern the
maximal parabolic subgroups given in the table~\ref{tab:Parabolics}, which are obtained by deleting the 
 left node,  the right node and  the upper node of the Dynkin diagrams shown in fig.~\ref{fig:dynkin}.
 \begin{table}[h]
   \centering
 \begin{tabular}[]{|c|c|c|c|c|c|c|}
 \hline
 deleted~node&$E_8$&$E_7$&$E_6$&$E_5=D_5$&$E_4= A_4$&$E_3=A_2A_1$\\
 \hline
 left & $D_7$& $ D_6$&$ D_5$& $ D_4$
 & $ D_3$& $ D_2$\\
 upper&$ A_7$&$ A_6$&$ A_5$&$  A_4$&$ A_3$&$ A_2$\\
right&$       E_7$&$       E_6$&$      D_5$&$
  A_4$&$ A_2A_1$&$A_1A_1$\\
 \hline
 \end{tabular}
   \caption{Maximal Parabolic subgroups  of $E_{d(d)}$ arising in string
     theory are of the form $GL(1)\times X_{d-1}$, where the rank-$(d-1)$ subgroups are listed. We use the notation $A_d=SL(d+1)$, $D_d=SO(d,d)$.
Each   parabolic   subgroup  can   be   decomposed   as  the   product
 $P_\alpha=N_\alpha L_\alpha$ of a
 unipotent radical $N_\alpha$ and a Levi factor $L_\alpha$.  The Levi factors determine the Lie  groups  generated by  the remaining nodes of the Dynkin diagram, which are listed in the table.}
   \label{tab:Parabolics}
 \end{table}

 There are several interesting coincidences.  
 \begin{itemize}
 \item  In $D=7$, where the U-duality group is $E_{4(4)}=SL(5)$,  the symmetry group of string perturbation theory is  $SL(4) = SO(3,3)$, which is also the symmetry of M-theory on $\calT^4$ in the decompactification to eleven dimensions.

 \item $E_{5(5)}=SO(5,5)$  arises in the $D=6$ theory, for  which the 
   group $SL(5)$ arises  both as the symmetry of  M-theory on $\cT^5$
   limit and as the U-duality group upon
   decompactification  to $D=7$.

 \item $SO(5,5)$ arises  both as the symmetry of  string perturbation  theory in the
   $D=5$ theory and as the  decompactification limit to the $D=5$
   theory, which has duality group $E_{5(5)}$.

    \item  $E_{6(6)}$ arises as the U-duality group  in $D=5$ and is symmetric under the interchange of nodes $1$ and $6$.  This symmetry interchanges the limit of decompactification to $D=6$ with the perturbative string theory limit.
       \end{itemize}

 \subsection{Eisenstein series for maximal parabolic subgroups and their constant terms.}

 The general Eisenstein  series    are automorphic functions  of $d$
 complex parameters, $s_i$ ($i=1, \dots, d$) associated with different
 parabolic subgroups  of the $E_{d(d)}$ groups.   Their definition may
 be  found  in~\cite{Harish,  Langlands}  and is briefly reviewed in appendix~\ref{sec:Slnseries}.
The construction of the minimal parabolic $SL(d)$ series, is also described
 in  appendix~\ref{sec:Slnseries},  based  closely on  notes  by Stephen
 Miller  and  extensions   of~\cite{MillerThesis}.   
 
 However,  we  are here 
 primarily   interested  in  very   special  cases   corresponding  to
 Eisenstein series for {\it maximal} parabolic subgroups, defined with
 respect  to  one particular  node  associated  with  the simple  root
 $\alpha_u$.  Such a series may  be obtained by taking residues of the
 minimal parabolic series  on the poles at $s_i=0$  for all $i$ except
 $i=u$, so  the series depends  on only one parameter,  $s\equiv s_u$.
 The series can be  indexed by the Dynkin label $[0^{u-1},1,0^{d-u}]$,
 where the  $1$ is in the  $u$'th position.  The  particular values of
 $u$ of  interest to us  will be determined  on a case by  case basis.
 Such a series for a maximal  parabolic subgroup of the group $G$ will
 be denoted $\bE^{G}_{[0^{u-1},1,0^{d-u}]; s}$.  
 
The simplest example is provided by the $SL(d)$   series with $u=1$ (the Epstein zeta function), which can be expressed as a sum over a single integer-valued $d$-component vector,
 \be
 \bE^{SL(d)}_{[1, 0^{d-2}]; s} = \sum_{m^i\in\mathbb Z^d\bs\{0\}} ( m^i g_{ij} m^j)^{-s}\,,
 \label{epsteindef}
 \ee
  where the sum is over all values of $m^i$ with the value $m^1=m^2 =\dots =0$ omitted.  The metric $g_{ij}$ is the metric on $SO(d)\bs SL(d)$.  Our conventions for labelling the $SL(d)$ Dynkin diagrams are shown in figure~\ref{fig:dynkin}(iii). A less trivial case that we will also need to consider is the $SL(d)$ Eisenstein series with $u=2$, which is given by
 \be
 \bE^{SL(d)}_{[0,1, 0^{d-3}]; s} = \sum^\prime_{m^i,n^i\in\mathbb Z^d} (m^{[i}n^{j]} g_{ik} g_{jl} m^{[k} n^{l]})^{-s}\,,
 \label{alphatwo}
 \ee
 where $\sum^\prime$  here indicates the sum is  over integers subject
 to  the constraint  that at  least one minor $\delta^{[ij]} =  m^{[i}n^{j]}$ is
 non-zero.   For $SL(3)$ this  series is  proportional to  the Epstein
 series,~(\ref{epsteindef}) with a shifted value of $s$, as we show in
 appendix~\ref{sec:Sl3Series}.  More generally, the $SL(d)$ series 
$\bE^{SL(d)}_{[0^{d-2}, 1]; s}$ is proportional to the Epstein series with a shifted value of $s$,  a simple consequence of the symmetry under $s \to d/2 -s$, which follows from  the Weyl symmetry of the weight lattice of $SL(d)$.    Some relevant properties of the $SL(d)$  series are deduced in appendix~\ref{sec:Slnseries}.

The other cases  that will be considered explicitly  in this paper are
particular cases  of Eisenstein series for  $SO(d,d)$.  In particular, these symmetries arise as T-duality groups of string perturbation theory in $10-d$ dimensions, and $SO(5,5)$ is  the full U-duality group  for  $D=6$.
We   will  discuss  the  maximal  parabolic
Eisenstein series of the form $\bE^{SO(d,d)}_{[1,0^{d-1}];s}$, where the
distinguished    node     is    the    one    on     the    left    in
figure~\ref{fig:dynkin}(ii)  --   i.e.,  associated  with   the  vector
representation.  A  number of properties of these  series are obtained
in  appendix~\ref{sec:Dnseries}  based   on  a  novel  representation
motivated  by  compactified   two-loop  Feynman  diagrams.  Although the series with more general Dynkin indices are relevant, we will not discuss them in this paper.  

\medskip
\noindent{\bf Constant terms.}
\smallskip

The  three   degeneration  limits~(i),~(ii)  and~(iii)   that  we  are
interested in  correspond to decompositions of  the Eisenstein series,
$\bE^{G}_{[0^{u-1},1,0^{d-u}];   s}$,   with   respect  to   parabolic
subgroups of the form, $P_v  \equiv GL(1) \times G_v$, associated with
one of  three distinct  nodes, $\alpha_v$, of  the Dynkin  diagram, as
described  earlier.  The  $GL(1)$ factor  is parameterised  by  a real
parameter $r$, which corresponds in limit (i) $(v=d)$ to the radius of
the  compact dimension,  $r_d$, in  limit (ii)  ($v=1$) to  the string
coupling in $D$  dimensions, $y_D$, and in limit  (iii) ($v=2$) to the
volume of the M-theory torus, $\calV_{11-D}$.  
 In considering  these limits  we will retain  all the terms  that are
 power behaved in $r$.  These are contained in the \lq constant terms'
 obtained  by  taking  the  zero  Fourier mode  with  respect  to  the
 components of the unipotent radical, $N_v$, associated with the parabolic subgroup $P_{\alpha_v}$ (defined in section~\ref{sec:minim-maxim-parab}).  This  is an integral over the entries, $\nu_i$, in the upper triangular matrix, $N_v$
\begin{equation}
A^G_s(u,v; g)  = \int_{N_v/G(\mathbb  Z)\cap N_v}   dn \,
\bE^{G}_{[0^{u-1},1,0^{d-u}]; s}(gn)\, ,
\label{constdef}
\end{equation}
where $dn = \prod_i d\nu_i$ is the Haar measure on $N_v$.
In order to avoid complicated notation, we
will replace $\int_{N_v/G(\mathbb Z)\cap  N_v} dn$ by   $\int_{P_v}$ so that
\begin{equation}
  A^G_s(u,v; g) \equiv \int_{P_v}\, \bE^{G}_{[0^{u-1},1,0^{d-u}]; s} \,.
\end{equation}
The angular integral~(\ref{constdef})  generalizes the $SL(2,\ZZ)$ case of~(\ref{constterm}).   The constant terms are expansions in powers of $r$ with coefficients that are Eisenstein series (or products of Eisenstein series, in the non-simple case) of the schematic form
\be
\label{expandcon}
A^G_s (u,v;g)=    \sum_i c^i_{uv}\, r^{p_i}\, \bE^{G_v}_{[\dots]_i; s_i}\,,
 \ee
 where the  values of  the parameters $s_i$,  $p_i$ depend on  $u$ and
 $v$, and $r$ is a  scale factor associated with the $GL(1)$ subgroup.
 This integration  projects out the  non-zero modes of  the Eisenstein
 series, which  are non-perturbative in $r$ and exponentially suppressed in the appropriate degeneration limit.  The coefficients $\calE^{(D)}_{(0,1)}$ of
 $\partial^6\cR^4$  are not  Eisenstein series  and their  constant  terms do
 contain    exponentially   suppressed    pieces    corresponding   to
 instanton---anti--instanton pairs.

The Eisenstein series for other maximal parabolic $SO(d,d)$ series, as well as those for the higher-rank $E_{d(d)}$ groups, are much more difficult to construct in terms of explicit sums over integers but their explicit properties can be obtained from their basic definition given in~(\ref{e:minparab}). Starting from that definition, the constant terms of their parabolic subgroups have been derived in~\cite{MillerNotes}, which is likely to be of use in developing these ideas further.

 \subsection{The expansion parameters.}\label{sec:unit}
 
  In considering  M-theory on a $(d+1)$-dimensional  torus, 
 $\calT^{d+1}$, length scales are measured in units of the eleven-dimensional Planck length, $\ell_{11}$, whereas for string theory compactified on a $d$-dimension  torus,  $\calT^d$, scales are measured in units of the string length,  $\ell_s$,  or the ten-dimensional
 Planck length scales of the IIA and IIB theories,  $\ell_{10}^A$, $\ell_{10}^B$.
 These length scales are related by the well-known relations,
 \be
   \label{e:unit}
   \ell_{11}= g_A^{1\over 3}\, \ell_s\,,\qquad
 \ell_{10}^A =g_A^{{1\over4}} \,\ell_s \,,\qquad
 \ell_{10}^B =g_B^{{1\over4}} \,\ell_s \,,
 \qquad R_{11}= g_A\,\ell_s\,,
 \ee
 where  $g_A$ and $g_B$ are the type~IIA and IIB coupling constants and $R_{11}$ is the radius of the extra M-theory circle.

Compactifying from $10$ to $D=10-d$ dimensions on $\calT^{d}$ leads to the relations
  \begin{equation} 
 \ell_{D}^{D-2}=y_D\, \ell_s^{D-2}\,,
  \end{equation}
 where 
  the quantity $y_D$   is defined by the $(10-d)$-dimensional T-duality invariant  dilaton,  which defines the $D$-dimensional coupling,
 \be
 y_{10-d}\equiv  e^{2\phi_D}= \frac{g_A^2\ell_s^d}{V^A_d}  =  \frac{g_B^2\ell_s^d}{V_d^B} \,,
 \label{diladef}
 \ee
 where $V^A_d$ is the volume of the $d$-torus in IIA string units while $V^B_d$ is the volume in IIB units.  
Note further that he relation between the Planck length in $D$ dimensions and $D+1$  dimensions is
\begin{equation}
    \ell_{D+1}^{D-1}=               \ell_{D}^{D-2}\,
  \, r_d\, ,
\end{equation}
where $r_d$ is the radius of the $(d=10-D)$'th direction of $\calT^d$ in IIB string units. 

The parameters that we will use to define the three degeneration limits will be the following.

 (i)  The decompactification of a single dimension is given by  the limit $r_d/\ell_s \to   \infty$ in the string frame. 
 We will be interested in expressing the result in the Einstein frame in $(D+1)$ dimensions at fixed coupling,  in which case we will need to consider $r_d/\ell_{D+1} \to \infty$ with $y_{D+1}$ fixed.
 It will also  be  useful to introduce the  U-duality invariant quantity defined in terms of the dimensionless volume of the string theory $d$-torus, 
 \begin{equation}\label{nudef}
 \nu_d^{-\half} = \frac{1}{\ell_{10}^d}V^B_d \, ,
 \end{equation}
where we have set $\ell_{10}^B \equiv \ell_{10}$ in this and all subsequent expressions since we will not need to use $\ell_{10}^A$.   It is easy to deduce the useful relations
 \be
 \frac{r_d}{\ell_{D+1}}    ={r_d\over\ell_s}\, y_{D+1}^{-{1\over D-1}}    =\nu_d^{-\frac12}\,     \nu_{d-1}^{D\over
   2(D-1)} \,.
    \label{decomtwo}
 \ee  

 (ii)  String perturbation theory is an expansion in powers of the $D$-dimensional string coupling, $e^{\phi^B_D} \equiv y_D^{1/2}$ when $y_D\to 0$.  

 (iii)   Decompactification to semiclassical eleven-dimensional supergravity arises in the limit of large volume of the  $(d+1)$-dimensional M-theory torus. This volume,
 $\calV_{d+1}$, is defined by
 \begin{equation}
   G_{M\,IJ}=\cV_{d+1}^{2\over d+1}\,\tilde G_{M\,IJ}\,,
  \label{gequalone}
   \end{equation}
 where $G_{M\,IJ}$ ($I,J = 1,\dots, d$) is the M-theory metric on $\calT^{d+1}$ and   $\tilde G_{IJ}$ has unit determinant.  The dimensionless volume, $\hat \calV_{d+1}$,  can be expressed as 
 \begin{equation}
   \hat\cV_{d+1}\equiv    \frac{1}{\LP^{d+1}}\cV_{d+1}=   \frac{1}{\LP^{d+1}} \sqrt{\det(G_{M\,IJ})} =  g_A^{2-d\over3}\,
   {V^A_d\over \ell_s^d}\,.
 \end{equation}
   This can be converted to type-IIB units by compactifying one dimension  of radius $r_A$ so that $V^A_d=  r_A\times V_{d-1}$
 and introducing the volume $V^B_d= r_B \times V_{d-1}$, where $r_B=\ell_s^2/r_A$, giving
  \begin{equation}\label{e:typeIIbunit}
   \hat\cV_{d+1}=    g_A^{2-d\over3}\,
   {V^A_d\over      \ell_s^d}     =     g_B^{2-d\over3}\,\left(r_B\over
     \ell_s\right)^{d-8\over3}\, {V^B_d\over \ell_{s}^d}= \left(r_B\over
     \ell_{10}\right)^{d-8\over3}\, {V^B_d\over \ell_{10}^d}\,.
 \end{equation}
 The M-theory decompactification limit is given by the limit $\hat\cV_{d+1} \to \infty$.

 \section{The $\cR^4$ interaction} \label{rfour}
  The   first  term   in  the   low  energy   expansion   of the maximally
  supersymmetric string theory amplitude beyond the tree-level term is the
  $\cR^4$ term in \eqref{analytic}, which is described by a term in the effective action of the form
  \be
  \label{rfoureffect}
 S_{R^4}=  \ell_{D}^{8-D}\, \int d^{D}x\, \sqrt{-G^{(10)}}\, \calE^{(D)}_{(0,0)}\, \cR^4\, .
  \ee
  
In $D=10$ dimensions the coefficient function is given by  \cite{Green:1997tv}
 \begin{equation}
\cE^{(10)}_{(0,0)} (\Omega)=  \bE^{SL(2)}_{[1];\threeh} (\Omega) \,,
 \label{e:R410D}
 \end{equation}
which is the standard Eisenstein series for $SL(2,\mathbb Z)$, that is conventionally denoted $\bE_{\threeh}(\Omega)$\footnote{We will follow the convention of writing $\bE^{SL(2)}_{[1];s}$ as $\bE_s$.} and satisfies the Laplace equation
\begin{equation}
  \label{e:LaplaR410D}
  \Delta^{(10)} \cE^{(10)}_{(0,0)}= {3\over4} \,\cE^{(10)}_{(0,0)}\,,
\end{equation}
where $\Delta^{(10)}$ is the $SO(2)\bs SL(2)$ Laplace operator,
\begin{equation}
  \label{e:LaplaSL2}
  \Delta^{(10)} \equiv\Omega_2^2\,(\partial^2_{\Omega_1}+\partial^2_{\Omega_2})\,.
\end{equation}
The string frame expression for this interaction involves the identification 
 \begin{equation}
 {1\over\ell_{10}^2}\cE^{(10)}_{(0,0)}(\Omega)= {1\over\ell_{s}^2} \Omega_2^{\half} \, \bE_{\frac{3}{2}}(\Omega)\,,
 \end{equation}
using the relation between the ten-dimensional Planck length and
the string scale $\ell_s= \ell_{10} \, \Omega_2^{1/4}$.   The perturbative expansion is associated with the constant term,
\be
{1\over\ell_{10}^2}\,\int_{-\half}^\half d\Omega_1\, \calE^{(10)}_{(0,0)}(\Omega) =  \frac{1}{\ell_s^2}\, \left(\frac{2\zeta(3)}{y_{10}}+ 4 \zeta(2)\right)\,,
\label{tenrfour}
\ee
where $y_{10}= g_B^2$.  This exhibits a tree-level term and a one-loop term.

We   will  here  discuss  the
  theory after compactification on $\calT^d$ for $d=1,2,3,4$.  In each
  case we will  present a candidate expression and  verify that it has
  the correct properties in the three degeneration limits described in section~\ref{sec:intro}.  Several aspects of this discussion reproduce earlier
  work, but our analysis will stress the framework that generalizes to
  other terms in the low energy expansion and to the larger U-duality groups. 

 \subsection{Nine dimensions}
The  coefficient  function  in the nine-dimensional effective action (\eqref{rfoureffect} with $D=9$) was  determined  in~\cite{Green:1997tv,Green:1997as} to be  
  \begin{equation}\label{9dR4}
 \mathcal{E}_{(0,0)}^{(9)}  =\nu_1^{-{3\over7}}\bE_{3\over2}(\Omega)+4\zeta(2)\,\nu_1^{{4\over7}}\, ,
 \end{equation}
 with $\nu_1   =  (r_B/\ell_{10})^{-2}$,  which  is   invariant  under  the
 U-duality group $SL(2,\mathbb Z)$. 
 This coefficient function can straightforwardly be seen to satisfy the $SO(2)\bs
 SL(2)\times \IR^+$ Laplace eigenvalue equation,
 \begin{equation}
   \label{eq:Diff9dR4}
   \big(\Delta^{(9)}-{6\over 7} \big)\, \mathcal{E}^{(9)}_{(0,0)} = 0\,,
 \end{equation}
 where the Laplace operator for the nine-dimensional compactification has the form given in~(\ref{e:9DLapla}),
 \begin{equation}\label{e:9}
 \Delta^{(9)} \equiv \Delta_\Omega
  +{7\over 4} \nu_1 \partial_{\nu_1}(\nu_1 \partial_{\nu_1}) +
 {1\over 2} \nu_1 \partial_{\nu_1 }\, .
 \end{equation}
 In order to see how the action behaves in various limits we write $\nu_1$ in terms of the other parameters as
  \begin{equation}
   \label{e:cVtonu}
 \nu_1 =  \hat\cV_2^{\threeh}\,,
 \end{equation}
 or
 \begin{equation}
   \label{eq:1}
   \nu_1=    \Omega_2^{-\frac12}   \,\left(\ell_s\over   r_B\right)^2= y_9^{\quart}\, \left(\frac{\ell_s}{r_B}\right)^{\frac{7}{4}}\,,
 \end{equation}
 where $y_9= \ell_s/(\Omega_2^2\,r_B)$, or
 \begin{equation}
 \nu_1 = \left(\frac{\ell_9}{\ell_{10}}\right)^{14} =    \left(\frac{r_B}{\ell_{10}}\right)^{-2}  \,.
   \label{e:ell9def}
 \end{equation}

 We will now review the manner in which the expression~(\ref{9dR4})  reproduces the expected expressions in the three degeneration limits of interest.

 \medskip
 {\bf (i)  Decompactification to $D=10$}
\smallskip

 This limit is  obtained by
 letting  $r_B/\ell_{10} \to  \infty$ in~(\ref{9dR4}) 
 \begin{equation}
   {1\over   \ell_9}  \cE^{(9)}_{(0,0)}=   {r_B\over  \ell_{10}^2}
   \cE^{(10)}_{(0,0)} +{4\zeta(2)\over r_B}\,.
 \end{equation}
 The term proportional to $r_B$ survives the limit to give the $D=10$ expression~(\ref{e:R410D}).

\medskip
{\bf  (ii) $D=9$ perturbative string theory.}
\smallskip

The perturbative expansion of~(\ref{9dR4})  in the string frame   is given by evaluating the constant term,
 \begin{equation}
 {1\over\ell_9}\,\int_{-\frac12}^{\frac12}d\Omega_1\,\cE^{(9)}_{(0,0)}={1\over
   \ell_s} \, \left( { 2\zeta(3)\over y_9} + 4\zeta(2) \,\left({r\over\ell_s}+{\ell_s \over r}\right) \right)\,,
\end{equation}
where $y_9=g_B^2\,\ell_s/r_B=g_A^2\,\ell_s/r_A$ is invariant under T-duality and $r=r_B$  or $r_A$ (where $r_B = \ell_s^2/r_A$). This expression is manifestly invariant  under $r\to \ell_s^2 /r$, as expected at this order in string perturbation theory\footnote{The IIA and IIB four-graviton scattering amplitudes are known to be equal up to at least genus-four \cite{Berkovits:2006vc}.}. The coefficients are the same as  those obtained directly from tree-level and one-loop string scattering amplitudes.

 \medskip
 {\bf (iii)   Semiclassical M-theory limit}
 \smallskip
 
 The coefficient~(\ref{9dR4}) is expressed in eleven-dimensional M-theory units by 
 \begin{equation}\label{e:MthR4}
{1\over \ell_9}\, \cE^{(9)}_{(0,0)}={1\over \ell_{11}}\, \left(\hcV_2^{-{1\over 2}} \bE_{3\over      2}(\Omega)+ 4\zeta(2)\,\hcV_2\right)\,.
 \end{equation}
 This expression coincides with that obtained by evaluating the one-loop contribution of eleven-dimensional supergravity compactified on $\calT^2$~\cite{Green:1997as}.  This calculation has a $\Lambda^3$ divergent piece (where $\Lambda$ is a momentum cutoff)  that is regularised by adding a counterterm, $c\,\cR^4$, where the value of $c=4\zeta(2)$ is determined by imposing the equality of the IIA and IIB one-loop contributions~\cite{Green:1997as}.
 Furthermore there  are no higher-loop corrections to  $\cR^4$, so the
 result~(\ref{9dR4})  is exactly  given by  the supergravity
 expression. 

 \subsection{Eight dimensions}

The effective action of the form \eqref{rfoureffect} with $D=8$ was considered in~\cite{Green:1997as,Green:2005ba}, based on evaluation of
 the   contribution   of   one-loop  eleven-dimensional   supergravity
 compactified on $\calT^3$.  This takes into account the
 effect of super-supergravitons winding around the torus and has a manifest
 invariance under the modular group of the three-torus, $SL(3,\mathbb Z)$.
 This was  completed to the full duality  group $E_{3(3)}=SL(3)\times SL(2)$
 by  extending  the  expression  to include the  effects  of  wrapped
 $M2$-branes, giving
  \begin{equation}\label{8dR4}
   \cE^{(8)}_{(0,0)}=  \hbE^{SL(3)}_{[10];\threeh}+2\hat \bE^{SL(2)}_{[1];1}\, ,
 \end{equation}
which is the form presented in~\cite{Kiritsis:1997em}.
 The   expressions  $ \hbE^{SL(2)}_{[1];1}  \equiv \hbE_{1}$  and   $\hbE^{SL(3)}_{[10];\threeh}$  are
 regularised Eisenstein series  (specifically, Epstein series) for the
 groups $SL(2)$  and $SL(3)$, respectively\footnote{The series
   $\hbE^{SL(3)}_{[10];s}$   was  denoted  $\hbE^{SL(3)}_{{\bf   3};s}$  in~\cite{Pioline:Automorphic}.}.   Some  properties  of these  series  are discussed in  appendix~\ref{sec:Slnseries} and may be summarised as follows.
 The series $\bE^{SL(2)}_{[1];s}=\bE_s$ and  $\bE^{SL(3)}_{[10];s}$ have poles at $s=1$ and
 $s=3/2$,   respectively,   which   correspond  to   the  presence of  logarithmic
 singularities in the one-loop  graviton scattering amplitude in $D=8$ 
 dimensions --  which may be expressed as  poles in $\epsilon$ in  dimensional regularisation, 
 where $D=8+2\epsilon$. The hat 
 \ $\hat{}$ \ indicates that the  pole part is subtracted, leaving only
 the finite  part. 
  
  The  Eisenstein series $\bE^{SL(3)}_{[10]; s}$  is a
 special case of the most general  minimal parabolic Eisenstein series
 for $SL(3)$ and is discussed in \eqref{e:DefMil}.  The
 general series  has two  parameters, $s_1$ and $s_2$,  corresponding to  the non-compact
 Cartan directions of  the quotient $SO(3)\bs SL(3)$,  but the series of interest
 here  has  $s_1=s$,  $s_2=0$.   
 Appendix~\ref{sec:Sl3Series} provides more details concerning this series, which is defined by  \eqref{e:Efund} 
 in the case $d=3$.  
 The expression for the series $\bE^{SL(3)}_{[10]; s}$ in \eqref{eEm} is  written with an explicit parameterisation of the metric in terms of the U-duality invariant mass for $D=8$~\cite{Kiritsis:1997em},
 \begin{equation}
  \bE^{SL(3)}_{[10];s} =                       \sum_{
     (m_1,m_2,m_3)\in\ZZ^3\backslash\{0\}}\, \nu_2^{-{s\over3}}\,   \left(    {|  m_1+  m_2\Omega + B m_3|^2\over
     \Omega_2}+ {m_3^2\over \nu_2}\right)^{-s}
\,.
 \end{equation}
The   divergence  at   $s=3/2$ is regularised by setting  
 $s=3/2+\epsilon$ and subtracting the pole
 (see  appendix~\ref{sec:Sl3Series}  for details), 
\begin{equation}
\label{e:332reg}
   \bE^{SL(3)}_{[10];{3\over   2}+\epsilon}=   {2\pi\over\epsilon}+
   4\pi(\gamma_E-1)  +\hbE^{SL(3)}_{[10];\frac32} +O(\epsilon)\,,
 \end{equation}
where the regularised  series $\hbE^{SL(3)}_{[10];\frac32}$ is derived
in \eqref{e:E32reg} and is given by
\begin{eqnarray}
\hbE^{SL(3)}_{[10];\frac32}=\nu_2^{-{1\over2}}\,\bE_{3\over2}(\Omega)+{4
   \pi\over3}\,\log(\nu_2)+O(e^{-\Omega_2^\half\,\nu_2^{-\half}}, e^{-(\Omega_2\nu_2)^{-\half}}) \,.
\end{eqnarray}

In  type~IIB  variables  the  $U$   modulus  is  acted  only  by  the
 $SL(2)$ factor of the U-duality group
 $SL(3)\times SL(2)$.   The  $SL(2)$ Eisenstein series has a pole at $s=1$ as shown in~\eqref{eisdiv},
\begin{equation}\label{e2def}
\bE_{1+\epsilon}(U)=                                 {\pi\over\epsilon}
-\pi\log(U_2|\eta(U)|^4)+2\pi(\gamma_E-\log(2))+O(\epsilon)\,,
\end{equation}
and the regularised series is obtained by subtracting the pole,
 \begin{equation}
 \hat  \bE_1(U)=-\pi\log(U_2|\eta(U)|^4)\,.
 \label{eonehat}
 \end{equation}

 So far we have discussed the singularities of the individual Eisenstein series $\bE_s(U)$ and $\bE^{SL(3)}_{[10];s}$.  However the coefficient $   \cE^{(8)}_{(0,0)}$~(\ref{8dR4}) is a linear sum of these functions.    A
 crucial factor (not discussed in past work) is
 that   the   singularities   of   the  separate   Eisenstein   series  should not be regularised independently.  In fact, the singularities 
 in~(\ref{8dR4}) cancel  each other when regularised in a manner consistent with the considerations that follow later later in this paper.  This
implies that~(\ref{8dR4}) should be  written as 
 \begin{equation}\label{8dR4reg}
   \mathcal{E}^{(8)}_{(0,0)}=  \lim_{\epsilon\to 0} \,\left(\bE^{SL(3)}_{[10];\threeh+\epsilon}+2\bE^{SL(2)}_{[1];1-\epsilon}\right)-\log\mu_{(0,0)}\, ,
 \end{equation}
 where the hats have been  removed since this expression is finite and
 $\log\mu_{(0,0)}=4\pi(2\gamma_E-1-\log(2))$          in         order
 for~\eqref{8dR4reg} to agree with~\eqref{8dR4}. We
 will later obtain this result from the decompactification limit
 for the  coefficient of the $\cR^4$ coefficient  in $D=7$ dimensions,
 which is
 finite and reduces  to~(\ref{8dR4reg}) when $r_3 \to  \infty$ to give
 the $D=8$ expression.  
 This is the  first of several  cases in which divergences in different contributions to a coefficient function cancel with a suitable regularsation.  

 The $SL(2)$ Eisenstein series at $s=1$ satisfies the Laplace equation \eqref{eslaplace}
 \be
     \Delta_{SO(2)\bs SL(2)}
   \hbE^{SL(2)}_{[1];1}= \pi\,,
  \label{e1hat}
  \ee
  while the $SL(3)$ series satisfies
  \be
 \Delta_{SO(3)\bs SL(3)} \, \hbE^{SL(3)}_{[10];\threeh} = 4\pi\,,
  \label{sl3ehat}
   \ee
   where the $SO(3)\bs SL(3)$ laplacian is given in~(\ref{sl3laplace}).
 Therefore, applying the total $SO(3)\bs SL(3) \times SO(2)\bs SL(2)$ Laplacian of the eight-dimensional theory gives
 \begin{equation}\label{e:eqDifE00}
 \Delta^{(8)}\,\mathcal{E}^{(8)}_{(0,0)}= \Delta_{SO(3)\bs SL(3)} \hbE^{SL(3)}_{[10];\threeh}+2\, \Delta_{SO(2)\bs SL(2)}  \hbE^{SL(2)}_{[1];1}    = 6\pi\ .
 \end{equation}
 We will now verify that the expression~(\ref{8dR4}) gives the correct expression in each of the three degeneration limits under consideration.

 \medskip
 {\bf (i)   Decompactification to $D=9$}  
 \smallskip

 The nine-dimensional limit is obtained by taking one of the radii of the two-torus to infinity, $r_2/\ell_9\to \infty$.
 This is seen by setting $T_2=r_1 \, r_2/\ell_s^2$,  $U_2=r_2/r_1$    and    
 \begin{equation}
 \label{nu8deff} \nu_2^{-1}=\Omega_2\,T_2^2 =   \nu_1^{-\frac67}\,\left(r_2\over\ell_9\right)^2
\,.
 \end{equation}
Using the expansions for $\bE^{SL(3)}_{[10];s}$  in~\eqref{e:EsExpaNu} and  $\bE_s(U)$ in~(\ref{fourieres}),  and the general definition of constant terms in~(\ref{constdef}), the constant term of the combination~(\ref{8dR4}) in the $GL(1)\times SL(2)$ subgroup has the form 
 \begin{equation}
\int_{-\frac12}^{\frac12} dB_{\rm RR} dB_{\rm NS}\,\cE^{(8)}_{(0,0)} ={r_2\over \ell_9}\, \cE^{(9)}_{(0,0)}
 -{14\pi\over3}\,\log\left(  \frac{r_2}{\ell_9\,\mu_8}\right)\, ,
 \label{eEm32}
 \end{equation} 
where the double integral is over the elements of the unipotent radical corresponding to this subgroup.  At large $r_2$ and fixed $r_1$  the nonpertubative contributions  are exponentially suppressed and only this constant term survives.  The term proportional to $r_2$ gives the contribution to the $D=9$ action,   in   agreement    with   those   in~(\ref{9dR4})   with  $r_1=r_B$. 
 The $\log (r_2/\ell_9)$ term in~(\ref{eEm32}) is an important contribution to the massless threshold behaviour of   the nonanalytic term in 
 the one-loop four-supergraviton amplitude in eight dimensions, which has the form 
 $\log(-\ell_s^2\,s)\,    \cR^4$.    The    $\log(r_2/\ell_9)$    term
 in~(\ref{eEm32})  combines with  this  contribution into  $\log(-r_2^2
 s)\,\cR^4$   which   is  part   of   the   infinite  series    $(r_2^2\,s)^k\,
 \log(-r_2^2s)\, \cR^4$  that resums  into the nine-dimensional massless
 threshold, $\sqrt{s}\, \cR^4$, as analyzed in~\cite{Green:2008uj}.
The term proportional to $\log(\mu_8)$ is a scale contribution.

\medskip
 {\bf (ii)  $D=8$ perturbative string theory}
 \smallskip

 The perturbative string expansion of the $\cR^4$ coefficient in $D=8$ is obtained from the expansion of~(\ref{8dR4})  in powers of $y_8^{-1}=\Omega_2^2T_2$, which is associated with the constant term
  \begin{equation}
\int_{-\frac12}^{\frac12}d\Omega_1\, dB_{\rm RR}\, \mathcal{E}_{(0,0)}^{(8)}= {2\zeta(3)\over y_8}
   +  2\, (\hbE_{1}(T)+\hbE_{1}(U))  +{2\pi\over3}
   \log (y_8/\tilde\mu_8)\,,
   \label{gen1}
    \end{equation}
after using  the expansion  of  the regularised  $SL(3)$ series $\hbE^{SL(3)}_{[10];\frac32}$ in
\eqref{e:E322reg}, 
\begin{eqnarray}
\hbE^{SL(3)}_{[10];\frac32} = {2\zeta(3)\over y_8}+2\hbE_{1}(T)+{2\pi\over3}\log(y_8)+O(e^{-(y_8\, T_2)^{-\half}},e^{-T_2^\half\,y_8^{-\half}})\,.
\end{eqnarray}
 The first  term is  the correctly normalized  tree-level contribution
 and   the  one-loop   contribution  is   given  by  
 \begin{equation}
     \lim_{\epsilon\to 0}(\bE_{1+\epsilon}(T)+\bE_{1-\epsilon}(U))=\hbE_{1}(T)+\hbE_{1}(U)-{2\pi\over3}\log(\tilde\mu_8)\,,
 \end{equation}
where $\log(\tilde\mu_8)$ is a constant scale determined in the appendices.
This expression matches the one derived from the analytic part of the string
 amplitude in~(\ref{e:finiteoneloop})  obtained by decompactifying the
 genus-one amplitude on a three-torus.  The presence of the $\log y_8$ term is important.  As explained earlier and in~\cite{Green:2008bf}, this logarithmic term arises from the
 Weyl rescaling of a $\cR^4\,\log(-\ell_s^2 s)$ contribution in passing
 from the string  frame to the Einstein frame.   This is the non-local
 contribution of  the massless states in  $D=8$ one-loop supergravity.
 More   generally,   the  presence   of   logarithms   of  moduli   is
 characteristic of  the presence of  infrared thresholds.  This expression can also be derived by making use of the regularisation
 of~\cite{Dixon:1990pc}. 
 
As with the complete $\cR^4$ coefficient, the genus-one part,~(\ref{gen1}), is finite without the need to regularise the divergent individual terms --  the poles at $s=1$ cancel between the two terms. This follows directly from an analysis of the string theory one-loop calculation as sketched in appendix~\ref{sec:G1T2}, and is a symptom of the finiteness of perturbative superstring amplitudes.

 \medskip
 {\bf (iii)   Semi-classical M-theory limit}
 \smallskip

 The one-loop four-supergraviton amplitude in eleven-dimensional  supergravity  compactified on $\calT^3$ was considered  in~\cite{Green:1997as,Russo:1997mk}  (see  appendix~\ref{sec:11d} for
 details).  This  is expected to reproduce  the $SL(3)$-dependent part
 of the amplitude on a three-torus.  The zero Kaluza--Klein mode
 contribution  in  the  loop  gives rise to  the  non-analytic  logarithmic  terms
 characteristic of   the onset of one-loop ultraviolet divergences in $D=8$  supergravity.   Using dimensional regularisation by evaluating the amplitude in $D=8+2\epsilon$ dimensions, and subtracting the $\epsilon$ pole, this has the symbolic form
 (which is reviewed in detail in~\cite{Green:1999pu}),
 \begin{equation}
 A^{nonan}_{L=1}=\pi\cR^4\,
 \left(\log(-S\,\LP^2)+\log(-T\,\LP^2)+\log(-U\,\LP^2)-2\log(\mu_8)\right)\,,
 \end{equation}
 where the Mandelstam invariants of the eleven-dimension theory are denoted by capital letters (and the invariants $T$ and $U$ should not be confused with the complex structure and the K{\"a}hler structure of the two-torus!).  
  Translating to  eight-dimensional units  this gives
 \begin{equation}\label{e:nonana}
 A^{nonan}_{L=1}=\pi\,\cR^4\, \left(\log(-s\,\ell_8^2)+\log(-t\,\ell_8^2)+\log(-u\,\ell_8^2)\right)+\pi\,\cR^4\,\log(\hcV_3/\mu_8^2)\,,
 \end{equation}
 where $\ell_8^6=\LP^6\, \hcV_3^{-1}$.

 The  analytic  part of  the  one-loop  supergravity  amplitude  is evaluated  in
  appendix~\ref{sec:11d}.  In order to regularise the ultraviolet divergence this contribution  
    is evaluated  in $D=8+2\epsilon$ dimensions  and is given  by
       \begin{equation}\label{8dR4eps}
  \int_{-\frac12}^{\frac12} dU_1\,  \cE^{(8+2\epsilon)}_{(0,0)}       =     \bE^{SL(3)}_{[10]; \threeh+\epsilon}\,
   \hcV_3^{-{2\epsilon\over3}}+
   4\zeta(2)\,\hcV_3\,.
 \end{equation}
 This only depends on the $\calT^3$  moduli, which form the ``geometrical'' part of the moduli space.  The ``stringy'' dependence on the K{\"a}hler structure, $U$, is due to $M2$-brane windings and is not apparent in the supergravity calculations.  More generally, this is consistent with the $SL(d)$ invariance of toroidal compactifications of perturbative supergravity on a $\calT^d$ torus.  However, the divergence of the $SL(3)$ expression  $\lim_{\epsilon \to 0}\bE^{SL(3)}_{[10]; \threeh+\epsilon}$ must be regularised by subtracting the pole at $\epsilon =0$ since it is no longer cancelled.  This reflects the presence of a one-loop logarithmic  ultraviolet divergence in supergravity.
  Therefore, 
 \begin{equation}
   \int_{-\frac12}^{\frac12} dU_1\, \cE^{(8+2\epsilon)}_{(0,0)} = {2\pi\over\epsilon}+ 
    \, \hbE^{SL(3)}_{[10];\frac32}+4\zeta(2)\,\hcV_3-2\pi \log(\hcV_3/\mu_8) 
+ O(\epsilon)\,.
\label{reflectdiv}
 \end{equation}
After subtracting the pole, the regularised  interaction is given by the $SL(3)$ invariant
 \begin{equation}
 \int_{-\frac12}^{\frac12} dU_1\,     \cE^{(8)}_{(0,0)} =\hbE^{SL(3)}_{[10];\threeh} 
 + 4\zeta(2)\, \hcV_3-2\pi\log(\hcV_3/\mu_8)\,,
 \label{regrfour}
 \end{equation}
 where  $\hbE^{SL(3)}_{[10];\threeh} $ is  the  regularised Eisenstein  series
 defined in appendix~\ref{sec:Sl3Series}.  
The  $\log(\cV_3/\ell_{11}^3)$ term in  this equation  cancels against
the one in \eqref{e:nonana}.

 The correspondence with string theory follows by using the string theory/M-theory dictionary, which implies 
 \begin{equation}
 m_1^2 R_{11}^2+ m_2^2 R_{10}^2+ ÷m_3^2 R_9^2
 =\nu_2^{1\over3}\,   \left(    {|  m_1+  m_2\Omega|^2\over
     \Omega_2}+ {m_3^2\over \nu_2}\right) \equiv   m_{SL(3)}^2\,,
 \end{equation}
so that $\hbE^{SL(3)}_{[10];\threeh}$ in~(\ref{regrfour}) is identified with the expression in~(\ref{8dR4}).
 Expressing the volume, $\cV_3$,  of the M-theory torus in terms of the string theory variables using~\eqref{e:typeIIbunit}
 we have
 \begin{equation}
   \hcV_3={V^A_2\over \ell_s^2}={V^B_2\over r_B^2}\,,
 \end{equation}
so $\hcV_3$ is   identified with the volume of the
 two-torus $T_2=r_Ar_2/\ell_s^2$ on
 the type~IIA side and  to the complex structure parameter $U_2=r_2/r_B$ on
 the type~IIB side. Thus \eqref{reflectdiv} is written as 
 \begin{equation}\label{e:EE1c}
 \int_{-\frac12}^{\frac12}dU_1\,    \cE^{(8)}_{(0,0)} = \hbE^{SL(3)}_{[10];\threeh} +
       4\zeta(2)\, U_2-2\pi\log (U_2/\mu_8) \, .
 \end{equation}
 In  type~IIB  variables  the  $U$   modulus  is  acted  only  by  the
 $SL(2,\mathbb Z)$
 group of the U-duality group
 $E_{3(3)}=SL(3)\times  SL(2)$.   The $U_2$-dependent  part  is  completed into  the
 $SL(2,\mathbb Z)$-invariant expression,  $\hat  \bE_1(U)=-\pi\log(U_2|\eta(U)|^4)$
 (see appendix~\ref{sec:Sl2Series}) by the $M2$-brane contributions in the full theory.

 \subsection{Seven  dimensions}
 \label{rfoursevend}

Compactification to  dimensions $D<  8$ raises a  new issue  since the
leading dependence on $s$, $t$,  $u$ no longer comes from the analytic
$\cR^4$ interaction.  The one-loop supergravity contribution in $4<D <
8$  dimensions   is  finite  and  gives   a  well-studied  nonanalytic
contribution,  symbolically  of  the  form determined  by  dimensional
analysis $A^{nonan} \sim s^{D/2-4}\, \cR^4$ (suppressing a plethora of logarithms depending on ratios of Mandelstam invariants)~\cite{Green:1982sw}.  Infrared divergences arise for $D\le 4$.  We are interested in subtracting this contribution in order to isolate the analytic $\cR^4$ interaction.

  After compactification of type II string theory the effective action, \eqref{rfoureffect} with $D=7$   is invariant under the $U$-duality group $SL(5)$.
 The natural conjecture is that the coefficient function, $\calE^{(7)}_{(0,0)}$,  is
 a   $SL(5)$-invariant    Epstein    series,   similar to the one  
 in~\cite{Kiritsis:1997em}.   According   to   this  conjecture   the coefficient of the 
 seven-dimensional $\cR^4$ interaction in the Einstein-frame action is
  \begin{equation}\label{e:R47dI}
   \cE_{(0,0)}^{(7)}=\bE^{SL(5)}_{[1000];\threeh}\,.
 \end{equation}
 As before, our notation implies that the series is given by 
the minimal parabolic Eisenstein series for $SL(5)$ at a special value
 of  the  parameters (see, \eqref{e:DefMil} in  appendix~\ref{sec:Slnseries}).   Setting
 $s_2=s_3=s_4=0$ gives the Epstein zeta function, which has the general form of \eqref{e:Efund} with $d=5$.  Using a familiar $U$-duality invariant  parameterisation of the metric in terms of the $SO(5)\bs SL(5)$ moduli gives 
 \begin{equation}
 \bE^{SL(5)}_{[1000];\threeh}=\sum_{(m_1,m_2,n_1,n_2,n_3)\neq0}
 \left(\nu_3^{2\over
         5}\,\left[{|m_1+m_2\Omega+B\cdot       n|^2\over
         \Omega_2}+{n^T\cdot       \tilde       g^{-1}\cdot      n\over
         \nu_3^{2\over3}}\right]\right)^{-\frac32}\,.\\
   \label{e:R47d}
 \end{equation}
 The   term in brackets  is   proportional   to   the
 $SL(5)$-invariant mass squared in a parametrisation that makes
 manifest  the  string theory  three-torus  with  $SL(3)$ metric $\tilde g_{ij}$ ($\tilde g = g\, (\det g)^{-1/3}$, where $g$ is the $GL(3)$ metric)  and  associated  Kaluza--Klein  charges, $n_i$.  The  three scalar fields 
  \be
  B^i=B_{\rm RR}^i+ \Omega B_{\rm NS}^i\qquad\quad   i=1,2,3\,,
  \label{bidef}
  \ee
 arise   from  the   reduction  of  the   complex  two-form
 $C^{(2)}+\Omega  B^{\rm   NS}$  on  the  three   two-cycles of the
 three-torus $\cT^3$.

 Although this series appears to be divergent and in need of regularsation, analyticity in $s$ guarantees that it is well defined by meromorphic continuation. In other words, it does not need to be regulated (which is a different interpretation from that of~\cite{Kiritsis:1997em}).   A detailed analysis  of  its
   behaviour is given in  appendix~\ref{sec:Sl5Series}.  Furthermore, as we will soon see, decompactification to $D=8$ leads to precisely the finite combination of terms that was determined in the previous section.
 
   \medskip
 {\bf (i)   Decompactification to $D=8$}  
 \smallskip

The  $r_3/\ell_8\to \infty$ limit is associated with the constant term in the maximal parabolic subgroup $P_{\alpha_4}=P(3,2)$  with   Levi  subgroup $GL(1)\times SL(3)\times SL(2)$, which is the U-duality group for $D=8$.
In considering this limit in   $\bE_{[1000];s}^{SL(5)}$ we will make
use of the relations
 \begin{equation}
   \nu_3^{-1}= \Omega_2^{3\over2}\, {1\over \ell_s^6}(r_1r_2r_3)^2= \nu_2^{-\frac56}\, \left(\frac{r_3}{\ell_8}\right)^2\,,
   \label{nuseveight}
   \end{equation}
 recalling that $\nu_2^{-1}= \Omega_2\, (r_1r_2)^2/\ell_s^4$.
 
 The    $SL(5)$-invariant  mass  that enters the exponent of~(\ref{e:R47d})  decomposes into the sum of a $SL(3)$-invariant term and $SL(2)$-invariant term under the decomposition
 $\calT^3(r_1,r_2,r_3)                    \supset\calT^2(r_1,r_2)\times
 \mathcal{S}^1(r_3)$, which is relevant for the $P(3,2)$ parabolic.
 The  quantity in brackets in  the definition  of the series  in
 \eqref{e:R47d} then becomes the sum of  the $SL(3)$ and  $SL(2)$-invariant
 mass squared, $m_{SL(5)}^2=m_{SL(3)}^2+m_{SL(2)}^2$, where
 \begin{eqnarray}
   m_{SL(3)}^2&=&\nu_2^{1\over3}\,\left( {|m_1+m_2\Omega+m_3  B|^2\over \Omega_2}+ {m_3^2\over
     \nu_2}\right)\,,\\
 \nn m_{SL(2)}^2&=& {1\over \nu_2}\, {|n_1+n_2 U|^2\over U_2 T_2}\,,
 \end{eqnarray}
with $T_2=r_1r_2/\ell_s^2$ and $U_2=r_1/r_2$.

Details of the evaluation of the 
constant  term  of  the  $SL(5)$  Eisenstein series  on  this  maximal
parabolic are  given in appendix~\ref{sec:Sl5Series},  with the result
 \begin{equation}
   \ell_7\, \int_{P(3,2)}\, \bE^{SL(5)}_{[1000];\frac32}
 =r_3 \, \left( \hbE^{SL(3)}_{[10];\threeh}+2\hat \bE_1(U) - 4\pi\log\left({r_3\over \ell_8\mu_7}\right)\right)\,,
 \end{equation}
where $\log\mu_7=\log(4\pi)-\gamma_E$.
 This shows that the  $\cR^4$ interaction in  $D=7$ dimensions decompactifies to the $D=8$ interaction
 \begin{eqnarray}
 \ell_7\,  
 \cE^{(7)}_{(0,0)}= r_3 \,\left( \cE^{(8)}_{(0,0)} - 4\pi\log\left({r_3\over \ell_8\mu_7}\right)\right)+O(e^{-r_3/\ell_8})\,.
 \end{eqnarray}
The term proportional to $r_3$ contains the requisite $D=8$ coefficient together with a $r_3 \log r_3$ term that is essential for cancelling a similar term in the sum of the infinite series of $(s\,r_3^2)^m$ terms that reproduces the eight-dimensional $s \log(-\ell_8^2\, s)\, \cR^4$ threshold behaviour (as described in~\cite{Green:2006gt,Green:2008uj} and the introduction).  

 \medskip
 {\bf (ii) $D=7$ perturbative string  theory.}
 \smallskip

 The $D=7$ perturbative expansion parameter is $y_7^{-1}=\Omega_2^2\, v_3$, where   $v_3=(r_1r_2r_3)/\ell_s^3$.
 The invariant  mass is given  in terms of  $y_7$ and $v_3$ by
 \begin{equation} 
   m^2_{SL(5)}=   y_7^{-{1\over5}}\,   \left( y_7 \, (m_1+B_{\rm   RR}\cdot
     n+\Omega_1 \, B_{\rm NS}\cdot n)^2 + m^2_{SL(4)} \right) \,,
 \end{equation} 
where we have introduced the $SL(4)$-invariant mass 
\begin{equation}
  \label{e:massSL4}
  m^2_{SL(4)}=  {|m_2+B_{\rm NS}\cdot n|^2\over v_3 }+ v_3^{1\over 3}\ {}^tn\cdot \tilde g\cdot n\,.
\end{equation}
In the perturbative string theory limit the U-duality group reduces to its maximal parabolic 
subgroup $P_{\alpha_1}=P(1,4)$ with Levi subgroup $GL(1)\times SO(3,3)$.

The results of appendix~\ref{sec:Slnseries} imply
 \begin{equation} 
   \label{e:R47dpertA} 
\int_{P(1,4)}\,\bE^{SL(5)}_{[1000];s}= y_7^{-{4s\over 5}}\, 2\zeta(2s)+ \pi^{\half}\,{\Gamma(s-\frac12) 
 \over\Gamma(s)}\, y_7^{{s\over5}-{1\over 2}} 
  \, \bE^{SL(4)}_{[100];s-\frac12}\,.
 \end{equation} 
 Setting $s=3/2$ this gives
 \begin{equation}\label{e:R47dpert}
  \ell_7\int_{P(1,4)}\, \cE^{(7)}_{(0,0)}=    \ell_s \left({2\zeta(3)\over y_7}+2 \, \bE^{SL(4)}_{[100];1} \right)\,.
 \end{equation}
 The overall normalisation  has been chosen so that  the first term is
 the standard tree-level contribution,  while the second term, which is
 independent of $y_7$, is the genus-one contribution. 
This agrees with the perturbative genus-one string theory contribution to $\cR^4$  evaluated
in~(\ref{e:R4oneloopT3}).

 \medskip
 {\bf (iii) Semiclassical M-theory limit}
 \smallskip

 We will now discuss  the relation  between  the  $\cR^4$ interaction  in $D=7$ dimensions
 and the interaction  obtained by considering the one-loop ($L=1$) amplitude
 of  eleven-dimensional  supergravity  on  a  four-torus  (derived  in
 appendix~\ref{sec:11d}).  This  limit   corresponds  to  the  maximal
 parabolic subgroup $P_{\alpha_2}=P(4,1)$ with Levi subgroup $GL(1)\times SL(4)$ of
 the U-duality group.

In this limit the $SL(5)$-invariant mass reduces to 
\begin{equation}
 m_{SL(5)}^2 =\hat\cV_4^{-3/10}\,
   m_{SL(4)}^2 +n_3^2\, \hat\cV_4^{6/5}\,,
 \end{equation}
 where we have used $\hcV_4=(R_{11}R_{10}/\LP^2)^{5/4}\,\nu_3^{-1/2}$
and $\ell_7=\LP\,\hcV_4^{-1/5}$.

Therefore  the  constant term of $SL(5)$ series evaluated
in appendix~\ref{sec:Sl5Series}  implies that the $\cR^4$ interaction is given by 
\begin{equation}
    \ell_7\,  \int_{P(4,1)}\,    \cE^{(7)}_{(0,0)}=\ell_{11}\,
   \left(\hat\cV_4^{1\over4}\,
  \bE^{SL(4)}_{[100];\threeh}+ 4\zeta(2)\, \hat\cV_4\right)\,.
 \end{equation}
 which  is invariant under  the $SL(4)$  symmetry associated  with the
 geometry of $\calT^4$ and precisely matches the expansion of the M-theory $L=1$ amplitude on a
 four-torus in appendix~\ref{sec:11d}.

 \subsection{Six dimensions}
 \label{sixfour}

 For $D=6$ the $U$-duality group is $E_{5(5)} \equiv SO(5,5)$ and the conjectured coefficient of the $\cR^4$ interaction is
   \begin{equation}
 \cE^{(6)}_{(0,0)}= \bE^{SO(5,5)}_{[10000];\threeh}\,,
 \label{sixmod}
 \end{equation}  
which corresponds to the suggestion in~\cite{Kiritsis:1997em,Pioline:Automorphic} although
our  analysis  will be  somewhat  different (in  particular  regarding  the
regularisation).
The Eisenstein series  depends on the  moduli parametrizing
the coset $SO(5)\times SO(5)\bs SO(5,5)$.
 The Dynkin diagram of  figure~\ref{fig:dynkin}(i) with $n=5$ is symmetric under the interchange
 of nodes  2 and 5, which  means that the  decompactification limit to
 $D=7$  and decompactification  to M-theory  are each  described  by a
 constant term associated with a $SL(5)$ maximal parabolic subgroup of
 $SO(5,5)$ (see table~\ref{tab:Parabolics}).
 
 \medskip
 {\bf (i)   Decompactification to $D=7$}  
 \smallskip

 Equation~(\ref{e:DntoAn})
 together with  the relation $V_{(5)}=
 (r_4/\ell_7)^{5/2}$        gives  the explicit
 relation between the $SO(5,5)$ Epstein series
  $ \bE^{SL(5)}_{[1000];\threeh}$  and the Epstein  series associated
  with one of the $SL(5)$ maximal parabolic subgroups.
The  decompactification limit is  obtained by  deleting the  last node
$\alpha_5$    of    the     Dynkin    diagram    for    $E_{5(5)}=D_5$
in figure~\ref{fig:dynkin}(i). The decompactification limit $r_4/\ell_7 \to \infty$  is associated with the constant term of the parabolic subgroup,  $P_{\alpha_5}$, which has the   form
  \begin{equation}
 \label{decomsixseven}
  \ell_6^2\,  \int_{P_{\alpha_5}}  \cE^{(6)}_{(0,0)}   =    \ell_7\,    r_4\,
   \left(4\zeta(2)\, \frac{r_4}{\ell_7}+\cE^{(7)}_{(0,0)}\right)\,,
 \end{equation}
where we  have used the relation between  the  Planck
lengths in six and seven dimensions $\ell_6= \ell_7^{5/4}\, r_4^{-1/4}$. The coefficient of the term proportional to $r_4$ is the expected $D=7$ $\cR^4$ coefficient and the term proportional to $r_4^2$ combines once more with terms in an infinite series of $(r_4^2 s)^n$ terms to build the threshold behaviour in the nonanalytic term in $D=7$.

 \medskip
 {\bf (ii) $D=6$ perturbative string  theory}
 \smallskip

 We may now check agreement  with the $D=6$ perturbative string theory
 expansion.  This is obtained by deleting first node $\alpha_1$ of the
 Dynkin  diagram,  resulting  in  a  series of  terms  with  $SO(4,4)$
 T-duality   invariance.  The   associated  parabolic
 subgroup is denoted $P_{\alpha_1}$.
Substituting the relation between the $SO(5,5)$ Eisenstein series, $\bE^{SO(5,5)}_{[10000];s}$ and  $\bE^{SO(4,4)}_{[1000];s'}$ (given in~\ref{e:D5toD4})) and transforming to string frame using $\ell_6 = \ell_s\, y_6^{\quart}$,  we obtain
 \begin{equation}
 \ell_6^2\,\int_{P_{\alpha_1}} \cE^{(6)}_{(0,0)}  = \ell_s^2\, \left({2\zeta(3)\over
    y_6}+ 2\bE^{SO(4,4)}_{[1000];1}\right)\,.
   \label{perturbinsix}
 \end{equation}
 The first term on  the right-hand side of~(\ref{perturbinsix}) is the
 tree-level string theory term and the second term gives the genus-one
 contribution,   in  agreement   with  the   explicit   string  theory
 calculation given in~\eqref{e:genusOned} evaluated for $d=4$. 
 
 \medskip
 {\bf (iii) Semiclassical M-theory limit}
 \smallskip

  Finally, we may check the  M-theory  limit,   $\hcV_5\to   \infty$,  where
 $\hcV_5$ is the dimensionless volume of the M-theory torus, $\calT^5$.  This limit is obtained by deleting node $\alpha_2$ of the
 Dynkin  diagram in figure~\ref{fig:dynkin}(i).   The  associated  parabolic   subgroup  is  denoted
 $P_{\alpha_2}$. In
 this  limit we can  use the  relation between  the Planck  lengths, $
 \ell_6^4= \ell_{11}^4\, \hcV_5^{-1}$, and the relation~(\ref{e:DntoAn}) to show that 
  \begin{equation}
     \ell_6^2\,  \int_{P_{\alpha_2}}  \cE^{(6)}_{(0,0)}    =    \ell_{11}^2\,\hcV_5\,
     \left(4\zeta(2)+\hcV_5^{-\frac 35}\,\bE^{SL(5)}_{[1000];\frac32}\right)\,.
  \label{mtheoryexp}
 \end{equation}   
 This equation agrees explicitly with the regularised  one-loop
 amplitude in eleven dimensions of appendix~\ref{sec:11d}.
 Note that the symmetry between   the nodes $\alpha_2$
and $\alpha_5$ of the Dynkin diagram for  $E_{5(5)}$  in figure~\ref{fig:dynkin}(i) means that the decompactification limit
in~(\ref{decomsixseven}) and  the M-theory limit in~(\ref{mtheoryexp})
take similar forms.

More generally, compactification of string theory on a higher-dimensional torus, $\calT^d$ (or M-theory on $\calT^{d+1}$) with $d >4$, leads to a $D=(10-d)$-dimensional theory with exceptional  U-duality group $E_{d+1(d+1)}$.
 Consideration of limits (i), (ii) and (iii)  should  again pin down the  details of  the $\cR^4$ coefficients,
 $\calE^{(D)}_{(0,0)}$, in these cases. Although we have not completed a detailed analysis of these coefficients, we have a sketchy understanding of some of their properties, including the Laplace eigenvalue equations that they satisfy, as will be described in the discussion section~\ref{sec:briefsummary}.

 \section{The $\partial^4\cR^4$ interaction}
 \label{sec:dfourrfour}
 The next contribution to the low-energy expansion of the local part of the four-supergraviton effective action (or, equivalently, to the analytic part of the low-momentum expansion of the four-supergraviton S-matrix) in the $D$-dimensional type IIB theory after the $\ell_s^{-1}\, \cR^4$ term  is of the form
 \be
S_{\partial^4R^4}=\ell_{D}^{12-D} \,\int d^{D}x\, \sqrt{-G^{(D)}}\, \cE_{(1,0)}^{(D)} \,  \partial^4\cR^4\,.
 \label{Ddimfour}
 \ee

 The duality-invariant coefficient function in $D=10$ dimensions is a familiar non-holomorphic
 Eisenstein series for $SL(2)$ evaluated at $s=5/2$, 
 \begin{equation}
   \label{e:D4R410D}
\cE_{(1,0)}^{(10)} =\frac12\, \bE_{5\over2}(\Omega)\,.   
 \end{equation}
This coefficient function was initially  obtained  directly by considering the two-loop ($L=2$) amplitude of eleven-dimensional supergravity compactified on $\calT^2$ in the limit in which the volume, $\calV_2$, vanishes~\cite{Green:1999pu}.  This follows from the nine-dimensional expression to be presented in~(\ref{D4R49d}).  Its perturbative expansion is given by the constant term,
\be
\ell_{10}^2\int_{-\frac12}^{\frac12}d\Omega_1\,
\mathcal{E}_{(1,0)}^{(10)}=  \ell_s^2\,\left({2\zeta(5)\over  y_{10}}+
  {8\over3}\zeta(4)\, y_{10}\right) \,,
\label{d4r4ten}
\ee
which contains the correct tree-level and two-loop terms (and the absence of a one-loop contribution also agrees with string perturbation theory).  
The expression~(\ref{e:D4R410D})  can also be strongly motivated by supersymmetry arguments~\cite{Sinha:2002zr} that extend those of \cite{Green:1998by}.  

The coefficient $\cE_{(1,0)}^{(10)}$  satisfies the $SO(2)\bs SL(2)$ Laplace equation
\begin{equation}
  \Delta^{(10)} \cE_{(1,0)}^{(10)}= {15\over4} \, \cE_{(1,0)}^{(10)}\,.
\end{equation}

In the following subsections we will discuss the generalisation of the $\partial^4\cR^4$ interaction to $D=9$, $8$ and $7$ dimensions.  Comments about the $D=6$ will be made in the discussion in section~\ref{sec:briefsummary}
with some more details in~\cite{Green:2010}.

\subsection{Nine dimensions}
The effective $\partial^4\, \cR^4$ action in $D=9$ dimensions (\eqref{Ddimfour} with $D=9$) has the coefficient function,
\begin{equation}\label{9dD4R4}
 \mathcal{E}_{(1,0)}^{(9)}=\frac12\,
 \nu_1^{-{5\over 7}}\,\bE_{5\over2}(\Omega)+{2\zeta(2)\over15}\, \nu_1^{{9\over7}} \,
 \bE_{3\over2}(\Omega)+  {4\zeta(2)\zeta(3)\over15}\, \nu_1^{-{12\over7}}\,.
 \end{equation}
 Making use of the laplacian on nine-dimensional moduli space~(\ref{e:9}) we see that $\mathcal{E}_{(1,0)}^{(9)}$
 satisfies the differential equation
 \begin{equation}
   \label{eq:Diff9dD4R4}
   \big(\Delta^{(9)}-{30\over 7} \big)\,  \mathcal{E}^{(9)}_{(1,0)} = 0\ .
 \end{equation}

 \medskip
{\bf (i)  Decompactification to ten dimensions.}\nobreak\smallskip
\smallskip

 In the $r_B/\ell_{10} \to \infty$ it is useful to write~(\ref{9dD4R4}) as
  \begin{equation}
 \ell_9^3 \,   \cE^{(9)}_{(1,0)}=\ell_{10}^2\,   r_B\ \left(  \cE^{(10)}_{(1,0)}
 +{2\zeta(2)\over 15}\, \left(\ell_{10}\over r_B\right)^4\, \cE^{(10)}_{(0,0)}+
 {4\zeta(2)\zeta(3)\over 15}\, \left(r_B\over \ell_{10}\right)^2\right)\,.
\end{equation}
 The term linear in $r_B$ gives the 
 finite ten-dimensional result.   The term proportional to $r_B^3$ is known to be necessary~\cite{Green:2008bf,Green:2008uj}  in order to account for the ten-dimensional normal threshold proportional to $s \log(-\ell_{10}^2 s)\, \cR^4$.  As described in the introduction, this arises from the interchange of limits needed in making the transition from the $D=9$ low energy limit $r_B^2 s \ll1$ and the $D=10$ low energy limit $1 \ll r_B^2 s \ll r_B^2\,\ell_s^{-2}\, s$ \footnote{The amplitude compactified  on a circle has an infinite series of massive square root  thresholds of the form $\sum_p c_p \,(s + p/r_B^2)^{1/2} \, \cR^4 \sim  \sum_n  d_{n}\,  (r_B^2 s)^n/r_B\ \cR^4$.  In the limit $r_B^2 s \gg 1$ this series sums to the logarithmic singularity.  However, this infinite series of powers of $r_B^2 s$ is relevant in the low energy limit $r_B^2 s \ll 1$ in  the $D=9$ interactions. The $r_B^3$ term in 
~(\ref{twobdfour}) is the $n=2$ term in this series.}.  The term proportional to $r_B^{-3}$ multiplies the modular invariant function $\cE^{(10)}_{(0,0)}$, which is the coefficient of $\cR^4$ in $D=10$.  This fits in with the general statement that  terms suppressed by powers of $r_B$ are coefficients of interactions with fewer derivatives.

 \medskip
{\bf (ii)  $D=9$ perturbative string theory.}
\smallskip

 The perturbative limit is simply obtained by expanding the Eisenstein series in powers of $y_9= g_s^2\ell_s/r$,
  giving
 \begin{equation}\begin{split}
 \ell_9^3\,\int_{-\frac12}^{\frac12}d\Omega_1
\,\cE^{(9)}_{(1,0)} =\ell_s^3\, \Big(& {\zeta(5)\over
y_9}+{4\over 15} \zeta(2)\zeta(3)\,\left({r^3\over\ell_s^3}
+ {\ell_s^3\over r^3}\right)\cr
&+ {4\over 3} \zeta(4) y_9\,\left({r^2\over\ell_s^2}
+
{\ell_s^2\over r^2}\right)\,.
\Big) \end{split} \label{twobdfour}
\end{equation}
 This reproduces the tree-level term proportional to $1/y_9$, the genus-one terms in~(\ref{gen1}), which are independent of $y_9$ and genus-two terms proportional to $y_9$.  The coefficients of all these terms are consistent with direct calculations in string perturbation theory. Furthermore, since $y_9$ is invariant under T-duality,
the expression exhibits the known equivalence of the perturbative IIA and  IIB  theories for genus less than or equal to four.

 \medskip
{\bf (iii) Semi-classical M-theory limit.}
\smallskip

 The M-theory limit is also easy to establish.  Indeed the complete expression~(\ref{9dD4R4}) can be obtained directly by adding together the $L=1$  and $L=2$ contributions to the four-supergraviton amplitude of eleven-dimensional supergravity compactified on a two-torus~\cite{Green:1999pu}, giving (in M-theory units),
 \begin{equation}
\ell_9^3\,\cE^{(9)}_{(1,0)}=\ell_{11}^3\, \left(\frac12\, {1 \over \hcV_2^{\frac32}}\bE_{5\over 2}(\Omega)+
 {4\over 15}{1\over\hcV_2^3} \zeta(2) \zeta(3)-8 \zeta(4) \hcV_2^{\frac32}\, \bE_{-{1\over 2}}(\Omega)\right)\,.
 \label{D4R49d}
 \end{equation}
 The last  term is the contribution of  one-loop supergravity ($L=1$),
 while  the second term  comes from  the finite  part of  the two-loop
 ($L=2$) supergravity  amplitude.  The  first term is  the sum  of the
 $L=2$ sub-divergences and the triangle diagram in which one vertex is
 a  $\cR^4$  one-loop counter-term.   The  divergences cancel  between
 these  terms   leaving  the  displayed   finite  contribution.   Upon
 converting from M-theory units  to nine-dimensional Planck units this
 expression coincides with~(\ref{9dD4R4}).  
    
 \subsection{Eight dimensions}

 Compactification on $\calT^2$ gives rise to the $\partial^4\cR^4$ effective action \eqref{Ddimfour} with $D=9$, which is invariant under the   $D=8$  duality group,
 $E_{3(3)}=SL(3)\times SL(2)$.  Since this is a product group the automorphic function is generally, by  separation   of
 variables,
 expected to be the sum of products of eigenfunctions of the $SO(2)\bs SL(2)$ and
 $SO(3)\bs SL(3)$ Laplacian operators.  As argued in~\cite{Basu:2007ru}, the modular function has the explicit form
 \begin{equation}\label{e:D4R48d}
 \mathcal{E}^{(8)}_{(1,0)}=\frac12\,\bE^{SL(3)}_{[10];\fiveh}
 -4\, \bE^{SL(3)}_{[10];-\frac12}  \,\bE_2(U)\, .
 \end{equation}
 Interestingly, we find by explicit computation that the total  interaction $\mathcal{E}^{(8)}_{(1,0)}$  is an
 eigenfunction of the total $SO(3)\bs SL(3)\times SO(2)\bs SL(2)$ Laplacian
 \begin{equation}
 \Delta^{(8)} \,\mathcal{E}^{(8)}_{(1,0)}=
 {10\over3} \, \mathcal{E}^{(8)}_{(1,0)}\ .
 \end{equation}
 However,  the total interaction is not an eigenfunction of the cubic Casimir (whereas the Eisenstein series are).
 The evidence that~(\ref{e:D4R48d}) is the correct expression is based on the fact that it reduces to the expected expressions in the three degeneration limits described earlier, as we will now demonstrate.

\medskip
 {\bf (i) Decompactification to $D=9$}
 \smallskip
 
This   is the constant term corresponding to the  $r_2/\ell_9\to \infty$ limit.  Using the expansions of $\bE^{SL(3}_{[10];s}$ and $\bE_s$ it is straightforward to obtain the constant term,
  \begin{eqnarray}
\ell_8^4\int_{-\frac12}^{\frac12} dB_{\rm RR}dB_{\rm NS}\, \cE_{(1,0)}^{(8)}= \ell_9^3 r_2\, \left(
\cE_{(1,0)}^{(9)}+\frac12\,
  \left(\ell_9\over  r_2\right)^3\,  \cE_{(0,0)}^{(9)}
  +{4\pi\zeta(4)\over 45}\,  \left(r_2\over\ell_9\right)^3\right)\,.
  \end{eqnarray}
 The term linear in $r_2$ reproduces the $D=9$ $\partial^4 R^4$ coefficient, while the term proportional to $r_2^{-2}$ is proportional to the $\cR^4$ coefficient.  The term proportional to $r_2^4$ is the expected contribution to the nonanalytic $\cR^4$ threshold term. 

\medskip
{\bf (ii) $D=8$ perturbative string theory.}
\smallskip

 The coupling constant associated with string perturbation theory, $y_8$ is a modulus in the $SO(3)\bs SL(3)$ part of the moduli space.    The weak coupling expansion can therefore be obtained using properties of the $SL(3)$ Eisenstein series described in~(\ref{e:EsExpyT})
 \begin{eqnarray}
\int_{-\frac12}^{\frac12}dB_{\rm RR}d\Omega_1 \bE^{SL(3)}_{[10];\fiveh} &=&{2\zeta(5)\over y_8^{5\over3}}+{4\over3}\,y_8^{1\over3}\,\bE_2(T)\,, \\
 \int_{-\frac12}^{\frac12}dB_{\rm RR}d\Omega_1\bE^{SL(3)}_{[10];-\half}&=&-{1\over6}\,y_8^{1\over3}
 -\,{1\over2\pi^3}\,{1\over y_8^{2\over3}}\,\bE_2(T)\, .
 \label{eEm12}\end{eqnarray}
The perturbative expansion in terms of $SL(2)\times SL(2)$ functions is given by the constant term,
 \begin{equation}
   \label{e:E01Weak}
 \ell_8^4\int_{-\frac12}^{\frac12}dB_{\rm RR}\, d\Omega_1\, \cE_{(1,0)}^{(8)}=\ell_s^4 \left({\zeta(5)\over y_8}\,
 +       {2\over       \pi^3}\,
   \bE_2(T)\bE_2(U)+{2\over3} \, y_8 \,
   \left(\bE_2(T)+\bE_2(U)\right) \right)\,,
 \end{equation}
  which contains tree-level, genus-one and genus-two contributions,
 All three of these terms can be verified directly from the low-energy expansion of the four-supergraviton scattering amplitude in string perturbation theory compactified on $\calT^{2}$.  The tree-level term is standard.  Higher loops are briefly discussed in appendix~\ref{sec:genusone}.  The   $\partial^4\cR^4$  interaction extracted by expanding  the genus-one integrand has  a factor of $\bE_2(\tau)$, where $\tau$ is the world-sheet modulus that has to be integrated over the fundamental domain, ${\mathcal{F}_{SL(2)}}$~\cite{Green:1999pv,Green:2008uj}.     Upon compactifying, the integrand is multiplied by the lattice factor,  giving
 \begin{equation}
   \label{e:D4R4G1}
   I^{(2)}_1=\int_{\mathcal{F}_{SL(2)}}\,{d^2\tau\over\tau_2^2}\,
   \bE_2(\tau)\, \Gamma_{(2,2)}(T,U)={2\over\pi^2}\, \bE_2(T)\bE_2(U)\,,
 \end{equation}
 in agreement with~(\ref{e:E01Weak}). We refer to appendix~\ref{sec:G1T2} for the
 evaluation of this integral.
 The two-loop amplitude given in~\cite{D'Hoker:2002gw,D'Hoker:2005jc}, when compactified on $\calT^{2}$  is proportional to $\partial^4 \cR^4$ multiplied by
 \begin{equation}
   \label{e:D4R4G2}
   I^{(2)}_2=\int_{\mathcal{F}_{Sp(4)}}\,{|d^3\tau|^2\over(\det\Im\textrm{m}\tau)^3}\,
  \, \Gamma_{(2,2)}\,,
 \end{equation}
 where $\Gamma_{(2,2)}$ is the genus two lattice sum.  This integral was evaluated in~\cite{Pioline:Automorphic}
 (also reviewed in   appendix~\ref{sec:genus2}),
 giving\begin{equation}
 I^{(2)}_2={4\over 3\pi}\,\left(\bE_2(T)+\bE_2(U)\right)\,.
 \end{equation}

 \medskip
 {\bf (iii)  Semiclassical M-theory limit}
\smallskip

 The  expression~(\ref{e:D4R48d})  may be  motivated by  analyzing the
 M-theory  limit  obtained by  compactification  of the  four-supergraviton
 amplitude in eleven-dimensional  supergravity on $\calT^{3}$ at one
 and two  loops.  This builds  in the $SL(3,\mathbb Z)$  invariance as
 the geometric  symmetry of $\calT^{3}$,  whereas compactification of
 perturbative supergravity does not build in the $SL(2,\mathbb Z)$ part of the
 duality  group,  which  is  sensitive  to the  effects  of  euclidean
 $M2$-branes  wrapped  around  $\calT^{3}$.   This  results  in  the
 following         expression         for         the         $\partial^4\cR^4$
 interaction~\cite{Green:1999pu,Green:2008bf} 
 \begin{equation}
  \ell_8^4\,\int_{-\frac12}^{\frac12} dU_1\, \cE^{(8)}_{(1,0)}= {1\over\ell_{11}}\, {1\over\hcV_3^{5\over3}} \left(\frac12\, \bE^{SL(3)}_{[10];\fiveh}+ {2\over\pi}\,  \bE^{SL(3)}_{[01];2} \,\left(2\zeta(4)
   \hcV_3^2+{\pi\zeta(3)\over 5}\,{1\over\hcV_3}\right)\right)\,,
 \end{equation}
 The  first   term  arises  from  the   two-loop  ($L=2$)  counterterm
 calculation  given   by  the   triangle  diagram  evaluated   in  the
 appendix~\ref{sec:11d}. The second term  arises from the the M-theory
 one-loop ($L=1$) and the last term arises from the finite part of the
 two-loop amplitude and is evaluated in appendix~\ref{sec:Twoloop11d}. 
 Transforming  to  the eight-dimensional  Einstein  frame using  $\LP=
 \ell_{8}\, \hcV_3^{1/6}$ and $\hcV_3=U_2$ and using 
 the relation $\bE^{SL(3)}_{[10];2}=-\pi^4
 \bE^{SL(3)}_{[01];-\frac12}$ given in~(\ref{e:EfundPois})
 gives
 \begin{equation}
  \ell_8^4\,\int_{-\frac12}^{\frac12} dU_1\, \cE^{(8)}_{(1,0)}   =\ell_8^4\,\left(\frac12\,\bE^{SL(3)}_{[10];\fiveh}+ {2\over\pi}\, \bE^{SL(3)}_{[01];2}  \,\left(2\zeta(4)
   U_2^{2}+{\pi\zeta(3)\over 5 \, U_2}\right)\right)\,.
   \label{dfoureightt}
 \end{equation}
 It  is   easy  to  see   that~(\ref{dfoureightt})  has   the  unique
 $SL(3,\mathbb Z)\times SL(2,\mathbb Z)$ completion given in (\ref{e:D4R48d}).

 \subsection{Seven dimensions}

In this  subsection we will  show that the  seven-dimensional $\partial^4\cR^4$ effective action,  \eqref{Ddimfour} with $D=7$, contains the coefficient function
 \begin{equation}\label{e:D4R4seven}
  \cE^{(7)}_{(1,0)}            =\frac12\,\hbE^{SL(5)}_{[1000];\frac52}+{3\over \pi^3}\hbE^{SL(5)}_{[0010];\frac52}\,.
 \end{equation}
The symbol $\ \hat{}\ $ signifies that each $SL(5)$ Eisenstein series is regulated by evaluating the series at $s=5/2 + \epsilon$  and subtracting the pole in the limit $\epsilon\to 0$.
These poles are a  signal of the
ultraviolet  divergence  of  the  supergravity two-loop  amplitude  in
$D=7$.  The detailed evaluation of  the series close to the pole in  appendix~\ref{sec:Sl5Series}
gives
\begin{equation}\begin{split}
  \bE^{SL(5)}_{[1000];\frac52+\epsilon}&= {4\pi^2\over3\epsilon} +\hbE^{SL(5)}_{[1000];\frac52}+{8\pi^2\over9}(3\gamma_E-4)+O(\epsilon)\,,\cr
  \bE^{SL(5)}_{[0010];\frac52+\epsilon}&=
{2\pi^5\over9\epsilon}+\hbE^{SL(5)}_{[0010];\frac52}+{2\pi^3\over27}\,\left(6\pi^2\gamma_E-11\pi^2+36\zeta'(2)\right)+O(\epsilon)\,.
\end{split}\end{equation}
It is significant that the poles cancel in the combination
 \begin{equation}
  \lim_{\epsilon\to0}\left(\hbE^{SL(5)}_{[1000];\frac52+\epsilon}+{6\over
    \pi^3}\hbE^{SL(5)}_{[0010];\frac52-\epsilon}\right)
=\hbE^{SL(5)}_{[1000];\frac52}+{6\over \pi^3}\hbE^{SL(5)}_{[0010];\frac52}+\log(\tilde\mu_7)\,,
 \end{equation}
 which  is therefore finite. The constant  
\begin{equation}
  \label{e:scaleD4R47d}
\log\tilde\mu_7=16\zeta'(2)+16\pi^2\gamma_E/3-76\pi^2/9\,, 
\end{equation}
can be absorbed into the definition of the scale of the logarithm in the nonanalytic part of the amplitude, leaving the combination of Eisenstein series on the right-hand side of the ansatz \eqref{e:D4R4seven}.

Using the properties of the $SL(5)$ Eisenstein series in appendix~\eqref{sec:Sl5Series} it follows that this combination of Eisenstein series satisfies
\be
\label{sevensl5laplace}
\Delta^{(7)} \calE^{(7)}_{(1,0)} = \frac{40\pi^2}{3}\,,
\ee
As with the coefficient $\calE^{(8)}_{(0,0)}$ in~\eqref{e:eqDifE00} the presence of the inhomogeneous term on the right-hand side of this equation implies the presence of an additive logarithm in $\calE^{(7)}_{(1,0)}$, which is in this case a sign that the low energy supergravity limit has a two-loop  logarithmic ultraviolet divergence.

 \medskip
 {\bf (i) Decompactification to $D=8$}
\smallskip

The $r_3/\ell_8 \to \infty$ limit again involves the constant term in the $P(3,2)$ parabolic.
Using  the relation  between  the  Planck length  in  seven and  eight
dimensions, $\ell_7^5 = \ell_8^6\,  r_3^{-1}$, and the  formulas of
appendix~\ref{sec:Slnseries},  
we have
 \begin{equation}\begin{split}
    \ell_7^5\,\int_{P(3,2)}\,  \cE_{(1,0)}^{(7)}&= \ell_8^4 r_3\, \Big(
\cE_{(1,0)}^{(8)}+
  \left(\ell_8\over  r_3\right)^2\,{\pi\over3}  \Big(\cE_{(0,0)}^{(8)}+{28\pi\over5}\log(\ell_8\tilde\mu_7/r_3)\Big)\cr
  &+ {2\pi\over15}\, \left(r_3\over\ell_8\right)^4\Big)\,.
\end{split}  \end{equation}
The term proportional to $r_3$
reproduces the eight-dimensional interaction~(\ref{e:D4R48d}) and the
 coefficient of the $1/r_3$ term is the $\cR^4$ interaction in $D=8$ dimensions.  The term with a positive power $r_3^4$  is needed to contribute to the series  of $(r_3^2\, s)^n$ terms that sums to give the $\cR^4 \log(-\ell_8^2\, s)$ threshold in eight dimensions.

   \medskip
 {\bf (ii) $D=7$ perturbative string theory}
\smallskip

Using the relation between the seven-dimensional Planck length and the
string scale $ \ell_7=\ell_s\, y_7^{1/5}$, 
 in $D=7$ the string  perturbative expansion, which is associated with
 the $P_{\alpha_1}=P(1,4)$ parabolic with Levi component $GL(1)\times SO(3,3)$, has the form
\begin{equation}
  \ell_7^5  \,\int_{P(1,4)}\, \cE_{(1,0)}^{(7)}=  \ell_s^5\left({\zeta(5)\over y_7}+{3\over\pi^3}\,
    \bE^{SL(4)}_{[010];\frac52}
+                                                         {2y_7\over3}
(\hbE^{SL(4)}_{[100];2}+\hbE^{SL(4)}_{[001];2})+{4\pi^2\over 15}\,y_7\log(y_7/\tilde\mu_7)\right)
\,,
\label{e:pert7d}
\end{equation}
 which matches  the direct string perturbation theory calculations of the   tree-level,  genus-one  terms  
in~(\ref{e:D4R4oneloopT3}) and the genus-two contribution  in~(\ref{e:Genus2Lat3d}).
The tree-level term and the first genus-two term come from the $P(4,1)$ parabolic of  $\hbE^{SL(5)}_{[1000];\fiveh}$ in~(\ref{e:D4R4seven}), while the  genus-one term and the second genus-two term come from the $P(4,1)$ parabolic of the series $\hbE^{SL(5)}_{[0010];\fiveh}$ in~(\ref{e:D4R4seven}).  Thew $\log y_7$ term is the genus-2 ultraviolet threshold, which has a coefficient that is proportional to the inhomogeneous term on the right-hand side of \eqref{e:D4R4seven}.
 
  \medskip
  {\bf (iii)  Semi-classical M-theory limit}
\smallskip

 As   before,   the   compactification   of   the   eleven-dimensional
 supergravity amplitude provides the data for the constant
 term for  the parabolic subgroup associated with  node $\alpha_2$ in fig.~\ref{fig:dynkin}(i), which gives a series
 of $SL(4)$-invariant terms.  

The validity of the ansatz  for the $\partial^4\, \cR^4$  coefficient, \eqref{e:D4R4seven}, can be checked  in this limit  by using  the relation between the seven-dimensional Planck length and the
eleven-dimensional Planck length $ \ell_7= \ell_{11}\, \hat\cV_4^{-1/5}$ the $\partial^4\cR^4$.  This leads to
 \begin{equation}\label{e:D4R4Mth}
 \ell_7^5\int_{P(4,1)}\, \cE^{(7)}_{(1,0)}                          =
 {\ell_{11}^{5}\over\hcV_4^2}\left(\frac12\,\hcV_4^{\frac34}\bE^{SL(4)}_{[100];\frac52}+{\pi\over30}\hcV_4^{\frac94}
  \bE^{SL(4)}_{[001];\frac52}
+{2\over
    \pi^4}\hbE^{SL(4)}_{[010];2}-{6\pi^2\over5}\log(\hcV_4/\tilde\mu_7)\right)
\end{equation}
 This series of terms again coincides with contributions from Feynman diagrams in eleven-dimensional supergravity.
 The  first term arises  from the  finite part  of the  two-loop $L=2$
 diagrams in $D=11$ supergravity on $\cT^4$.  This   finite  contribution   is  given   by  the   integral   of  the
 $\Gamma_{(4,4)}$ lattice over the fundamental domain of the torus,
 which  leads  using the  techniques  of  appendix~\ref{sec:11d} to  the  series
 $\zeta(4)\bE^{SO(3,3)}_{[100];5/2}=\bE^{SL(4)}_{[010];5/2}$. 
 The second term in~(\ref{e:D4R4Mth})  arises from the one-loop 
 $L=1$ diagrams and 
 the last term from the triangle diagram that contains the one-loop counterterm.

 In order  to understand  the coefficients in  dimensions $D\le  6$ in
 detail we  need to make use  of the properties of  the constant terms
 that have not yet been obtained in detail.  However, we have pinned down the combination of two Eisenstein series that arises in $D=6$ (with U-duality group $SO(5,5)$) although we have not determined their relative coefficient.  Further comments will be made in the discussion in section~\ref{sec:briefsummary}, where we will also present the Laplace eigenvalue equations that we believe these series should satisfy for all $D\ge 3$.
 
 \section{The $\partial^6\cR^4$ interaction}
 \label{sec:dsixrfour}

The next order in the analytic part of the momentum expansion of the amplitude is encoded into the local effective action,
\be
S_{\partial^6R^4}=\ell_{D}^{14-D} \,\int d^{D}x\, \sqrt{-G^{(D)}}\, \cE_{(0,1)}^{(D)} \,  \partial^6\cR^4\,.
 \label{Ddimsixfour}
 \ee 
At  this  order in  the  low energy  expansion  the  structure of  the
equation satisfied by the coefficient functions changes, as is evident
from the $D=10$  $SL(2,\ZZ)$ case~(\ref{inhom}), which  has a source  term on
the right-hand side~\cite{Green:2005ba}
\begin{equation}
  (\Delta_{SO(2)\bs SL(2)}-12)\cE^{(10)}_{(0,1)} =- (\cE^{(10)}_{(0,0)})^2\ .
  \label{laplacersix}
\end{equation}
  Although this  has not been
derived explicitly  from supersymmetry,  it is easy  to argue  for the
qualitative structure of the equation 
based on a generalisation of the arguments of~\cite{Green:1998by} used
to determine the coefficient of the $\cR^4$ interaction.
The constant term is given by
\begin{equation}
 \ell_{10}^4           \int_{-\frac12}^{\frac12}           d\Omega_1\,
 \cE^{(10)}_{(0,1)}=\ell_s^4\,\Big({2\zeta(3)^2\over3}\Omega_2^2+{4\zeta(2)\zeta(3)\over
   3}+{8\zeta(2)^2\over5}\Omega_2^{-2}
+{4\zeta(6)\over27}\Omega_2^{-4}+O(e^{-4\pi\Omega_2})\Big)\, ,
\end{equation}
which  has  perturbative  contributions  up  to  genus  three  and  has
contributions  from D-instanton/anti-D-instanton  pairs with  zero net
instanton number.

Once again, we will see  that the generalisation to higher-rank groups
does not change the structure of the equation although the eigenvalues
of the homogeneous equation  change.  The structure of the coefficient
$\cE^{(D)}_{(0,1)}$  was determined  for  $D=10$ in~\cite{Green:1998by}
and  generalisations to $D=9,8$ were suggested by  Basu~\cite{Basu:2007ck}.   We will
demonstrate that in each case
$\cE^{(D)}_{(0,1)}$  satisfies  an  inhomogeneous  Laplace  eigenvalue
equation.  In $D=8$ dimensions subtle effects due to the regularisation of the
$\cR^4$ term in  the source imply additional contributions  to the solution
given in~\cite{Basu:2007ck}.  We will later determine the $D=7$ equation and properties of
its solution. The $D=6$ $\partial^6\cR^4$, which  is of particular interest since it contains the three-loop ultraviolet logarithm characteristic of the ultraviolet divergence in maximal supergravity~\cite{Bern:2008pv}, will not be discussed here although a few comments will be made in the concluding discussion section~\ref{sec:briefsummary}
(and in~\cite{Green:2010}).

 \subsection{Nine dimensions}
 
In this case the effective action, \eqref{Ddimsixfour} with $D=9$, contains the coefficient function determined in~\cite{Basu:2007ck} to be
 \begin{equation}\label{9dD6R4}
 \mathcal{E}_{(0,1)}^{(9)}=\nu_1^{-{6\over7}}\,
 \mathcal{E}_{(0,1)}^{(10)}  +     {2\zeta(2)\over3}    \,    \nu_1^{1\over7}\,
 \bE_{3\over2}
 + {2\zeta(2)\over 63}\, \nu_1^{15\over7} \, \bE_{5\over2}
 +{4\zeta(2)\zeta(5)\over63}\,\nu_1^{-{20\over7}}
 + {8\zeta(2)^2\over5}\,\nu_1^{8\over7}\ .
 \end{equation}
 The function $\cE^{(10)}_{(0,1)}$ is the ten-dimensional coefficient that satisfies the inhomogeneous Laplace equation,  \ref{Ddimsixfour}.

It is readily checked that $\mathcal{E}_{(0,1)}^{(9)}$  satisfies
 \begin{equation}
   \label{eq:Diff9dD6R4}
   \big(\Delta^{(9)}-{90\over 7} \big)\, \mathcal{E}^{(9)}_{(0,1)}=
  -\big( \mathcal{E}^{(9)}_{(0,0)} \big) ^2 \, .
 \end{equation}
 The  source term  is again  quadratic in the modular function that arises for the coefficient of the $\cR^4$ interaction, as it was for  $D=10$ in~(\ref{inhom}).

\medskip
{\bf (i)  Decompactification to ten dimensions.}
\smallskip

The contribution~(\ref{9dD6R4}) can be reexpressed in ten-dimensional units
 recalling that $ \ell_9 = \ell_{10}^{\frac87}\, r_B^{-\frac17}$  and  $\nu_1   =  (r_B/\ell_{10})^{-2}$, giving 
 \bea
\ell_9^5\, \cE^{(9)}_{(0,1)} &=&     \ell_{10}^4\,\,r_B \, \left(\cE^{(10)}_{(0,1)}
 +     {2\zeta(2)\over3}   \left(\frac{\ell_{10}}{r_B}\right)^2\,
 \cE^{(10)}_{(0,0)}
 + {4\zeta(2)\over 63}\, \left(\frac{\ell_{10}}{r_B}\right)^6\, \cE^{(10)}_{(1,0)}\right.\nn\\
 &+& \left.
 {4\zeta(2)\zeta(5)\over63}\, \left(\frac{r_B}{\ell_{10}}\right)^4 
 + {8\zeta(2)^2\over5}\,\left(\frac{\ell_{10}}{r_B}\right)^4+ O(e^{-r_B})\right)
 \,.
 \label{effrsixBis}
 \eea 
 The term proportional to $r_B$  gives the ten-dimensional expression in the $r_B\to \infty$ limit.  Once again, there
 is a growing term with the expected power of $r_B^5$, which contributes a term proportional to   $(s\, r_B^2)^2\, \cR^4$ to the expansion of the ten-dimensional $s\, \cR^4  \log(-\ell_{10}^2\, s)$  threshold in the limit $s\, r_B^2 \to \infty$.

\medskip
{\bf (ii)  Perturbative string theory.}  
\smallskip

 The perturbative expansion of this coefficient is given by expanding in powers of the string coupling,
 \be\begin{split}
 \ell_9^5\int_{-\frac12}^{\frac12}d\Omega_1&\,\cE^{(9)}_{(0,1)}
 =\ell_s^5\,r_B\, \bigg( {\zeta(3)^2\over 3g_B^2} 
 +   
 {\zeta(2)\zeta(3)\over 9}\, \left(1+{\ell_s^2\over r_B^2}\right)
+       {\zeta(5    )\zeta(2)\over   189}\,\left({r_B^2\over\ell_s^4}+
{\ell_s^6\over r_B^6}\right)
 \cr 
 &
 + {5\zeta(4)g_B^2\over 9}  \, {\ell_s^2\over r_B^2}
+{\zeta(4)g_B^2\over 3}\,\left(1+{\ell_s^4\over r_B^4}\right) 
 + {7\zeta(6)\over 576} g_B^4\,\left(1+{\ell_s^6\over r_B^6}\right)
 + O(e^{-1/g_B})
 \bigg)\,.
 \label{dos}
 \end{split}\ee
 This expression is symmetric under the T-duality transformation
  $r_B\to 1/r_A$ and $g_B\to g_A/r_A$.
 The  genus-three term  proportional to  $g_B^4$ comes  from expanding
 $\EE    $   and    was    shown   to    match    the   IIA    results
 in~\cite{Green:2006gt}. The symbol $O(e^{-1/g_B})$ indicates schematically the presence of
instanton/anti-instanton pairs in the zero D-instanton sector.

 \medskip
{\bf  (iii)  Semi-classical M-theory limit.}
 \smallskip
 
 The contributions to the $\partial^6\cR^4$ interaction obtained by compactifying the one-loop and two-loop Feynman diagrams of eleven-dimensional supergravity on $\calT^{2}$ were evaluated in~\cite{Green:2005ba}.
 Collecting the $L=2$ and $L=1$ modular functions along with the genus-one  terms of~(\ref{gen1}), we find the modular invariant expression,
 \begin{equation}\label{D6R49d}
 \ell_9^5\cE^{(9)}_{(0,1)}=\ell_{11}^5\, \hcV_2 \Big( {\EE \over 12}{1\over\hcV_2^3}+
 {\zeta(5 )\zeta(2)\over 189}{1\over\hcV_2^6} 
+ {\zeta(4)\over 3}\ +\hcV_2^{7\over2}{\zeta(2)\over 378}
 \bE_{5\over2}+ {\zeta(2)\over 9}\hcV_2^{1\over2}\,\bE_{3\over2} \Big)\,.
 \end{equation}
 This  expression sums  all  the contributions  determined  from the
 analysis of  the $L=1$ and $L=2$  loop amplitude on a  torus, to which
 has been added the  contribution $\zeta(5)\zeta(2)/\hcV_2^6$, which
 arises  from a  $\Lambda^3$ divergence  of the  $L=3$  amplitude. This
 contribution has been regularised by matching the string-theory genus-one  contribution determined
 in~(\ref{gen1}),  and is a prediction for  the three-loop supergravity
 contribution to the $\partial^6\,\cR^4$ interaction.
 
 In the next sub-section we will see how  this  nine-dimensional  interaction arises by decompactifying the  eight-dimensional term   proposed in~\cite{Basu:2007ck} and discuss further properties of this expression.

 \subsection{Eight dimensions}
 \label{sec:d6r4eight}

In this section we analyze  the eight-dimensional $\partial^6\,\cR^4$  interaction, which has an effective action
\eqref{laplacersix} that  is invariant under the U-duality group
 $E_{3(3)}=SL(3)\times SL(2)$.
 We will  show that the modular function proposed in~\cite{Basu:2007ck},
 satisfies the differential equation
 \be\label{e:DiffD6R4Bis}
 \Delta^{(8)}\,\mathcal{E}^{(8)}_{(0,1)}
 =12\,\mathcal{E}^{(8)}_{(0,1)}  
 -\,(\mathcal{E}^{(8)}_{(0,0)})^2\ .
 \ee
 where $\Delta^{(8)}$ is the $SL(3)\times SL(2)$ Laplacian.
 The source  term appearing in  this equation  again involves the square of the eight-dimensional
 $\cR^4$ coefficient.

The systematic solution of this equation will be obtained  in  appendix~\ref{sec:solutionD6R4}, where we will  see that it is uniquely specified by matching the known properties of string perturbation theory.
 The solution is close to the one argued for in~\cite{Basu:2007ck} on the basis
 of consistency  with the higher-dimensional  interaction (our normalisation differs by a factor 2/3 from \cite{Basu:2007ck}), 
  \begin{equation}\label{e:E01}\begin{split}
 \mathcal{E}^{(8)}_{(0,1)}&= \mathcal{E}_{(0,1)}^{SL(3)}
 +{40\over9}           \,           \bE^{SL(3)}_{[10]; -{3\over2}}\,\bE_3(U)+ {1\over 3}
 \hbE^{SL(3)}_{[10];   {3\over2}}   \,\hbE_1(U)   +f(U)\cr
&+{\pi\over 36} \hbE^{SL(3)}_{[10]; {3\over 2}}+
{\pi\over 9}  \hbE_1(U) +{\zeta(2)\over 9}\, ,
 \end{split}\end{equation}
 where the function $f(U)$ is defined as the solution of the equation
 \be 
( \Delta_U - 12)\, f(U) = -4\, \hbE_1^2(U)\,,
 \label{fdefeq}
 \ee
 where $\Delta_U = U_2^2 \,(\partial_{U_1}^2+\partial_{U_2}^2)$.  It  is  straightforward to  extract  the  power-behaved  terms in  its
expansion (see~(\ref{fexpand})).
We have also introduced $\cE^{SL(3)}_{(0,1)}$ satisfying 
 \begin{equation}
   (\Delta_{SO(3)\bs SL(3)}-12) \cE^{SL(3)}_{(0,1)} = - (\hbE^{SL(3)}_{[10];\frac32})^2\,.
 \end{equation}
 The   last three terms  in~\eqref{e:E01}  (absent   in  the   solution  presented
 in~\cite{Basu:2007ck}) arises from the regularisation of the $\cR^4$ interaction.

 We will now consider the limits~(i) and~(ii), but since we have
not  evaluated the  derivative  expansion of  the  $L=2$ amplitude  on
higher-dimensional tori the limit~(iii) will not be discussed.

\medskip
 {\bf (i) Decompactification  to $D=9$} 
\smallskip

 In the decompactification limit  $r_2/\ell_9\to\infty$ the  $SL(3,\mathbb{Z})$   modular  functions  in \eqref{e:E01} have  the
 form 
 \begin{eqnarray}
 \int_{-\frac12}^{\frac12}dB_{\rm RR}dB_{\rm NS}\,\bE_{[10];-{3\over2}}^{SL(3)}&=&{9\over16\pi^4}\,\nu_2^{1\over2}\,\bE_{5\over2}(\Omega)+{\pi\over315}\,\nu_2^{-2}\,,
\\
\int_{-\frac12}^{\frac12}dB_{\rm RR}dB_{\rm NS} \, \hbE_{[10];{3\over2}}^{SL(3)}&=&\nu_2^{-{1\over2}}\, \bE_{3\over2}(\Omega)+\pi\,\log\nu_2\, .
 \label{eE32}\end{eqnarray}
 Substituting the latter expansion into  the source term in~(\ref{e:Alapla}),
 one finds that the interaction coefficient becomes
 \begin{equation}\begin{split}
  \int_{-\frac12}^{\frac12}dB_{\rm RR}dB_{\rm NS}\,\cE_{(0,1)}^{SL(3)}&= {1\over\nu_2}\,
 \cE_{(0,1)}^{(10)} +\left({2\pi\over9}\nu_2^{-\frac12}\log(\nu_2)+c_1\nu_2^{3\over2}+c_2\nu_2^{-\frac52}\right)\,\bE_{\frac32}(\Omega)\cr
&+{\zeta(2)\over9}\left(5+4\log(\nu_2)+8\log^2(\nu_2)\right)+O(e^{-\Omega_2^\half\nu_2^{-\half}},e^{-(\Omega_2\nu_2)^{-\half}})  \ ,
 \end{split}\end{equation}
 where $c_1, \ c_2$ are integration constants. They are determined by taking at the same time the perturbative string limit and
comparing with the  expressions of appendix~\ref{sec:solutionD6R4}.  We find
$c_1=\zeta(5)/(12\pi ) $ and $c_2=0$.  In this case the zero instanton
sector   contains   instanton/anti-instanton   pairs   consisting   of
D-instantons and  wrapped $(p,q)$-string world-sheets  as indicated by
the last term.
 
 The $SL(2,\mathbb{Z})$ modular functions have the expansions
 \begin{eqnarray}
 \int_{-\frac12}^{\frac12}dU_1 \bE_3(U)&=& 2\zeta(6)\, U_2^3+{3\pi\zeta(5)\over4}\, U_2^{-2}\,,\\
\int_{-\frac12}^{\frac12}dU_1 \hbE_1(U)&=& 2\zeta(2) \, U_2-\pi \log(U_2)\, ,
 \end{eqnarray}
 and the expansion of  the function $f(U)$ given in~\cite{Basu:2007ck}
 and in~\eqref{fexpand} is\footnote{We correct
  a missing $1/\pi $ factor in the $1/U_2^3$ term in \cite{Basu:2007ck}.}
 \begin{equation}
   \begin{split}
      6 f(U)&=     {\pi^2 \over 180}\,    \left(65-20\pi    U_2+48\pi^2
 U_2^2\right)+ {\zeta(3)\zeta(5)\over \pi U_2^3}\cr
& -2\zeta(2) \log U_2\, \left(4\pi U_2-6\log
 U_2+1\right) + O(e^{-U_2})\, .
   \end{split}
  \end{equation}
Therefore, the constant term associated with decompactifying to nine dimensions is
\begin{eqnarray}
\nn && \ell_8^6\int_{-\frac12}^{\frac12}dB_{\rm RR} dB_{\rm
    NS}\,\cE^{(8)}_{(0,1)}=  \ell_9^5\,  r_2
   \cE^{(9)}_{(0,1)} \\
\nn &+& \ell_9^6\, \left( {\pi\over36}\,
  \cE^{(9)}_{(0,0)}+
\left(\frac{\ell_9}{r_2}\right)^4 \, {15\zeta(5)\over4\pi^3}\, \cE^{(9)}_{(1,0)}
 +{16\pi\zeta(6)\over 567}\left( {r_2\over \ell_9}\right)^6 \right)
\nn \\
&-& \ell_9^6 \, {\pi\over 9} \, \log\Big(
{r_2\over \ell_9} \Big)  \left(  7  \cE^{(9)}_{(0,0)} -4\zeta(2)\,
  \nu_1^{\frac47}\right)-\ell_9^6\,  \nu_1^{\frac47} {4\pi\zeta(2)\over21}\,\log(\nu_1)\, \\
 \nn
&+& \frac{\ell_9^7}{r_2} \zeta(2) 
\left( {37\over 36}+{86\over9}\, \log^2 \left(\frac {r_2}{\ell_9}\right)-{20\over 9}
\log \left(\frac {r_2}{\ell_9}\right)\right)\\
\nn&-&\frac{\ell_9^7}{r_2} \,  {\zeta(2)\over 21} \,\log (\nu_1)\Big( 1+4 \log \left(\frac {r_2}{\ell_9}\right)-{48\over 7}
\log (\nu_1)\Big)+O(e^{-r_2})\, .
\end{eqnarray}
The term linear in $r_2$ reproduces the nine-dimensional $\partial^6\cR^4$ interaction,
the term independent of $r_2$  is proportional to the nine-dimensional
$\cR^4$ interaction, and the term proportional to $r_2^{-4}$ is proportional to the nine
dimensional $\partial^4\cR^4 $ interaction.  The term proportional to $r_2^2$ is needed to reproduce the $D=9$ threshold of the form $(-s)^\half\, \cR^4$.

\break\medskip
 {\bf (ii) $D=8$ perturbative string theory}
 \smallskip

 The  perturbative  expansion  of   the  coefficient
 $\cE^{(8)}_{(0,1)}$ in increasing powers of $y_8 =(\Omega_2^2 T_2)^{-1}$ is performed
 in appendix~\ref{sec:solutionD6R4}.  We may summarise the result in terms terms of the functions $I^{(2)}_h(j_h^{(p,q)})$ that would be obtained by evaluating the appropriate terms  at genus-$h$ in string perturbation theory. 
 The function $j_h^{(p,q)}$ is the expansion of the integrand of the genus-$h$ string loop diagram to order $\sigma_2^p\, \sigma_3^q\, \cR^4$
 (the notation is explained in appendix~\ref{sec:genusone}).  
 \begin{equation}
   \begin{split}
&\ell_8^6 \int_{-\frac12}^{\frac12}d\Omega_1dB_{\rm RR}\cE^{(8)}_{(0,1)}=\ell_s^6\,\Big({2\zeta(3)^2\over 3\,y_8}+ {64\pi\over3}\, I^{(2)}_1(j_1^{(0,1)})+{2\pi\zeta(3)\over 9}\log(y_8) \cr
&+{2\over3} y_8\, I^{(2)}_2(j_2^{(0,1)})
+ {\pi\over 9}\,\left({\pi\over2} + I^{(2)}_1(j_1^{(0,0)})\right)\,y_8\, \log(y_8)+ {\pi^2\over 27}\,y_8\, \log (y_8)^2\cr
&
+   20y_8^2\,  I^{(2)}_3(j_3^{(0,1)})  +   O(e^{-(T_2  y_8)^{-\half}},
e^{-T_2^\half \,y_8^{-\half}}) \Big)\,.
   \end{split}
\label{xpert}\end{equation} 
The genus-one contribution to this expression has the form 
\begin{equation}
 I^{(2)}_1(j_1^{(0,1)})= {10\over 32\pi^6}
  \,\bE_3(T)\,\bE_3(U)+{\zeta(3)\over 32\pi}\, (\hbE_1(T)+\hbE_1(U) +\log\mu)\,.
\end{equation}
This follows both from the expansion of the coefficient $\calE^{(8)}_{(0,1)}$ and  from the direct evaluation  of the genus-one string theory amplitude in~(\ref{e:D6R4oneloop}).  

There is also a logarithmic correction to the genus-one term of the form $\log y_8$ in \eqref{xpert}.  This is a manifestation of a logarithmic ultraviolet divergence in supergravity that originates from the one-loop $\cR^4$ subdivergence of the two-loop supergravity diagram.    As before, the origin of the $\log y_8$ is in the transformation of $\log(-\ell_s^2\, s)$ from string frame to Einstein frame.
 
Comparing~(\ref{xpert}) with the expansion of $\calE^{(8)}_{(0,1)}$ in appendix~\ref{e:DiffD6R4}  we see that  the genus-two contribution is given by 
\begin{equation}
  I^{(2)}_2(j_2^{(0,1)})= {2\over3}\hbE_1(T)\,\hbE_1(U) +{\pi\over 9}\,
  (\hbE_1(T)+ \hbE_1(U))
+f(T)+f(U)
+ {11\zeta(2)\over 36}\,.
\label{gentwosix}
\end{equation}
In principle it should be possible to check~(\ref{gentwosix}) with the expansion of the genus-two string theory amplitude of~\cite{D'Hoker:2002gw,D'Hoker:2005jc} at order $\partial^6\cR^4$, but this has not been done.

There is also a logarithmic term of the form $y_8\, \log y_8$ in~\eqref{xpert}.  As described earlier, such a term signifies the presence of a two-loop supergravity logarithmic ultraviolet divergence. In other words, there is a $\ell_s^6\, s^3\, \cR^4\, \log(-\ell_s^2\, s)$ contribution to the amplitude in string frame, which generates the $y_8 \log y_8$  term in~\eqref{xpert} upon transforming to the Einstein frame.

The genus-three contribution in~(\ref{xpert}) extracted from the expansion of $\calE^{(8)}_{(0,1)}$ in appendix~\ref{e:DiffD6R4} is 
\begin{equation}
  I^{(2)}_3(j_3^{(0,1)})={1\over270}\,(\bE_3(T)+\bE_3(U))\,.
\end{equation}
 Little is known in detail about the genus-three superstring amplitude apart from the fact that its leading low energy behaviour contributes to $\partial^6\cR^4$~\cite{Berkovits:2006vc}.  However, it is interesting to note that this genus-three expression is given by the evaluation of the two-dimensional
lattice   integrated   over   the   Siegel  fundamental   domain   for
$Sp(3,\mathbb Z)$     evaluated    in   appendix~\ref{sec:highergenus}.

\subsection{Seven dimensions}
\label{sec:sevendimsix}

The construction of the   coefficient of the $\partial^6\cR^4$ interaction in the effective action~\eqref{laplacersix} with $D=7$, follows the same logic as in $D=8$, so this section will be brief.
The modular function multiplying the $\partial^6\cR^4$ interaction in $D=7$ is determined by
\be
\big(\Delta^{(7)}-{42\over5} \big)\, \cE^{(7)}_{(0,1)}=              -(\cE^{(7)}_{(0,0)})^2\ ,
\label{laplaceseven}
\ee
where
\be
\cE^{(7)}_{(0,0)} =  \bE^{SL(5)}_{[1000];{3\over 2}} \,.
\ee
As in the $D=8$ case, the solution can be written as  
\begin{equation}\label{e:D6R47D}
\cE^{(7)}_{(0,1)} =\cE_{(0,1)}^{SL(5)}  +{25\over 2\pi^5}\bE^{SL(5)}_{[0010];{7\over 2}} \ ,
\end{equation}
where    $\cE_{(0,1)}^{SL(5)} $   is    a   particular    solution   and
$\bE^{SL(5)}_{[0010];7/2} $  is the  only solution of  the homogeneous
equation that  has perturbative  terms consistent with  string theory.
The relative coefficient in~(\ref{e:D6R47D})
will now be confirmed by studying the decompactification limit.

\medskip
{\bf (i)   Decompactification to eight dimensions}   
\smallskip

In the limit  $r_3/\ell_8\to \infty $
the $(3,3)$ entry in the matrix in~(\ref{tab:A4}) (after setting $r_3=r^2$) becomes 
\be
\int_{P(3,2)} \bE^{SL(5)}_{[0010];{7\over 2}} = 2\zeta(6)\zeta(7)\ \left(r_3\over\ell_8\right)^{42\over 5}+ {\pi^2\zeta(2)\over 5}\ 
 \left(\ell_8\over         r_3\right)^{{8\over         5}}        {\bf
   E}^{SL(3)}_{[10];{5\over 2}} + {8\over 15}\ \left(r_3\over \ell_8\right)^{{12\over 5}} {\bf E}^{SL(3)}_{[01];3} {\bf E}^{SL(2)}_{3} \,.
\label{cuatro}
\ee
{}From this expression we recognise the term ${\bf E}^{SL(3)}_{[01];3} {\bf E}^{SL(2)}_{3} $ that decompactifies to eight dimensions. 
The other possible solutions to the homogeneous equation (with Dynkin labels $[1000]$ and $[0100]$) are ruled out because in the perturbative string limit
they give rise to terms that cannot be identified with  perturbative string theory (i.e. they give wrong powers of the string coupling).   The  $r_3^{42/5}$ term in \eqref{cuatro}  contributes to the $D=8$ threshold.

Comparing with the eight-dimensional expression for $\cE^{(8)}_{(0,1)}$ given in section~\ref{sec:d6r4eight}, and using $\bE^{SL(3)}_{[01];3} =2\pi^5/3\, \bE^{SL(3)}_{[01];-3/2}$, 
fixes the relative coefficient in~(\ref{e:D6R47D}), as follows.
In addition, we recognise the term $ {\bf E}^{SL(3)}_{[10];{5\over 2}} $ in~(\ref{cuatro}),  multiplied by  $r_3^{-8/5}$, which is part of the
$\partial^4 \cR^4$ interaction in eight dimensions.
The  other  part of  the  $\partial^4  \cR^4$  interaction is  a  term
$r_3^{-8/5} {\bf E}^{SL(3)}_{[01];2} {\bf E}^{SL(2)}_{2}$, 
which does not show up in~(\ref{cuatro}), but arises from  $\cE^{SL(5)}_{(0,1)}$, as follows.
The large-$r_3$ limit of the source term is obtained with the use of 
\be
\int_{P(3,2)}\bE^{SL(5)}_{[1000];{3\over   2}}    =   \left(r_3\over\ell_8\right)^{6\over   5}
\cE^{(8)}_{(0,0)} - 4\pi \left(r_3\over\ell_8\right)^{6\over   5}\log\left(r_3\over \ell_8\mu_7\right)\,.
\ee
In this limit, the   constant term of the particular solution $\cE_{(0,1)}^{SL(5)} $ contains the contributions 
\be
\int_{P(3,2)}       \,\cE_{(0,1)}^{SL(5)}       =\left(r_3\over\ell_8\right)^{12\over       5}
\,\left(\cE^{SL(3)}_{(0,1)}+{1\over3}\hbE_{[10];\frac32}^{SL(3)}\hbE_1(U)+f(U)
  + \left(\ell_8\over r_3\right)^{4} \cE_h+\cdots\right)\,.
\ee
The first three terms reproduce the eight-dimensional result (once added to the contribution of $\bE^{SL(5)}_{[0010];7/ 2}$).
Since the source term does not contain the power $r_3^{-8/5}$,  $\cE_h$ solves a homogeneous equation for the
$SL(3)\times SL(2)$ Laplacian with eigenvalue 10/3, which is the same as the eigenvalue of $\bE^{SL(3)}_{[10];5/2} $ in~(\ref{cuatro}).  
The term  we are expecting is of the form $k \,\bE^{SL(3)}_{[01];2} \bE^{SL(2)}_{2}$, where the coefficient $k$ is fixed by comparing 
with the $\partial^4 \cR^4$ interaction, which gives $k = -8\pi^2\zeta(2)/ 5$.

\medskip
{\bf (ii)   Perturbative string theory}   
\smallskip

We will now find the  constant part of the particular solution,  $\cE_{(0,1)}^{SL(5)}$, in the parabolic subgroup of relevance to limit (ii), the limit of perturbative string theory. In this limit, the result is expressed in terms of functions invariant under $SO(3,3)\sim SL(4)$,  the T-duality group.
We will need the expansions
\bea
 \int_{P(4,1)}\bE^{SL(5)}_{[1000];{3\over 2}} &=& 2\zeta(3)\ y_7^{-{6\over 5}}+ 2\ y_7^{-{1\over 5}}\bE^{SL(4)}_{[100];1}\,,
\label{siete}
\\
 \int_{P(4,1)} \bE^{SL(5)}_{[0010];{7\over 2}} &=& y_7^{-{7\over 5}}\bE^{SL(4)}_{[010];{7\over 2}}+ {8\pi \zeta(4)\over 15}\ y_7^{{3\over 5}}\bE^{SL(4)}_{[001];3}\,,
\label{ocho}
\eea
which can be found in entries $(1,1)$ and $(1,3)$ of~(\ref{tab:A4})  (setting $y_7=1/r^4$). Thus the homogeneous solution provides part of the genus-one and genus-three contributions.

In order to study the perturbative string theory limit we will also need the decomposition of the $SL(5)$ Laplace operator into the $SL(4)$ Laplace operator plus the second-order differential operator
associated with $y_7$,
\be
\Delta^{(7)}=\Delta_{SO(5)\bs SL(5)} \to \Delta_{SO(4)\bs SL(4)}+ {5\over 2} (y_7\partial y_7)^2 +5 (y_7\partial y_7)\,.
\ee
The coefficients $5/2$  and $5$ in this equation  have been determined
by using the known $D=8, 7$ $\cR^4$ and $\partial^4 \cR^4$ interaction coefficients. The $\cR^4$ coefficient is given in~(\ref{siete}),
whereas the $\partial^4\cR^4$ case can be checked using
\bea
 \int_{P(4,1)} \bE^{SL(5)}_{[1000];{5\over 2}} &=& 2\zeta(5)\ y_7^{-2}+ {4\over 3} \bE^{SL(4)}_{[100];2} \,,
\\
 \int_{P(4,1)}\bE^{SL(5)}_{[0010];{5\over    2}}    &=&   y_7^{-1}\bE^{SL(4)}_{[010];{5\over 2}}+ {4\pi\zeta(2)\over 3} \bE^{SL(4)}_{[001];2}\,.
\label{diez}
\eea
The constant term  of the particular solution associated  with the parabolic subgroup of relevance to the perturbative expansion is a series of the form
\be
\ell_7^7\int_{P(4,1)}\cE_{(0,1)}^{SL(5)}  =\ell_s^7\,  \sum_{n=0}^3 \cE^{SL(4)}_{n} y_7^{n-1}\ ,
\label{genusexp}
\ee
The coefficient functions $\calE^{(SL(4)}_n$ can be determined by 
substituting  this genus  expansion into  the  Laplace
equation~\eqref{laplaceseven}   and  using~(\ref{e:D6R47D}),   which gives
\bea
&& 6  \cE^{SL(4)}_{0} = 4\zeta(3)^2\,,
\label{sevone}\\
&& \left( \Delta_{SO(4)\bs SL(4)} -{21\over 2} \right) \cE^{SL(4)}_{1} = - 8 \zeta(3) \bE^{SL(4)}_{[100];1}\,,
\label{sevtwo}\\
&& \left( \Delta_{SO(4)\bs SL(4)} -10 \right) \cE^{SL(4)}_{2} = - 4 (\bE^{SL(4)}_{[100];1})^2\,,\label{sevthree}
\\
\label{sevfour} && \left( \Delta_{SO(4)\bs SL(4)} -{9\over 2} \right) \cE^{SL(4)}_{3} = 0\,.
\eea
Equation~\eqref{sevone}  gives the tree level contribution. The genus-one coefficient is determined 
by~\eqref{sevtwo}, which is solved by
\be
\cE^{SL(4)}_{1} =  a \bE^{SL(4)}_{[100];1+2\sqrt{2}}+a' \bE^{SL(4)}_{[001];1+2\sqrt{2}}+ b \bE^{SL(4)}_{[010];\frac72} +{2\zeta(3)\over 3}   \bE^{SL(4)}_{[100];1}\,,
\ee
for  any $a,a',b$.  The constants  $a,a'$ must  be zero  to  match the
genus-one contribution in $D=8$, and $b$ can be fixed by the decompactification limit.
Equation~\eqref{sevthree}  defines  the genus-two function  $ \cE^{SL(4)}_{2}$  which, by
construction, in  the decompactification  limit becomes the  genus-two
contribution 
$\hbE_1(T)\hbE_1(U)+f(T,\bar   T)+f(U,\bar  U)$   of   the  $\partial^6\cR^4$
interaction in eight dimensions.
Finally, \eqref{sevfour}  has two independent admissible solutions $\bE^{SL(4)}_{[001];{3}}$ and $\bE^{SL(4)}_{[100];{3}}$. The first one combines with
the solution of the homogeneous  equation, see~(\ref{ocho}). 

Thus, the complete perturbative expansion of the modular function $\cE^{(7)}_{(0,1)}$
is given by
\begin{equation}\begin{split}
\ell_7^7\,\int_{P(4,1)}   \cE^{(7)}_{(0,1)}   &=\ell_s^7\,\Big(
{2\zeta(3)^2\over 3}{1\over y_7}+ ({2\zeta(3)\over 3} \bE^{SL(4)}_{[100];1}+(1+b) \bE^{SL(4)}_{[010];{7\over 2}}) + y_7 \cE^{SL(4)}_{2} 
\cr
&+  2 y_7^{2}\left( \bE^{SL(4)}_{[001];3} + \bE^{SL(4)}_{[100];3}\right)  + n.p.\Big)\,,
\end{split}\end{equation}
where $n.p.$  indicates non-perturbative contributions.
By construction  this reproduces~(\ref{xpert}) in the decompactification limit since, as discussed above, in this limit 
the differential equation becomes the eight-dimensional one.
The genus-one contribution in string perturbation theory is given by $I^{(3)}_1(j_1^{(0,1)})$ 
evaluated in~(\ref{e:D6R4oneloopT3}) is given by
\begin{equation}
     I^{(3)}_1(j_1^{(0,1)}) ={25\over 8!}
  \,\bE^{SL(4)}_{[010];\frac72}+{\zeta(3)\over                 16\pi}\,
  \bE^{SL(4)}_{[100];1}\, ,
\end{equation}
which   determines  the value  of
$b=5\pi/756-1$.  It would be  interesting to  determine the  genus-two
coefficient by expanding the string theory amplitude~\cite{D'Hoker:2002gw,D'Hoker:2005jc}.

Interestingly, as in $D=8$, the  value  of the  genus-three
contribution  is  given by integrating the  three-dimensional
lattice  factor  over   the   Siegel  fundamental   domain   for
$Sp(3,\mathbb Z)$ 
evaluated in appendix~\ref{sec:highergenus},
\begin{equation}
 \int_{\cF_{Sp(3,\mathbb        Z)}}        {|d^{6}\tau|^2\over       (\det
      \Im\textrm{m}\tau)^{5}}\,\Gamma_{(3,3)}= {1\over
  270}\, \left(\bE^{SL(4)}_{[100];3}+\bE^{SL(4)}_{[001];3}\right)\, .
\end{equation}

 \section{Discussion}
 \label{sec:briefsummary}
 In this paper we have extended earlier analyses of the nonperturbative structure of the coefficients of terms in the low energy expansion of the four-supergraviton amplitude to the higher-rank duality groups that arise in toroidal compactifications of maximally supersymmetric string theory or M-theory.  We have considered terms up to order $\partial^6\cR^4$ in the derivative expansion of the effective action and compactification on $\calT^d$ to $D=10-d$ dimensions. 
 The $\cR^4$ coefficient  has been understood in cases  with $d\le 7$.
 The $\partial^4\cR^4$  coefficient has been understood  in detail for
 $d\le  3$,   with  partial  results  for  $d=4$   (see  below).   The
 $\partial^6\cR^4$ coefficient,  which has the  richest structure, has
 been understood for $d\le 3$. 
  
 The derivation of the coefficient functions necessarily followed a rather tortuous path since the aim is to discover the modular invariant coefficients for low-dimension string theory (high-rank duality groups) from information in  higher dimensions (low-rank duality groups), which involves checking many limits.  Nevertheless the results may be stated compactly.   The three terms in the low energy expansion of the four-supergraviton amplitude can be expressed as local terms in the effective action of the form
\begin{eqnarray}
 S_{\partial^{2k}\,R^4}&=& \ell_D^{2k+8-D}\, \int d^Dx \sqrt{-G^{(D)}}\, \cE^{(D)}_{(p,q)}\,\partial^{2k}\, \cR^4\,,
\label{sumint} 
\end{eqnarray}
where $(p,q)=(0,0)$, $(1,0)$ and $(0,1)$ and $k=2p+3q=0,2,3$.   
 The coefficient functions $\calE^{(D)}_{(p,q)}$ are automorphic functions of the coset space coordinates that transform as scalars under the appropriate duality groups.  Starting from the known structure of these functions we have determined their form in the compactified theory by demanding consistency in the three limits described in the introduction:  (i) decompactification from $D$ to $D+1$ dimensions; (ii) known properties of string perturbation theory in the limit of small string coupling; (iii)  The limit of large volume of the M-theory torus, $\calT^{d+1}$, which is described by loop diagrams of eleven-dimensional supergravity.   
 
Clearly many, if not all, of the properties of the coefficients are highly constrained by maximal supersymmetry combined with the dualities.  In particular we have found that they   satisfy Laplace eigenvalue equations, with or without source terms, which are known to be  consequences of supersymmetry in the simplest examples~\cite{Green:1998by,Sinha:2002zr}, although we do not have a general proof.  Given such an equation for $\cE^{(D)}_{(p,q)}$ it is easy to derive similar equations satisfied by the constant terms for maximal parabolic subgroups of any given duality group.  These follow from the decomposition of the Laplace operator with respect to the same subgroups as described in appendix~\ref{sec:laplacian}.    In summary, we found that 
the coefficients are solutions of 
\begin{eqnarray}
  \left( \Delta^{(D)}                  -{3(11-D)                  (D-8)\over
    D-2}\right)\,\cE^{(D)}_{(0,0)}&=&  6\pi\,\delta_{D-8,0} 
\label{laplaceeigenone}\\   
\left( \Delta^{(D)}  -{5(12-D) (D-7)\over D-2}\right)\,\cE^{(D)}_{(1,0)}&=&{20\pi^2\over3}\, \delta_{D-7,0}
\label{laplaceeigentwo}\\   
\left( \Delta^{(D)} -{6(14-D) (D-6)\over D-2} \right)\,\cE^{(D)}_{(0,1)}&=&-(\cE_{(0,0)}^{(D)})^2 + c\,  \delta_{D-6,0}\,,
\label{laplaceeigenthree}
\end{eqnarray}
where the Laplace operators are defined on the appropriate moduli space and $c$ is a constant that remains to be determined (see below).
The overall scale of the Laplace operators (and hence, the eigenvalues) of any one of the above equations  is convention-dependent\footnote{
The formula for  the $\cR^4$ eigenvalues differs by a factor
of 2 from equation~(4.11) in~\cite{Pioline:Automorphic}, since our conventions differ. }, but the relative normalisations in the three equations is convention-independent

The           coefficients          satisfying~(\ref{laplaceeigenone})-(\ref{laplaceeigenthree}) were discussed in  detail in the body of
this paper for various values of $D$.  In particular, the inhomogeneous Kronecker delta terms on the right-hand side of these equations contribute in
 the `critical' dimensions, $D=D_c=4+6/L$  -- the lowest dimensions in which
 the  $L$-loop diagrams  of low-energy  supergravity  have logarithmic
 ultraviolet  divergences.   These  are  $L=1$, $D_c=8$  for  $\cR^4$  (see \eqref{e:eqDifE00}) and 
 $L=2$, $D_c=7$ for $\partial^4 \cR^4$ (see~\eqref{sevensl5laplace}). In addition,  \eqref{laplaceeigenthree} gives the $L=3$ $D_c=6$  case for $\partial^6\cR^4$, which was not discussed here but will be described in~\cite{Green:2010}.   It
 is also notable that the eigenvalues in all these cases
 vanish in  the critical dimensions.  This structure  implies  that the solutions
 have  logarithmic  terms   characteristic  of  the
  ultraviolet divergences  of  maximal supergravity.  The coefficients of these logarithms, suitably normalised, should equal the residues of the epsilon poles in dimensionally regularised supergravity, up to convention-dependent normalisations.
This is straightforward to verify for the  $D_c=8$ and $D_c=7$ cases ($L=1$ and $L=2$, respectively),  where the analysis has been carried out in detail. The value of the constant $c$ in the $D_c=6$ case determines the coefficient of the genus-three logarithmic term in $\calE^{(6)}_{(0,1)}$.  This has to be consistent with
the residue of the $\epsilon$ pole in the three-loop supergravity calculation in \cite{Bern:2008pv}, which is proportional to $\zeta(3)$.
A preliminary study indicates this is the case \cite{Green:2010}.

  Although our considerations are for the most part limited to $D\ge 6$, 
 in appendix~\ref{sec:eigenD} we argue that~(\ref{laplaceeigenone})-(\ref{laplaceeigenthree}) probably apply
 for  all  $D \ge 3$.     This follows simply by requiring that the Eisenstein series continue to satisfy a Laplace 
 eigenvalue equation for all $D\le 6$.  
 
 Having obtained a coefficient function in $D$ dimensions, all     results in
 dimensions  greater  than  $D$  follow,  after some  work,  by  expanding in  the radius, $r$, of a compact
 dimension.    Importantly we find that   potentially  divergent terms cancel in this process, once account is taken of terms of the form $(r^2 s)^n$, which diverge in the large-$r$ limit in a manner associated with the presence of non-analytic thresholds of the scattering amplitude.  It appears to be very nontrivial that  whenever a coefficient function contains divergent Eisenstein series the divergences cancel between different terms.  The presence of such cancelling divergences is indicated by logarithms of the moduli that are signals of logarithmic ultraviolet divergences in the low energy field theory.
 
  As   a   detailed   example    of   these   results,   consider   the
 $SL(5)$-invariant coefficients  of the $D=7$  interactions, which was
 the  lowest dimension considered  in full  detail.  The  solutions we
 obtained were as follows,
 \bea
 \cE^{(7)}_{(0,0)}& = &\bE^{SL(5)}_{[1000];\threeh}\,,
 \label{rfoursummary}\\
 \cE^{(7)}_{(1,0)} &=&
                          \frac12\,      \hbE^{SL(5)}_{[1000];\fiveh}+{3\over \pi^3}\hbE^{SL(5)}_{[0010];\fiveh}\,,
 \label{d4rfoursummary}\\
 \cE^{(7)}_{(0,1)} &=& \bE^{SL(5)}_{[0010];\sevenh} + \cE^{SL(5)}_{(0,1)}\,.
 \label{d6rfoursummary}
\eea
In   particular,  the  coefficient   $\calE^{(7)}_{(1,0)}$  multiplies
$\partial^4\cR^4$,  which  has a  non-analytic  two-loop threshold  in
$D=7$      supergravity,     accompanied     by      a     logarithmic
divergence. This is manifested in the string expression 
in~(\ref{d4rfoursummary}), which   illustrates  the   cancellation  of
divergences mentioned earlier.   We have subtracted the constant
  $\log \mu_{(1,0)}$ from the
   epsilon  regularised $\cE^{(7)}_{(1,0)}$  because this  quantity is
    the scale factor of the
   threshold contribution $s^2\cR^4\,\log(-\ell_7^2\,s/\mu_{(1,0)})$. 
   The higher-dimensional interactions can be deduced by considering the sequence of decompactifications corresponding to limit~(i).    
   
 We can also make  some  comments  about  Eisenstein  series  for  the  groups $G_d = E_{d+1(d+1)}$ with $4\le d\le 7$ (of relevance to $3\le D\le 6$, where $D=10-d$). These are more difficult to analyze by elementary methods, but
  by making use of some  relations derived by Miller~\cite{MillerNotes} we find 
  the following  in dimensions $3\leq D\leq 6$:

\begin{itemize}

\item The $D=6$ $\cR^4$ interaction with symmetry $SO(5,5)$ has a coefficient  $\calE^{(6)}_{(0,0)}= \bE^{SO(5,5)}_{[10000]; 3/2}$, as described in section~\eqref{sixfour}, but the analysis for $3\le D \le 5$ has not been completed.  However, the eigenvalues in~\eqref{laplaceeigenone} 
coincide with those of the Eisenstein series  $\calE^{(D)}_{(0,0)}= \bE^{G_d}_{[1,0,\dots, 0]; 3/2}$, as can be seen directly from \eqref{e:LaplaEisenstein} setting $\lambda=[3/2,0,\dots,0]$.  This  strongly suggests that the $\cR^4$ coefficient is given by $\calE^{(D)}_{(0,0)}= \bE^{G_d}_{[1,0,\dots, 0]; 3/2}$ for all $D\ge 3$, 
 as suggested in~\cite{Pioline:Automorphic}.
 
 \item  Although  the $D=6$ $\partial^4 \cR^4$ interaction  has not been determined in detail, by looking at the decompactification limit it can be inferred
that  it must  be  of  the form  $\hbE^{SO(5,5)}_{[10000];  5/2} +  c\,
\hbE^{SO(5,5)}_{[00001]; 3}$, where our  knowledge of the second series
is based on~\cite{MillerNotes}.  The value of $c$ is determined by the
cancellation of the poles of these series at $s=5/2$ and $s=3$ respectively.

 \item The $D=6$ $\,\partial^6 \cR^4$ interaction coefficient is uniquely determined from~\eqref{laplaceeigenthree}
 by matching the different limits, in the same manner as in earlier sections.  
In particular, this determines the constant $c$, which arises as the coefficient of a genus-three logarithmic term.
This is of special interest since it is proportional to the coefficient of the ultraviolet divergence of three-loop maximal supergravity in $D=6$ dimensions.

\item As argued above, in $D=3,4,5$ we expect that the modular functions multiplying  the $\partial^4 \cR^4$  and $\partial^6 \cR^4$  interactions are still determined by (\ref{laplaceeigentwo}) and (\ref{laplaceeigenthree}), 
but these equations alone do not determine the  Dynkin labels of the possible Eisenstein series with the same eigenvalue. These must be found by matching with the different limits, as done in this paper for the higher $D$ cases.
This is an issue that we will return to using more powerful methods.
 \end{itemize}

Finally, we remark that that the analysis of interactions of higher order that $\partial^6\, \cR^4$ raise interesting new issues. 
In particular, it was shown in~\cite{Green:2008bf} that the coefficient functions for the $\p^8\cR^4$,
$\p^{10}\cR^4$ and $\p^{12}\cR^4$ interactions in $D=9$ dimensions consist of sums of 
modular functions with different eigenvalues. The generalisation to higher-rank duality groups should be interesting but is beyond the considerations of this paper.

  \acknowledgments
 We are  very grateful to Stephen  Miller for  help in evaluating
 the $SL(5)$  Eisenstein series on maximal parabolic  subgroups and for many insights into those of higher-rank groups. We  would  also like to  thank  Laurent Lafforgue  for  patient and  detailed
 explanations of the  work of Langlands and the  theory of automorphic
 forms. In addition we  would like to thank  Thibault  Damour, Luc Illusie
 and Michael Harris for discussions.

 \appendix
 
 \section{Applications of the unfolding method }\label{sec:Rankin-Selberg}

This section will present some applications of the unfolding method to
the computation of integrals of modular functions that are used in the
main  body  of  the paper.  At  several  points  we need  to  evaluate
integrals of the type 
 \begin{equation}
   \label{e:RSI}
 I[\bE_s,f]=           \int_{\mathcal{F}_{SL(2,\ZZ)}}\,{d^2\tau\over\tau_2^2}\,
 \bE_s(\tau)\, f(\tau)\,,
 \end{equation}
where  $f(\tau)$ is a modular function,  $\calF_{SL(2)}$ is  a fundamental domain  for $SL(2,\ZZ)$
and $\bE_s(\tau)$ is the $Sl(2,\mathbb Z)$ Eisenstein series defined
by
\begin{equation}
   \bE_s(\Omega)= \sum_{(m,n)\neq(0,0)} {\Omega_2^s\over |m+n\Omega|^{2s}}\,.
 \end{equation}
The integral~(\ref{e:RSI}) can be evaluated by means of   the    standard unfolding   method using the fact that 
$\bE_s(\tau)=\zeta(2s)\sum_{\gamma\in                           \Gamma_\infty\bs
  SL(2,\ZZ)}(\Im\textrm{m}(\gamma\cdot\tau))^s$, with $\Gamma_\infty=\{\pm
\begin{pmatrix}
  1&n\cr 0& 1
\end{pmatrix}, n\in \ZZ
\}$
 is an incomplete Poincar{\'e} series, leading to
\begin{equation}
  \label{e:RSIb}
  I[\bE_s,f]     =2\zeta(2s)     \,    \int_0^\infty     {d\tau_2\over
    \tau_2^{2-s}}\, \int_{-\frac12}^{\frac12} d\tau_1 f(\tau)\,.
\end{equation}

 A second type of integral that we need to consider is integration of a modular function $f(\tau)$ multiplied by a  Lattice sum,
 \begin{equation}
   \label{e:RSII}
 I[\Gamma_{(d,d)},f]=           \int_{\mathcal{F}_{SL(2,\ZZ)}}\,{d^2\tau\over\tau_2^2}\,
\Gamma_{(d,d)}\, f(\tau)
 \end{equation}
where $\Gamma_{(d,d)}$ is the (even) Lattice sum
 \begin{eqnarray}\label{e:LatDef}
   \Gamma_{(d,d)}&=&\sqrt{\det g}\, \sum_{(m^i,n^i)\in\ZZ^d\times
     \ZZ^d}  \,  \exp(-{\pi\over\tau_2}\, (g_{ij}+b_{ij}) (m^i -\tau  n^i)  (m^j
   -\bar\tau n^j) ) \\
\nn &=&\tau_2^{d\over2}\, \sum_{(p_L,p_R)\in \Lambda_{(d,d)}} \, \exp(-\pi \tau_2\,(p_L^2+p_R^2)+i\pi\tau_1\,(p_L^2-p_R^2))
 \end{eqnarray}
 where $p_L=(n+m.(b+g)).e^*$ and $p_R=(n+m.(b-g)).e^*$ with $e$ defined
 by $g=e^T e$ provide a basis
 of the lattice  $\Lambda^d$ so that $\calT^d=\IR^d/(2\pi\Lambda^d)$,
 and $e^*$ is a basis of the dual lattice.

 This  type   of  integral   can  be  evaluated   by  the   method  of
 orbits~\cite{Dixon:1990pc,Kiritsis:1997em,Bachas:1997mc,Kiritsis:1997hf,Pioline:Automorphic,Pioline:2001jn}, as follows.
The exponent in~(\ref{e:LatDef}) can be rewritten as 
 \begin{equation}
    {1\over \tau_2} \,(g+b)_{ij} (m^i -\tau n^i) (m^j
   -\bar\tau n^j) ={1\over \tau_2}
   \begin{pmatrix}
     1 &-\bar\tau
   \end{pmatrix}
 M^T \, (g+b)\, M
   \begin{pmatrix}
     1\\
 -\tau
   \end{pmatrix}\,,
 \end{equation}
 where $M$ is the $d\times 2$-rectangular matrix with integer entries
 \begin{equation}\label{e:Mmn}
   M=
   \begin{pmatrix}
     m_1&n_1\\
     \vdots&\vdots\\
 m_d&n_d
   \end{pmatrix}\,.
 \end{equation}
 The   $SL(2,\mathbb  Z)$   action,  $\tau\to  (a\tau+b)/(c\tau+d)$ 
 represented by the matrix $A\in Sl(2,\ZZ)$ transforms the matrix $M$ on the right
 \begin{equation}
M\to M  A= \begin{pmatrix}
     m_1&n_1\\
     \vdots&\vdots\\
 m_d&n_d
   \end{pmatrix}\begin{pmatrix}
     d&-c\\
 -b&a
   \end{pmatrix}\,.
 \end{equation}

 Therefore  the integral can  be decomposed  into various  orbits with
 respect to the $Sl(2,\mathbb Z)$ action. The orbits
 are i) the singular orbit that 
 corresponds to  $m^i=n^i=0$ for all $i=1,\dots,d$; ii) the degenerate
 orbit  where  all  the  sub-determinants of  the  $2\times2$  matrices
 defined by the  $i$th and $j$th line of the  matrix $M$ are vanishing
 $d^{ij}=m^in^j-m^jn^i=0$, which reduces to  $n^i=0$ for  all $1\leq
 i\leq d$;   iii) the non-degenerate orbit where at
 least one determinant $d^{ij}$ is non-zero.  
 Up to relabelling, the representative of the
 orbit can always be taken to have the  form 
 \begin{equation}
 \label{e:Rep}
   M_{0,k}=
   \begin{pmatrix}
     m_1&j_1\\
     m_2 &j_2\\
 \vdots&\vdots\\
     m_k&j_k\\
     0&n_{k+1}\\
 \vdots&\vdots\\
 0&n_d
   \end{pmatrix}, \qquad 0\leq j_k< m_k, \qquad 2\leq k\leq d\,.
 \end{equation}
Therefore the integral  in~(\ref{e:RSII}) can  be
expanded as
\begin{equation}\begin{split}
I[\Gamma_{(d,d)},f] &=\int_{\cF_{SL(2,\ZZ)}}{d^2\tau\over\tau_2^2}\, f(\tau)\cr
 &+\sum_{m^i\in\ZZ^d\backslash\{0\}}\int_0^\infty{d\tau_2\over\tau_2^2} \, e^{-\pi {m^i g_{ij} m^j\over \tau_2}}\,
 \int_{-\frac12}^{\frac12} d\tau_1 f(\tau)\cr
&+
 2\sum_{(m^i,n^i)\in\ZZ^d\times \ZZ^d\backslash\{0\}^2}\int_{\IC^+} {d^2\tau\over \tau_2^2} \, f(\tau)\,
 e^{-{\pi\over\tau_2} (g_{ij}+b_{ij}) (m^i-\tau n^i) (m^j- \bar\tau n^j)}\,.
  \label{e:RSIIb}
\end{split}\end{equation}
We remark that the unfolding has been expressed in terms of the matrix
$(g+b)_{ij}$, which implies that  the last line of \eqref{e:RSIIb}  contains exponentially
suppressed effects  of order $\exp(- g_{ij})$.  If it is necessary to  consider an
expansion in which exponentially  suppressed terms  are of  order $\exp(-
g^{-1}_{ij})$, then one would apply  the same formula starting from the
lattice  expressed  in terms  of  $g^{-1}$  after  a complete  Poisson
resummation over all $m^i$ and $n^i$ integers in~(\ref{e:LatDef}).

 \section{Eisenstein series for $SL(d)$}\label{sec:Slnseries}
 The minimal parabolic Eisenstein series for a group $G$ is defined by~\cite{Langlands}
 \begin{equation}
  \label{e:minparab}
  \bE^G_\lambda(g)= \sum_{\gamma\in G(\mathbb Q)/ B(\mathbb Q) }
  e^{\langle \lambda+\rho, H(g\gamma)\rangle}\,,
\end{equation}
 where $\langle  \cdot,\cdot\rangle$ is  the inner
product  on the  root system  of  $G$. Any  $g\in G$  can be  uniquely
decomposed according the Iwasawa  decomposition as $g=kan$ where $n\in
N$ in the unipotent subgroup, $a$ is in the maximal Abelian subgroup and $h$ is in the maximal compact
subgroup $K$. We have identified $a$ with $\exp(H(g))$. Finally, $\rho$ is half the sum of the positive roots and
$\lambda$ is a  vector in the weight space of the  lie algebra $\fr g$
of $G$ and $B$ is a Borel subgroup of $G$.\footnote{ Because the
  function  $g\to \exp(\langle  \lambda+\rho, H(g)\rangle)$ is
  defined on $G(\mathbb A)$, where $\mathbb A$ is the ring of Adeles of
$\mathbb Q$, it is common to consider the sum defined on the group of Adeles although this will not be necessary for the considerations of this paper.}  Eisenstein series  are eigenfunctions of the invariant differential operators of $K\bs G$.  In 
particular,    they    are    eigenfunctions      of    the
Laplacian,\footnote{
Invariance under
  $K$ implies that the eigenvalue  of the Laplacian is the same as
  the  value of the  second-order Casimir  of
  $G$ $\langle\lambda,\lambda\rangle-\langle\rho,\rho\rangle$.}
\begin{equation}
  \label{e:LaplaEisenstein}
  \Delta_{K\bs                  G}                 \,\bE^G_\lambda(g)=
  2(\langle\lambda,\lambda\rangle-\langle \rho,\rho\rangle)\,\bE^G_\lambda(g)\,.
\end{equation}
They are also eigenfunctions of higher-order Casimir operators of $G$.
 
 However, we will only need this general definition in order to discuss the special low-rank cases of interest here.  For  large part we are interested in  Eisenstein series for $SL(d)$, which can be analyzed relatively easily in terms of their definitions as multiple sums (see, for example,~\cite{Terras}), as we will see in this appendix.   Although we will not need to explicitly consider the most general $SL(d)$ series in this paper, it is nevertheless illuminating to review their construction since the maximal parabolic series can be obtained from it..  
 The following  treatment is  based  closely on  notes  by Stephen
 Miller  and  extensions   of his thesis~\cite{MillerThesis}.

To begin,   we consider  $H=\gamma\,  g  \,
 \gamma^T$, where $\gamma \in SL(d,\mathbb  Z)$ and $g$ is the $SL(d)$
 matrix parametrizing the coset space $SO(d)\bs SL(d)$.  Letting $H_k$ be the bottom right $k\times k$ minor of $H$ the general minimal parabolic
 Eisenstein   series~\cite{MillerNotes}  associated  with   the  minimal
 parabolic subgroup $P(1,\dots,1)$,
 \begin{equation}\label{e:DefMil}
   \bE^{SL(d)}_{[\epsilon_1,\dots,\epsilon_{d-1}];s_1,\dots,s_d}=
   \sum_{\gamma\in SL(n,\mathbb Z)/B(\mathbb Z)} \, \prod_{k=1}^{d-1} (\det
   H_k)^{\lambda_{d-k+1}-\lambda_{d-k}-1\over 2} \,,
 \end{equation}
 which is a special case of the general formula~(\ref{e:minparab}). 
  Here   we      have     set  $2s_k=\lambda_{d-k+1}-\lambda_{d-k}-1$ for     $1\leq     k\leq      d-1$, and $\epsilon_k=1$ if $s_k\neq 0$
 and 
 $\epsilon_k=0$ if $s_k=0$.

 The $SL(d)$ series  that are studied in this paper are
 \begin{itemize}
 \item The series  $\bE^{SL(d)}_{[1,0^{d-2}];s}$ given by $\lambda_d
   =1+\lambda_{d-1}+ 2s$ and for $2\leq i\leq d-1$ we have $\lambda_{d-i}=\lambda_{d-i-1}-1$.
 \item The series  $\bE^{SL(d)}_{[0,1,0^{d-3}];s}$  given by $\lambda_d
   =1+\lambda_{d-1}$, $\lambda_{d-1}=1+\lambda_{d-2}+2s$ and for $3\leq
   i\leq d-1$ we have $\lambda_{d-i}=\lambda_{d-i-1}-1$.
 \item The series  $\bE^{SL(d)}_{[0^{d-2},1];s}$  given by $\lambda_{2}=1+\lambda_{1}+2s$ and for $1\leq
   i\leq d-2$ we have $\lambda_{d-i}=\lambda_{d-i-1}-1$.
 \item  Since $H=\gamma\,  g\,\gamma^T$,  $\det H_k=m^{[i_1}\cdots
   m^{i_k]} m^{[j_1}\cdots
   m^{j_k]} \,  \prod_{r=1}^k g_{i_r j_r}$,  where $(m_1,\dots,m_d)$ is
   the last row of  $\gamma\in SL(2,\mathbb Z)$
 \end{itemize}

Since  $\det H_d=1$  in the  definition~(\ref{e:DefMil}) one  does not
need  to introduce  $2s_d=\lambda_{1}-\lambda_0-1$. However, in order to make the symmetry more explicit we introduce  such  variables and  consider  the change  of
variables~\cite{Terras} $s_j=z_{j+1}-z_j+1/2$ for $j\leq d$ and $s_d=-z_d+1/2$, i.e.,
$z_i=-\sum_{j=i}^d  s_j  +  {d-i+1\over2}$.  The variables  $z_i$  are
related to the $\lambda_i$ variables by $\lambda_{d-i}=2z_i+1$ for $1\leq
i\leq d$. We define 
\begin{equation}
  \Xi(z)=       {1\over\pi^{2\sum_{j=1}^d       j       z_j}}       \,
  \bE^{SL(d)}_{[\epsilon_1,\dots,\epsilon_{d-1}];s_1,\dots,s_{d-1}}\,
  \prod_{1\leq i<j\leq d} \, \Gamma(z_j-z_i+\frac12)\,.
\end{equation}
Then 
\begin{equation}
 \bE^{SL(d)}_{[\epsilon_1,\dots,\epsilon_{d-1}];s_1,\dots,s_{d-1}}\, \prod_{1\leq i<j\leq d} (z_j-z_i+\frac12)
\end{equation}
can be analytically continued to  a holomorphic function for all $z\in
\mathbb  C^n$ and  $\Xi(z)$ satisfies  the $d!$  functional equations~\cite{Langlands}
\begin{equation}\label{e:FunRelSLd}
  \Xi(\omega(z))=\Xi(z)  \,,
\end{equation}
where   $\omega(z)=\{z_{\omega(1)},\cdots   ,z_{\omega(d)}\}$   is   a
permutation of the $z$ elements of the Weyl group of $SL(d)$.

The             poles             of            the             series
$\bE^{SL(d)}_{[\epsilon_1,\dots,\epsilon_{d-1}];s_1,\dots,s_{d-1}}$
are located at  $s_i=0$ or $s_i=1$ and the
residue at $s_i=0$  is given by the Eisenstein  series associated with
the parabolic subgroup $P_i=P(1,\dots,1,2,1,\dots,1)$ evaluated at the value of $(s_1,\dots,s_{i-1},s_{i+1},\dots,s_n)$.  Furthermore, the residue at $s_i=1$ is given by the Eisenstein series associated with
the parabolic subgroup $P_i$ evaluated  at the value of the parameters
$(s_1,\dots,s_{i-2},s_{i-1}+1/2,s_{i+1}+1/2,s_{i+2},\dots,s_n)$~\cite{Terras}. 
All the  series discussed in the  main text and  the following subsections
can be deduced by extracting residues of poles of the minimal parabolic series (although we shall not exploit this procedure).

We will first present general features of the series $\bE^{SL(d)}_{[1,0^{d-2}];s}$ and $\bE^{SL(d)}_{[0,1,0^{d-3}];s}$ and then specialise to the particular cases of  the  $SL(2)$, $SL(3)$ and  $SL(5)$ series that are of specific interest in the main text.

 \subsection{The series $\bE^{SL(d)}_{[1,0^{d-2}];s}$}

 The   series   $\bE^{SL(d)}_{[1,0^{d-2}];s}$   is defined   by setting
 $\lambda_d=1+\lambda_{d-1}+2s$  and  $\lambda_{d-i}=\lambda_{d-i-1}-1$
 for $2\leq i\leq  d-2$. These Epstein  series can be written  in the usual form
 as
 \begin{equation}
   \label{e:Efund}
   \bE^{SL(d)}_{[1,0^{d-2}];s}   =   \sum_{(m^1,\dots,m^d)\in  \ZZ^d\bs
     (0,\dots,0)} \, {1\over ( m^i g_{ij} m^j)^s}\,,
 \end{equation}
 where $g_{ij}$ is the metric with unit determinant $\det g=1$. 
Since $\det g=1$, the inverse metric $g^{-1}= {\rm adj}(g)^T$ is given by the
 transpose of the  adjugate matrix. The
 elements of  the adjugate matrix are  the determinant of  the minors of
 order  $d-1$ of  the  matrix $g$.    If  we introduce the dual integers 
 $n_i=  \epsilon_{i j_1\cdots  j_{d-1}} m^{j_1}\cdots  m^{j_{d-1}}$ we
 can express the series $\bE^{SL(d)}_{[0^{d-2},1];s}$  in terms of
 the inverse of $g$ as 
\begin{equation}
  \label{e:Edual}
 \bE^{SL(d)}_{[0^{d-2},1];s}=   \sum_{(n_1,\dots,n_d)\in  \ZZ^d\bs
     (0,\dots,0)} \, {1\over ( n_i (g^{-1})^{ij} n_j)^s}\,.
\end{equation}
Applying the  general functional equation~(\ref{e:FunRelSLd})  we find
the  relation
\begin{equation}
   \label{e:EfundPois}
  {\Gamma(s)\over    \pi^s     }\,    \bE^{SL(d)}_{[1,0^{d-2}];s}    =
  {\Gamma({d\over2}-s)\over \pi^{{d\over2}-s}}\, \bE^{SL(d)}_{[0^{d-2},1];{d\over2}-s}\,.
 \end{equation}
The Epstein  series $\bE^{SL(d)}_{[1,0^{d-2}];s}$ has a single pole at
$s=d/2$  and converges  absolutely  for large  values of  $\Re{\rm
  e}(s)$.  It is defined by meromorphic continuation for other values of $s$~\cite{Terras}. 
These series do not have poles at the values $s=k/2$ for $1\leq k\leq d-1$, which agrees with the expectation from the string theory arguments given in the main text.  Note particularly that it follows, using analytic continuation and  $2\zeta(0)=-1$, that
 \begin{equation}
   \label{e:Efunds0}
   \bE^{SL(d)}_{[1,0^{d-2}];0}=-1\,. 
 \end{equation}
Using the integral representation of the series in~(\ref{e:Efund}),
\begin{equation}
{\Gamma(s)\over\pi^s}                       \bE^{SL(d)}_{[1,0^{d-2}];s}
=\sum_{(m^1,\dots,m^d)\in\mathbb  Z^d\bs\{0\}}  \int_0^\infty {dt\over
  t^{1+s}} e^{-{\pi\over t} \, m^i g_{ij} m^j}\,,
\end{equation}
it follows that the constant term on  the parabolic subgroup
$P_{\alpha_{d-1}}=P(d-1,1)$ with  Levi component $GL(1)\times SL(d-1)$
characterized by the matrix of the form\break $g=\textrm{diag}(
    r^{-(d-1)/ d} \, \tilde g, r^{(d-1)^2/ d})$ 
contains the explicit perturbative terms 
\begin{equation}\label{e:SLtoSL}
 \int_{P(d-1,1)} \bE^{SL(d)}_{[1,0^{d-2}];s} =r^{s(d-1)\over d} \bE^{SL(d-1)}_{[1,0^{d-3}];s}+2\pi^{d-1\over2}
      r^{(d-2s)(d-1)^2\over2d}\, {\Gamma(s-{d-1\over2})\over \Gamma(s)}\, \zeta(2s-d+1),
\end{equation}
This    implies    by    recursion    that    the    Epstein    series
$\bE^{SL(d)}_{[1,0^{d-2}];s}$ has a single pole at $s=d/2$, so that
\begin{equation}\begin{split}
    \bE^{SL(d)}_{[1,0^{d-2}];{d\over2}+\epsilon}&={\pi^{d\over2}\over\Gamma({d\over2})\,
    \epsilon}+\hbE^{SL(d)}_{[1,0^{d-2}];{d\over2}}\cr
 &+{\pi^{d\over2}\over\Gamma({d\over2})} \,\left(\gamma_E-\log(4)-{\Gamma'({d\over2})\over\Gamma({d\over2})}\right)+O(\epsilon)
\end{split}\end{equation}
where  $\gamma_E$ is  Euler's  Gamma constant  and  we introduced  the
regularized series $\hbE^{SL(d)}_{[1,0^{d-2}];{d\over2}}$.

Using  the  expression  for   the  $SO(d)\bs  SL(d)$  Laplacian  given
in~\cite{Pioline:Automorphic}  it is  straightforward to  verify  that these
series satisfy the following Laplace equations
\begin{eqnarray}
  \label{e:SlneigA}
   \Delta_{SO(d)\bs          SL(d)}         \bE^{SL(d)}_{[1,0^{d-1}];s}=
   s(s-{d\over2})\,{2(d-1)\over d}\,
   \bE^{SL(d)}_{[1,0^{d-1}];s}\\
  \Delta_{SO(d)\bs    SL(d)}    \bE^{SL(d)}_{[0^{d-1},1];s}=    s(s-{d\over2})\,{2(d-1)\over d}  \,
  \bE^{SL(d)}_{[0^{d-1},1];s}
 \end{eqnarray}
 These    equations    are    particular    cases
of~(\ref{e:LaplaEisenstein}) for the value of the weight vector $\lambda$
specified by the Dynkin labels $[s,0,\dots,0]$ and $[0,\dots,0,s]$.

For $s=d/2$ the eigenvalue vanishes and the Epstein series satisfy the
differential equation

\begin{equation}
     \Delta_{SO(d)\bs          SL(d)}         \hbE^{SL(d)}_{[1,0^{d-1}];{d\over2}}=
   {(d-1)\pi^{d\over2}\over \Gamma({d\over2})}
\end{equation}

 \subsection{The Series $\bE^{SL(d)}_{[0,1,0^{d-3}];s}$}\label{e:Eanti}

 The series  $\bE^{SL(d)}_{[0,1,0^{d-3}];s}$ is obtained  by substituting
 the values $\lambda_d
   =1+\lambda_{d-1}$, $\lambda_{d-1}=1+\lambda_{d-2}+2s$ and, for $3\leq
   i\leq d-2$,  $\lambda_{d-i}=\lambda_{d-i-1}-1$ in in~(\ref{e:DefMil}).
This gives
 \begin{equation}
   \label{e:Eant}
   \bE^{SL(d)}_{[0,1,0^{d-3}];s}  =  \sum_{1\leq k\leq d-1\atop [M_{0,k}]}  {1\over  (  g_{ij}     g_{kl} d^{il} d^{jk})^s}\,,
 \end{equation}
 where  $d^{ij}=m^i  n^j-m^j  n^i$,  which  can be  interpreted  as  the
 determinants  of the  order two  minors of  the rectangular  $d\times 2$
 matrix introduced in~(\ref{e:Mmn}). 
 Setting         $n^T=(n_1,\cdots,n_d)$     and
 $m^T=(m_1,\cdots,m_d)$,   we can introduce the matrix
 \begin{equation}\label{e:calM}
   {\cal M}=
   \begin{pmatrix}
     (n.g.n) & (n.g.m)\\
     (n.g.m) & (m.g.m)
   \end{pmatrix}\,,
 \end{equation} 
 such that
 \begin{equation}
 \label{e:detM} 
  2\det {\cal M} =2\left( (n.g.n) (m.g.m)-(n.g.m)^2\right)
 = g_{ij}g_{kl} d^{il} d^{jk}\,.
 \end{equation}
 The series in~(\ref{e:Eant}) can then be represented as
 \begin{equation}\label{e:EisAnt}
 \bE^{SL(d)}_{[0,1,0^{d-3}];s}= \sum_{1\leq k\leq d-1\atop [M_{0,k}]}\,{1\over (\det{\cal M})^s}
 \end{equation}
 We recognize  here the conditions characterizing  the non-degenerate 
orbit   when   unfolding   the   lattice   $\Gamma_{(d,d)}$   in  
appendix~\ref{sec:Rankin-Selberg}.

 The expression~\eqref{e:EisAnt} is a generalization of the $s=2$ case
 that  arises  in  the  evaluation  of the  two-loop  contribution  to
 four-supergraviton  scattering  in  compactified  supergravity,  which  is
 evaluated in~\eqref{e:I22Result}.  This motivates the introduction of
 the following integral representation when $s>-1$, 
 \begin{equation}
   \label{e:IntDef}
 I^d_s(\Lambda) = \int_0^{\Lambda} dV \, V^{2s-1}\, \int_{\mathcal{F}_{SL(2,\mathbb Z)}}\,{d^2\tau\over\tau_2^2}\, \sum_{(m^i,n^i)\in\ZZ^d\times
     \ZZ^d}  \,  e^{-V\,{\pi\over\tau_2}\, g_{ij} (m^i -\tau  n^i)  (m^j
   -\bar\tau n^j) }\,,
 \end{equation}
 where $\mathcal{F}_{SL(2,\mathbb  Z)}$ is  a  fundamental
 domain for $SL(2,\mathbb Z)$, so modular invariance is explicit.
  Evaluating   this    integral   with   the    unfolding   method   of
 appendix~\ref{sec:Rankin-Selberg}   the    finite   part  that arises from   the
 non-degenerate orbit leads  to the $\Lambda$-independent contribution
  \begin{eqnarray}
 \nn    I^d_s(\Lambda)|_{\Lambda^0}&=&     2\int_0^\infty    dV    \,    V^{2s-1}\,\int_{\mathbb
   C^+}\,{d^2\tau\over\tau_2^2}\,     \sum_{1\leq    k\leq    d-1\atop
   [M_{0,k}]} \, e^{- {\pi\over\tau_2}\,V\,g_{ij} (m^i -\tau n^i) (m^j
   -\bar\tau n^j) }\\
 &=&2\sum_{1\leq k\leq d-1\atop[M_{0,k}]} \,{1\over \sqrt{\det \mathcal{M}}}\,\int_0^\infty dV \,
 V^{2s-2}\, e^{-2\pi V \sqrt{\det \mathcal{M}}}\\
 \nn &=&2{\Gamma(2s-1)\over (2\pi)^{2s-1}}\, \sum_{1\leq k\leq d-1\atop [M_{0,k}]} \,{1\over
   (\det \mathcal{M})^s}=2{\Gamma(2s-1)\over (2\pi)^{2s-1}}\, \bE^{SL(d)}_{[0,1,0^{d-3}];s}
   \label{e:I22Result2}\end{eqnarray}
Therefore
 \begin{equation}
   I^d_s(\Lambda)  = 2\zeta(2)\, {\Lambda^{2s}\over  2s} +{1\over\pi}\,
   {\Lambda^{2s-1}\over                                          2s-1}\,
   \bE^{SL(d)}_{[1,0^{d-2}];1}+2{\Gamma(2s-1)\over (2\pi)^{2s-1}}\, \bE^{SL(d)}_{[0,1,0^{d-3}];s}
 \end{equation}
 where  the  series $\bE^{SL(d)}_{[1,0^{d-2}];1}$    is
 finite for $d>2$ and is defined by
 analytic continuation from the region where $\Re\textrm{e}(s)> d/2$. 

For    the   $d=3$    case   the    normalisation   of    the   series
$\bE^{SL(d)}_{[0,1,0^{d-3}];s}$ is different and we have
\begin{equation}
  I^3_s(\Lambda)= 2\zeta(2)\, {\Lambda^{2s}\over  2s} +{1\over\pi}\,
   {\Lambda^{2s-1}\over                                          2s-1}\,
   \bE^{SL(3)}_{[10];1}+2{\Gamma(2s-1)\zeta(2s-1)\over (2\pi)^{2s-1}}\, \bE^{SL(3)}_{[01];s}
\end{equation}

In order to   evaluate the  constant term on the  $P_{\alpha_{d-1}}=P(d-1,1)$ parabolic
subgroup characterized by the matrix of the form $g=\textrm{diag}(r^{-(d-1)/d} g_{d-1},r^{(d-1)^2/d})$, it is useful to split the lattice sum in~(\ref{e:IntDef}) into the
product   of  two   lattice factors,  $\Gamma_{(d,d)}=   \Gamma_{(1,1)}(r^{(d-1)^2/d})  \,
\Gamma_{(d-1,d-1)}(r^{-(d-1)/d}  g_{d-1})$.   Unfolding the $\Gamma_{(1,1)}$ factor~\cite{Bachas:1997mc}  
leads to the constant term
\begin{equation}\begin{split}
&{2\Gamma(2s-1)\over\pi^{2s-1}}\, \int_{P(d-1,1)}\, 
\bE^{SL(d)}_{[0,1,0^{d-3}];s}={2\Gamma(2s-1)\over\pi^{2s-1}}\,
\bE^{SL(d-1)}_{[0,1,0^{d-4}];s}\cr
&+
\int_0^\infty        dV        V^{2s-1}\int_0^\infty{d\tau_2\over
  \tau_2^2}\, \sum_{m\in\mathbb Z\bs\{0\}} e^{-\pi r^{(d-1)^2\over d} {V
      m^2\over             t}}\,             \int_{-\frac12}^{\frac12}
  d\tau_1\Gamma_{(d-1,d-1)}\,.
\end{split}\end{equation}
The $\tau_1$  integral projects on  the sector $p\cdot w=0$  where $p$
and $w$  are the  Kaluza-Klein and winding  modes of the  lattice. The
piece independent of $\Lambda$ arises\footnote{See  
  section~\ref{sec:decomp-seri-ea40100} for  detailed example on the
  $SL(5)$ series.} from the zero winding sector $w^2=0$,
leading     to\footnote{\label{foot:conj5} Conjecture~5 of~\cite{Pioline:Automorphic}
  states           that           $\pi\bE^{SL(d)}_{[0,1,0^{d-3}];1/2}=
  \bE^{SL(d)}_{[1,0^{d-2}];1}$.  Comparison of~(\ref{e:SLtoSL})  
  and~(\ref{e:SLtoSLBis})                  implies                  that
  $\bE^{SL(d)}_{[1,0^{d-2}];1}=-2\pi\bE^{SL(d)}_{[0,1,0^{d-2}];1/2}$
  for all values of $d\geq4$.}
\begin{equation}\begin{split}
\int_{P(d-1,1)}\,                      \bE^{SL(d)}_{[0,1,0^{d-3}];s}&=
r^{2s(d-1)\over d}\, \bE^{SL(d-1)}_{[0,1,0^{d-4}];s}\cr
& +r^{(d-2s)(d-2)(d-1)\over2d}\pi^{{d\over2}-1}\,{\Gamma(s+1-{d\over2})\over \Gamma(s)}\, \zeta(2s+2-d)\bE^{SL(d-1)}_{[1,0^{d-3}];s-\frac12}
\end{split}\label{e:SLtoSLBis}\end{equation}
 We note in particular the $d=4$ case, with our normalisations for the
 $SL(3)$ series, we find
\begin{equation}
  \label{e:SL4toSL3}
    \int_{P(3,1)}\,                      \bE^{SL(4)}_{[010];s}=
r^{3s\over2}\,\zeta(2s-1) \bE^{SL(3)}_{[01];s}
 +r^{3(2-s)\over 2}\pi\,{\Gamma(s-1)\over \Gamma(s)}\, \zeta(2s-2)\bE^{SL(3)}_{[10];s-\frac12}\,,
\end{equation}
which is used in various places in this paper.

Therefore the series $ \bE^{SL(d)}_{[0,1,0^{d-3}];s}$ has single pole
at $s=d/2$ so that
\begin{eqnarray}
 \bE^{SL(d)}_{[0,1,0^{d-3}];{d\over2}+\epsilon}&=&         {(2\pi)^d\over
   24\Gamma(d-1)\, \epsilon}+\hbE^{SL(d)}_{[0,1,0^{d-3}];{d\over2}}\\
\nn&+&
 {(2\pi)^d\over24\Gamma(d-1)}\,
 \left(\gamma_E+\log(2\pi)+12\zeta'(-1)-1-{\Gamma'(d-1)\over\Gamma(d-1)}\right)+O(\epsilon) \,,
\end{eqnarray}
where       we      introduced       the       regularized      series
$\hbE^{SL(d)}_{[0,1,0^{d-3}];{d\over2}}$ and similarly
\begin{eqnarray}
 \bE^{SL(d)}_{[0^{d-3},1,0];{d\over2}+\epsilon}&=&         {(2\pi)^d\over
   24\Gamma(d-1)\, \epsilon}+\hbE^{SL(d)}_{[0^{d-3},1,0];{d\over2}}\\
\nn&+&
 {(2\pi)^d\over12\Gamma(d-1)}\,
 \left(\gamma_E+\log(2\pi)+12\zeta'(-1)-1-{\Gamma'(d-1)\over\Gamma(d-1)}\right)+O(\epsilon) \,.
\end{eqnarray}

The antisymmetric rank-two  $d^{ij}$  representation can be converted into the antisymmetric  rank-($d-2$)  representation, $d_{r_1\cdots   r_{d-2}}=\epsilon_{ij r_1\cdots r_{d-2}}\, d^{ij}$ representation, so that
 \begin{equation}
   2\det \mathcal{M}= g_{ij}g_{kl} \epsilon^{ikr_1\cdots r_{d-2}}
   \epsilon^{jls_1\cdots          s_{d-2}}\,          \, d_{r_1\cdots
     r_{d-2}}\, d_{s_1\cdots s_{d-2}}\,.
 \end{equation}
Since $g_{ij}g_{kl}-g_{ik}g_{jl}$ are the rank-two minors of the
 matrix  $g$, it follows (for matrices  with $\det  g=1$) that 
 $g_{ij}g_{kl} \epsilon^{ikr_1\cdots r_{d-2}}
   \epsilon^{jls_1\cdots  s_{d-2}}$  are   the  rank  $d-2$  minors  of
   $g^{-1}$.  Therefore 
 \begin{equation}
   4\det \mathcal{M}=(d-2)! \prod_{i=1}^{d-2} (g^{-1})^{r_i s_i} \, d_{r_1\cdots
     r_{d-2}}\, d_{s_1\cdots s_{d-2}}\,.
 \end{equation}
This leads  to the  series with label  $[0^{d-3},1,0]$ evaluated  for  the metric
 $g^{-1}$. By Poisson resummation this sum can be brought back to a sum
 over  $g$, giving the following  functional equation,
 which is a particular case of~(\ref{e:FunRelSLd})
 \begin{equation}
{\Gamma(s)\Gamma(s-\frac12)\over \pi^{2s-\frac12}}
\,
\bE^{SL(d)}_{[0,1,0^{d-3}];s}={\Gamma({d\over2}-s)\Gamma({d-1\over2}-s)\over
\pi^{d-2s-\frac12}}\, \bE^{SL(d)}_{[0^{d-3},1,0];\frac{d}{2}-s}\,,
 \end{equation}
where use has been made of the replicating formula
$2\Gamma(2s-1)/(2\pi)^{2s-1}= \Gamma(s-1/2)\Gamma(s)/\pi^{2s-3/2}$. 

Using  the  expression  for   the  $SO(d)\bs  SL(d)$  Laplacian  given
in~\cite{Pioline:Automorphic}  it  is  easy  to  verify that  the
integral representation implies
\begin{eqnarray}
  \label{e:SlneigA2}
   \Delta_{SO(d)\bs    SL(d)}    \bE^{SL(d)}_{[0,1,0^{d-3}];s}&=&    s(s-{d\over2})\,{4(d-2)\over d}  \,
  \bE^{SL(d)}_{[0,1,0^{d-3}];s}\,,\\ 
 \Delta_{SO(d)\bs    SL(d)}    \bE^{SL(d)}_{[0^{d-3},1,0];s}&=&    s(s-{d\over2})\,{4(d-2)\over d}  \,
  \bE^{SL(d)}_{[0^{d-3},1,0];s}\,.
 \end{eqnarray}
 These    equations    are    particular    cases
of~(\ref{e:LaplaEisenstein}) for the value of the weight vector $\lambda$
specified by the Dynkin labels $[0,s,0,\dots,0]$ and $[0,\dots,0,s,0]$.

For the value $s=d/2$ this gives 
\begin{eqnarray}
  \label{e:SlneigA3}
   \Delta_{SO(d)\bs   SL(d)}  \hbE^{SL(d)}_{[0,1,0^{d-2}];{d\over2}}&=&
  {(2\pi)^d\over 12\Gamma(d-2)}\,.
 \end{eqnarray}

 \subsection{The $SL(2)$ Eisenstein series}\label{sec:Sl2Series}

Non-holomorphic $SL(2)$ Eisenstein series are defined by
 \begin{equation}\label{e:Esl2Def}
   \bE_s(\Omega)= \sum_{(m,n)\neq(0,0)} {\Omega_2^s\over |m+n\Omega|^{2s}}\,,
 \end{equation}
  with
 $\Omega=\Omega_1+i\Omega_2\in\mathfrak{h}=\{\Omega_2>0,
 \Omega_1\in\mathbb R\}$ in the complex upper-half plane.
 The modular function
 \begin{equation}
   \tilde \bE_s(\Omega) ={\Gamma(s)\over \pi^s}\, \bE_s(\Omega)
 \end{equation}
 has an analytic  continuation for all complex  $s$ and has simple poles
 at  $s=0$  and  $s=1$.  It  satisfies the  functional  equation  $\tilde
 \bE_{s}(\Omega)=\tilde \bE_{1-s}(\Omega)$ which  is a particular case
 of the general functional equation satisfied by the Eisenstein series~(\ref{e:FunRelSLd}).

 The Fourier expansion with respect to $\Omega_1$ is given by
 \begin{eqnarray}
 \nn  \bE_{s}(\Omega)                &    =   &                 2\zeta(2s)\,
   \Omega_2^s+2\sqrt{\pi}{\Gamma(s-\frac12)\over\Gamma(s)}\,\zeta(2s-1)   \,
   \Omega_2^{1-s} \\
 &+&{2\pi^s\over\Gamma(s)} \Omega_2^{\frac12}  \sum_{n\neq0} |n|^{s-\frac12}
   \sum_{0<d\atop    n/d\in    \mathbb    N}\,   {1\over    d^{2s-1}}\,
   K_{s-\frac12}(2\pi |n|\Omega_2)\, e^{2i\pi\,n\,\Omega_1}\,,
   \label{fourieres}
 \end{eqnarray}
where $K_s(x)$ is a modified Bessel function of the second-kind. These series are eigenfunctions of the Laplacian,
 \begin{equation}\label{e:lalpaSl2}
   \Delta_\Omega=
   \Omega_2^2(\partial_{\Omega_1}^2+\partial_{\Omega_2}^2)= 4\Omega_2^2\partial_\Omega\bar\partial_{\bar\Omega}\,,
 \end{equation}
 \begin{equation}
   \Delta_\Omega \bE_s(\Omega) = s(s-1)\, \bE_s(\Omega)\,.
   \label{eslaplace}
 \end{equation}

\medskip
\noindent{\bf Eisenstein series evaluated at special values}
\smallskip

$\bullet$ The $SL(2)$ Eisenstein  series has a pole at $s=1$.  Setting $s=1+\epsilon$ and expanding for small $\epsilon$ gives
 \begin{equation}
\bE_{1+\epsilon}(z)=                                 {\pi\over\epsilon}
-\pi\log(\Omega_2|\eta(\Omega)|^4)+2\pi(\gamma_E-\log(2))+O(\epsilon)\,,
\label{eisdiv}
\end{equation}
where $\gamma_E$ is Euler's  constant.
The regulated series,  $\hbE_1(\Omega)$, is defined by subtracting the pole and a constant to give
\begin{equation}\label{e:E1Def}
 \hbE_1(\Omega)     = -\pi\log(\Omega_2|\eta(\Omega)|^4)
 \end{equation}
 where $\eta(\Omega)$ is the  Dedekind function,
 \begin{equation}
   \eta(\Omega)=   e^{{i\pi   \Omega\over   12}}\,   \prod_{n=1}^\infty
   (1-e^{2i\pi n\Omega})\, .
 \end{equation}
Since
 $  \Delta \bE_{1+\epsilon}(\Omega) = \epsilon(1+\epsilon) \bE_{1+\epsilon}(\Omega)$\,,
  for any $\epsilon$ it follows that  
   \begin{equation}\label{e:Exp1}
   \Delta \hbE_1= \pi\,.
 \end{equation}

$\bullet$ The series with  $s=1/2$ appears to diverge, but is finite when defined in terms of a limit, 
\begin{eqnarray}\label{e:EHalf}
  \bE_{\half}(\Omega)&=&\lim_{\epsilon\to0} \bE_{\frac12+\epsilon}(\Omega)\\
\nn &=& 2 \,\Omega_2^{1\over2}(\gamma_E+\log(\Omega_2/(4\pi))+2\Omega_2^{1\over2}
   \sum_{(m,n)\in\mathbb Z^2}\, K_0(2\pi|mn|\Omega_2)\, e^{2i\pi mn\,\Omega_1}\,.
 \end{eqnarray}

$\bullet$\ The series with  $s=0$ is defined by analytic continuation to have the finite value
 \begin{equation}
 \bE_\epsilon(\Omega)=-1+\epsilon\, (\pi^{-1}\hbE_1-2\log(2\pi))+O(\epsilon^2)\,,
 \end{equation}
 which  is compatible  with  functional equation  of Eisenstein  series
 $\tilde \bE_{1+\epsilon}(\Omega)= \tilde \bE_{-\epsilon}(\Omega)$.

 \subsection{$SL(3)$ Eisenstein series}\label{sec:Sl3Series}

 For  the $d=3$ case it is useful to introduce the  integers $p_i = \epsilon_{ijk}
 d^{jk}$, where $\epsilon_{ijk}$  is the completely antisymmetric symbol ($\epsilon_{123}=1$),
and  \eqref{e:detM} becomes
 \begin{equation}\label{e:detMSl3}
 2\,\det\mathcal{M}= \epsilon^{ilm} \epsilon^{jkn}\, g_{ij}g_{kl}\,
 p_m p_n
 = (g^{-1})^{mn} \, p_m p_n\,,
 \end{equation}
which uses the fact that  $\epsilon^{ilm} \epsilon^{jkn}\, g_{ij}g_{kl}$ are
 the elements of the adjugate of the matrix $g_{ij}$ and that $g^{-1}= (\det g)^{-1} \,
 \textrm{adj}(g)^T$,   where   $\det    g=1$.    Therefore   the
 definition~\eqref{e:EisAnt} gives the functional relation between
 Eisenstein series
 \begin{equation}
 \label{sl3rel}
{\Gamma(s)\over\pi^s}\,      \bE^{SL(3)}_{[01];s}={\Gamma(\frac32-s)\over
  \pi^{\frac32-s}}\, \bE^{SL(3)}_{[10];\frac32-s}\,.
 \end{equation}

 \subsubsection{Fourier expansions}

 Using the parametrisation of $SO(3)\backslash SL(3)$ given in the main
 text the Eisenstein series $\bE^{SL(3)}_{[10];s}$ is defined by 
 \begin{equation}\label{eEm}
  \bE^{SL(3)}_{[10];s}=\sum_{(m_1,m_2,m_3)\neq (0,0,0)} \, {\nu_2^{-{s\over3}}\over \left({|m_1+m_2\Omega+m_3 B|^2\over \Omega_2}+{m_3^2\over\nu_2}\right)^s}\,,
 \end{equation}
 where $\nu_2^{-1}=\Omega_2 \, T_2^{2}$ is the inverse volume of the two-torus of
 compactification defined in~(\ref{nudef}) expressed in terms of the string variables, and
 $B=B_{\rm RR}+\Omega B_{\rm NS}$ is the usual combination of the RR and NS
 $B$-field (in the construction from the $L=1$ and $L=2$ supergravity loops
 there  is  no dependence  on  the  three-form  of  eleven -dimensional
 supergravity therefore we have to set $B=0$.)

  The $SO(3)\backslash SL(3)$ laplacian is given by~\cite{Kiritsis:1997em}
 \begin{equation}
 \Delta_{SO(3)\bs SL(3)}   =  4\Omega_2^2\partial_\Omega\bar\partial_{\bar\Omega}+
 {|\partial_{B_{\rm NS}}-\Omega\partial_{B_{\rm RR}}|^2\over\nu_2\Omega_2}+
 3\partial_{\nu_2}(\nu_2^2\partial_{\nu_2})\, ,
 \label{sl3laplace}
 \end{equation} 
which gives
 \begin{equation}
  \Delta_{SO(3)\bs SL(3)}\,  \bE^{SL(3)}_{[10];s} ={2s(2s-3)\over 3}\,  \bE^{SL(3)}_{[10];s}\,.
 \end{equation}

 For  $s\neq 3/2$ these Eisenstein series can be expanded
 using $T_2^{-2}=\nu_2\Omega_2$~\cite{Kiritsis:1997em,Basu:2007ru,Basu:2007ck}
 \begin{eqnarray}
   \label{e:EsExpaNu}
 \bE^{SL(3)}_{[10];s} &=&    \nu_2^{-{s\over3}}    \,      \bE_s(\Omega)+2
   \pi\,{\Gamma(s-1)\over\Gamma(s)} \,\zeta(2s-2) \nu_2^{2s-3\over
       3}\\
 \nn&+&
   {2\pi^s\over   \Gamma(s)}\,    \nu_2^{s-3\over6}\,   \Omega^{1-s\over2}_2\,
 \sum_{(m_1,m_2)\neq(0,0)\atop          m_3\neq0}\,         \left|m_2-m_1\Omega\over
   m_3\right|^{s-1}\times\\
\nn&\times&K_{s-1}(2\pi|m_3(m_2-m_1\Omega)|T_2)\,     e^{2i\pi
   m_3(m_1 B_{\rm RR}+m_2 B_{\rm NS})}\,.
 \end{eqnarray}
 Using the variables $(y_8,T)$  (where $y_8^{-1}=\Omega_2^2T_2$) this can be rewritten as
 \begin{eqnarray}
   \label{e:EsExpyT}
  \bE^{SL(3)}_{[10];s}&=&  2\zeta(2s)\,  y_8^{-{2s\over3}} +
   \sqrt\pi{\Gamma(s-\frac12)\over\Gamma(s)}\,               y_8^{2s-3\over
       6}\,
   \bE_{s-\frac12}(T)\\
 \nn&+&{2\pi^s\over\Gamma(s)}\, T_2^{2s-1\over4} y_8^{-{2s+3\over12}}\,
 \sum_{m_1\neq0,m_2\neq0}     \,     \left|m_1\over     m_2\right|^{s-\frac12}\,
 K_{s-\frac12}(2\pi\Omega_2|m_1m_2|)\,e^{2i\pi m_1m_2\Omega_1}\\
 \nn &+& {2\pi^s\over \Gamma(s)}\, \sqrt{T_2}\, y_8^{2s-3\over 6}\,
 \sum_{(m_1,m_2)\neq(0,0)\atop          m_3\neq0}\,         \left|m_2-m_1\Omega\over
   m_3\right|^{s-1}\times\\
\nn&\times&K_{s-1}(2\pi|m_3(m_2-m_1\Omega)|T_2)\,     e^{2i\pi
   m_3(m_1 B_{\rm RR}+m_2 B_{\rm NS})}\,.
 \end{eqnarray}

\medskip
\noindent{\bf Series evaluated at special values}
\smallskip

$\bullet$ For $s=3/2$ the expression has a logarithmic divergence associated
 with the one-loop divergence in eight dimensions discussed in the main
 text. The expression needs
 to be regulated, leading (in the $(\nu_2,\Omega)$ variables) to
 \begin{equation}
 \label{e:E321reg}
   \bE^{SL(3)}_{[10];{3\over   2}+\epsilon}=   {2\pi\over\epsilon}+
   4\pi(\gamma_E-1)  +\hbE^{SL(3)}_{[10];\frac32} +O(\epsilon)\,,
 \end{equation}
where   the regularised series $\hbE^{SL(3)}_{[10];\frac32}$  can be expanded in limit~(i) as
\begin{eqnarray}
  \label{e:E32reg}
\hbE^{SL(3)}_{[10];\frac32} = \nu_2^{-{1\over2}}\,\bE_{3\over2}(\Omega)+{4
   \pi\over3}\,\log(\nu_2)+O(e^{-\Omega_2^{\half}\nu_2^{-\half}},e^{-\Omega_2^{-\half}\nu_2^{-\half}})
\end{eqnarray}
 or in limit (ii) as
 \be
\hbE^{SL(3)}_{[10];\frac32}
 = {2\zeta(3)\over y_8}+2\hbE_{1}(T)+{2\pi\over3}\log(y_8)+O(e^{-(T_2y_8)^{-\half}},e^{-T_2^\half\,y_8^{-\half}})
  \label{e:E322reg}
  \ee
 Since
 \begin{equation}
   \Delta    \bE^{SL(3)}_{[10]; {3\over2}+\epsilon}    ={2\over    3}\epsilon
   (3+2\epsilon)\, \bE^{SL(3)}_{[10]; {3\over2}+\epsilon}  
 \end{equation}
 we deduce that
 \begin{equation} \label{e:Exp2}
     \Delta \hbE^{SL(3)}_{[10]; {3\over2}} =4\pi\, .
  \end{equation}

$\bullet$ For $s=1$ the expression using the $(\Omega,\nu_2)$ variables
 in~(\ref{e:EsExpaNu})    appears to diverge because    it   involves
 $\bE_1(\Omega)$ and  $\Gamma(s-1)$ and so seems to have a
 pole in $s$. But the pole cancels between the first two terms and no explicit
 subtraction  is  needed.  This is obvious from the  expansion  given
 in~(\ref{e:EsExpyT})  where  no  divergences  are  met  at  $s=1$.  The
 resulting expression is therefore 
 \begin{equation}\begin{split}\label{e:E32s1}
 \bE^{SL(3)}_{[10]; 1}  &= \lim_{\epsilon\to 0}\bE^{SL(3)}_{[10]; 1+\epsilon}\cr
     &=2\zeta(2)\,y_8^{-{2\over3}}        +y_8^{-{1\over6}}\,
 \bE_{\half}(T)+O(e^{-\sqrt{\Omega_2/\nu_2}},e^{-1/\sqrt{\Omega_2\nu_2}})\cr
&=\nu_2^{-\frac13}\left(\hbE_1(\Omega)-\pi\log(\nu_2)+
  2\pi(\gamma_E-\log(4\pi))\right)+O(e^{-(T_2y_8)^{-\half}},e^{-T_2^\half
\,y8^{-\half}})
\end{split} \end{equation}
 where we have used the expression for $\bE_{\half}(T)$ given in~(\ref{e:EHalf})
 Using  the duality  relation between  Eisenstein series  this  gives a
 definition of $\pi \bE^{SL(3)}_{[01]; \half}= \bE^{SL(3)}_{[10];1} $.

$\bullet$ For $s=1/2$ we get
\begin{equation}\begin{split}
  \bE^{SL(3)}_{[10];\frac12}   &=  \lim_{\epsilon\to  0}\bE^{SL(3)}_{[10];
    1+\epsilon}\cr
 &= \nu_2^{-\frac16} \, \bE^{SL(2)}_{\frac12}(\Omega)
 +{\pi\over3} \nu_2^{-\frac23} +O(e^{-\sqrt{\Omega_2/\nu_2}},e^{-1/\sqrt{\Omega_2\nu_2}})\cr
&=y_8^{-\frac13}\left(    {1\over\pi}     \hbE_1(T)    -    \log(y_8)+
  2(\gamma_E-\log(4\pi))\right)+O(e^{-(T_2y_8)^{-\half}},e^{-T_2^\half
y_8^{-\half}})
\label{newshalf}
\end{split}\end{equation}
The two set of equations~(\ref{e:E32s1}) and~(\ref{newshalf}) are compatible with the functional equation $\bE^{SL(3)}_{[10];1}=
\pi\bE^{SL(3)}_{[01];1/2}$.

 \subsection{$SL(5)$ Eisenstein series}\label{sec:Sl5Series}

In  the  following subsections  we will  determine  the entries in  the matrix
$A^{SL(5)}_s(u,v;r)$ defined in~(\ref{constdef}).  Recall that the columns of the matrix are labelled by $u$, which specifies the root, $\alpha_u$, which labels which of the $s_i$'s is non-zero.   The series associated with a particular $u$ is 
$\bE^{SL(5)}_{[0^{u-1}, 1, 0^{4-u}];s}$.
The rows, labelled by $v$, specify the node $\alpha_v$ that defines a  particular parabolic subgroup of the $SL(5)$ series. 

The detailed discussion of each entry will be given in subsections~(\ref{subone}) and~(\ref{subtwo}).   Since this is fairly complicated we will first summarize the results.  First note a simple consequence of the  symmetries of the  Weyl group  is the set of relations
\begin{eqnarray}
  A_s^{SL(5)}(u,1;r)&=&\pi^{2s-\frac52}\,{\Gamma(\frac52-s)\over
   \Gamma(s)}\, A_{\fiveh -s}^{SL(5)}(u,4;r)\\ 
\nn  A_s^{SL(5)}(u,2;r)&=&\pi^{4s-5}\,{\Gamma(\frac52-s)\Gamma(2-s)\over \Gamma(s-\frac12)\Gamma(s)}\, A_{\fiveh-s}^{SL(5)}(u,3;r)\,.
  \end{eqnarray}
 The explicit expressions for the entries are as follows
\begin{equation}
  \label{tab:A4}
  A_s^{SL(5)}(u,v;r)=
  \begin{pmatrix}
  ~(\ref{e:A11})&~(\ref{e:A12})&~(\ref{e:A13})&~(\ref{e:A14})\\
  ~(\ref{e:A21})&~(\ref{e:A22})&~(\ref{e:A23})&~(\ref{e:A24})\\
  ~(\ref{e:A31})&~(\ref{e:A32})&~(\ref{e:A33})&~(\ref{e:A34})\\
  ~(\ref{e:A41})&  ~(\ref{e:A42})&  ~(\ref{e:A43})&  ~(\ref{e:A44})\\
  \end{pmatrix}\,,
\end{equation}
where the entries number the equations where the constant terms can be
found.

\medskip
\noindent {\bf Constant terms of Eisenstein series at the special values in main text}
\smallskip

Since we are interested in the values of the constant terms at particular values of $s$ we will here summarize properties of the entries in~(\ref{tab:A4}) at those values.

$\bullet$ The $SL(5)$ series has a single pole at $s=5/2$.  Explicitly, setting $s=5/2 + \epsilon$ gives
\begin{equation}
  \label{e:SL5fundPole}
\bE^{SL(5)}_{[1000];\frac52+\epsilon} = {4\pi^2\over3\epsilon} +\hbE^{SL(5)}_{[1000];\frac52}+{8\pi^2\over9}(3\gamma_E-4)+O(\epsilon)
\end{equation}
The  constant  terms of $\hbE^{SL(5)}_{[1000];\frac52}$  for the parabolic subgroups considered in the main text are  
\begin{eqnarray}
 \int_{P(1,4)}\,         \hbE^{SL(5)}_{[1000];\frac52}&=&2r^8\zeta(5)+
 {4\over3} \hbE^{SL(4)}_{[100];2}- {16\pi^2\over 15}\,\log(r) \,,\\
 \int_{P(4,1)}\,  \hbE^{SL(5)}_{[1000];\frac52}&=&r^2 \bE^{SL(4)}_{[100];\frac52}-{64\pi^2\over15}\log(r)\,,\\
 \int_{P(3,2)}\,                       \hbE^{SL(5)}_{[1000];\frac52}&=&
 r^4\,\bE^{SL(3)}_{[10];\frac52}+{4\pi\over3}\, \hbE^{SL(2)}_{[1];1}-{16\pi^2\over5}\,\log(r)\,.
\end{eqnarray}

The series $\bE^{SL(5)}_{[0010];s}$ also has a pole when $s=5/2 +\epsilon$,
\begin{equation}
\bE^{SL(5)}_{[0010];\frac52+\epsilon}=
{2\pi^5\over9\epsilon}+{2\pi^3\over27}\,\left(6\pi^2\gamma_E-11\pi^2+36\zeta'(2)\right)+\hbE^{SL(5)}_{[0010];\frac52}+O(\epsilon)
  \label{e:SL5antiPole}
\end{equation}
and the relevant constant terms are 
\begin{eqnarray}
 \int_{P(1,4)}\,                      \hbE^{SL(5)}_{[0010];\frac52}&=&  
 r^4\bE^{SL(4)}_{[010];\frac52}+{2\pi^3\over9}\hbE^{SL(4)}_{[001];2} -{8\pi^5\over15}\log(r)\,,\\
 \int_{P(4,1)}\,      \hbE^{SL(5)}_{[0010];\frac52}&=& \zeta(4) r^6\bE^{SL(4)}_{[001];\frac52}+{2\pi\over3}\hbE^{SL(4)}_{[010];2} -{16\pi^5\over45}\log(r)\,,\\
 \int_{P(3,2)}\,      \hbE^{SL(5)}_{[0010];\frac52}&=&   10\zeta(4)\,\hbE^{SL(3)}_{[10];\frac32}-{32\pi^5\over45}\log(r)+{2r^4\over3}
  \bE^{SL(3)}_{[01];2}\bE^{SL(2)}_{[1];2}+ 2\zeta(4)r^{12}\,.
\end{eqnarray}

$\bullet$  The $SL(5)$ series $\bE^{SL(5)}_{[1000];s}$ is finite when $s=3/2$.  The constant
terms of interest to us are given by
\begin{eqnarray}
\int_{P(1,4)}\,  \bE^{SL(5)}_{[1000];\frac32}&=& r^{6\over5} \,
\bE^{SL(4)}_{[001];\frac32}+ 4\zeta(2) r^{16\over5}\, \\
\int_{P(3,2)}\,  \bE^{SL(5)}_{[1000];\frac32}&=&   r^{12\over5}  
\left(\hbE^{SL(3)}_{\frac32}+2\hbE_{[1];1}^{SL(2)}+8\pi\log(r)\right)
\end{eqnarray}

Furthermore, using   the   functional   equation   for   the   $SL(3)$   series ~(\ref{sl3rel})
 $\bE^{SL(3)}_{[01];1}=\pi    \,   \bE^{SL(3)}_{[10];1/2}$,      one    sees   that
$\bE^{SL(3)}_{[10];1/2}$ also contains a logarithmic term in its $P(2,3)$ constant term.

 \subsubsection{Parabolic subgroups $P(1,4)$ and $P(4,1)$}
 \label{subone}

 For the  maximal parabolic subgroup  $P_{\alpha_1}=P{(1,4)}$ obtained
 by   deleting   the   first   node   of   the   Dynkin   diagram  in
 figure~\ref{fig:dynkin}(iii) the matrix $g_{ij}$ has the block diagonal form 
 \begin{equation}\label{e:gP14}
 g_5=
 \begin{pmatrix}
   r^{-{16\over 5}} & 0\\
   0 &r^{4\over 5}\,  g_4
 \end{pmatrix}\,,
 \end{equation}
   where $ g_4$ is a  $4\times 4$ square matrix of unit determinant
   so  that $\det  g_5=1$. The  parabolic  subgroup  $P_{\alpha_4}=P(4,1)$ is
   obtained by deleting the last node of the Dynkin diagram in
 figure~\ref{fig:dynkin}(iii) and is 
   characterized by the matrix of the form
   \begin{equation}
     \label{e:gP41}
   g_5=
 \begin{pmatrix}
   r^{-{4\over 5}}\,  g_4&0\\
0&   r^{{16\over 5}} 
 \end{pmatrix} \,.
   \end{equation}
For these parabolic subgroups the Levi subgroup is $GL(1)\times SL(4)$.

 \paragraph{Constant term of the series $\bE^{SL(5)}_{[1000];s}$}
 \label{sec:decomp-seri-ea4100}

The constant term for the parabolic $P(1,4)$ is given by
 \begin{equation}
{\Gamma(s)\over\pi^s}\int_{P(1,4)} \, \bE^{SL(5)}_{[1000];s}=\!\!\!\!\!\!\!\!\!\!\!\!\!\!\!\!\!\!\!\! \sum_{(m,n_1,\dots,n_4)\in\mathbb Z^5\backslash\{(0,\dots,0)\}}\int_0^\infty
   {dt\over  t^{1+s}}   \,  \exp\left(-{\pi\over  t}   [m^2r^{-{16\over5}}+r^{4\over5}  \,
     n^T\cdot  g_4 \cdot n]\right)\,.
 \end{equation}
 Performing a Poisson resummation on $m$ one gets 
 \begin{equation}\label{e:A11}
     \int_{P(1,4)} \, \bE^{SL(5)}_{[1000];s}=2\zeta(2s)\, r^{16s\over 5}+ \sqrt\pi
      \,{\Gamma(s-\frac12)\over\Gamma(s)}\, r^{2-{4s\over5}}\, \bE^{SL(4)}_{[100];s-\frac12}\,,
 \end{equation}
which gives the element $A^{SL(5)}_s(1,1;r)$ of the $ A^{SL(5)}_s$ matrix in~(\ref{tab:A4}).

 The constant term in the $P(4,1)$ parabolic takes the form
 \begin{equation}
{\Gamma(s)\over\pi^s}\int_{P(4,1)} \, \bE^{SL(5)}_{[1000];s}=\!\!\!\!\!\!\!\!\!\!\!\!\!\!\!\!\!\!\!\! \sum_{(m,n_1,\dots,n_4)\in\mathbb Z^5\backslash\{(0,\dots,0)\}}\int_0^\infty
   {dt\over  t^{1+s}}   \,  \exp\left(-{\pi\over  t}   [m^2r^{{16\over5}}+r^{-{4\over5}}  \,
     n^T\cdot  g_4\cdot n]\right)\,.
 \end{equation}
Performing the Poisson resummation on the integers $(n_1,\dots,n_4)$ gives
 \begin{equation}\label{e:A41}
    \int_{P(4,1)} \,    \bE^{SL(5)}_{[1000];s}=  r^{{4s\over  5}}\,  \bE^{SL(4)}_{[100];s}+2\pi^2\zeta(2s-4)\,
     {\Gamma(s-2)\over \Gamma(s)}\, r^{8-{16s\over 5}}\,.
 \end{equation}
This gives the element $A^{SL(5)}_s(4,1;r)$ of the $ A^{SL(5)}_s$ matrix in~(\ref{tab:A4}).

 \paragraph{Constant term of the series $\bE^{SL(5)}_{[0001];s}$}
 \label{sec:decomp-seri-ea4001}

The constant terms for the parabolic $P(1,4)$ is given by
 \begin{equation}
{\Gamma(s)\over\pi^s}\int_{P(1,4)} \, \bE^{SL(5)}_{[0001];s}=\!\!\!\!\!\!\!\!\!\!\!\!\!\!\!\!\!\!\!\! \sum_{(m,n_1,\dots,n_4)\in\mathbb Z^5\backslash\{(0,\dots,0)\}}\int_0^\infty
   {dt\over  t^{1+s}}   \,  \exp\left(-{\pi\over  t}   [m^2r^{{16\over5}}+r^{-{4\over5}}  \,
     n^T\cdot  g_4^{-1} \cdot n]\right)\,.
 \end{equation}
 Performing a Poisson resummation on   $(n_1,\dots,n_4)$ gives
 \begin{equation}\label{e:A14}
     \int_{P(1,4)}  \, \bE^{SL(5)}_{[0001];s}= r^{{4s\over  5}}\,  \bE^{SL(4)}_{[001];s}+2\pi^2\zeta(2s-4)\,
     {\Gamma(s-2)\over \Gamma(s)}\, r^{8-{16s\over 5}}\,,
 \end{equation}
which gives the entry $A^{SL(5)}_s(1,4;r)$ of the $ A^{SL(5)}_s$ matrix in~(\ref{tab:A4}).

 The constant term in the $P(4,1)$ takes the form
 \begin{equation}
{\Gamma(s)\over\pi^s}\int_{P(4,1)} \, \bE^{SL(5)}_{[0001];s}=\!\!\!\!\!\!\!\!\!\!\!\!\!\!\!\!\!\!\!\! \sum_{(m,n_1,\dots,n_4)\in\mathbb Z^5\backslash\{(0,\dots,0)\}}\int_0^\infty
   {dt\over  t^{1+s}}   \,  \exp\left(-{\pi\over  t}   [m^2r^{-{16\over5}}+r^{{4\over5}}  \,
     n^T\cdot  g_4^{-1} \cdot n]\right)\,.
 \end{equation}
Performing the Poisson resummation on $m$ gives
 \begin{equation}\label{e:A44}
    \int_{P(4,1)}  \,    \bE^{SL(5)}_{[0001];s}= 2\zeta(2s)\, r^{16s\over 5}+ \sqrt\pi
      \,{\Gamma(s-\frac12)\over\Gamma(s)}\, r^{2-{4s\over5}}\, \bE^{SL(4)}_{[001];s-\frac12}\,,
 \end{equation}
which gives the entry $A^{SL(5)}_s(4,4;r)$ of the $ A^{SL(5)}_s$ matrix in~(\ref{tab:A4}).

 \paragraph{Constant term of the series $\bE^{SL(5)}_{[0100];s}$}
 \label{sec:decomp-seri-ea40100}
 
 To evaluate the constant terms for the parabolic $P(4,1)$ specified by the metric in~(\ref{e:gP41}) we will
write  the  lattice  sum in~(\ref{e:IntDef})  in the
 factorized form
 \begin{equation}
 \Gamma_{P(4,1)}= \sum_{(p,q)\in\mathbb Z^2} e^{-\pi V \, r^{{16\over5}} \, {|p+q\tau|^2 \over \tau_2} }\,
   \sum_{(m,n)\in  \mathbb Z^8} e^{-\pi  V\, r^{-{4\over5}}  {(m -\tau
       n)^T\cdot g_4
\cdot       (m-\bar\tau n) \over \tau_2}}\, .
 \end{equation}

 Starting   from the  representation in~(\ref{e:IntDef})  and unfolding
 the  $\Gamma_{(1,1)}$  lattice gives
 \begin{equation}\label{e:Int1}
   I^{(4,1)}_s(\Lambda)=    I^4_s(\Lambda)+   \int_0^\Lambda   dV\,V^{2s-1}\,
  \int_0^\infty {d\tau_2\over\tau_2^2}\, \sum_{m\neq0} e^{-\pi \, r^{\frac{16}5}V\,{
     m^2\over \tau_2}}\, \int_{-\frac12}^{\frac12}d\tau_1 \Gamma_{(4,4)}\,.
 \end{equation}
We are particularly interested in the finite part (order $\Lambda^0$) of this integral, which is given
by 
\begin{equation}
  I_s(\Lambda)^{(4,1)}|_{\Lambda^0}        =       2 {\Gamma(2s-1)\over
    (2\pi)^{2s-1}} \, \int_{P(4,1)} \,  \bE^{SL(5)}_{[0100];s}\,.
\end{equation}

The finite part of the first  term on the right-hand-side of~(\ref{e:Int1}) is given
by 
\begin{equation}
      I_s(\Lambda)^{4}|_{\Lambda^0}        =  2  r^{8s\over5}\,    {\Gamma(2s-1)\over
    (2\pi)^{2s-1}} \, \bE^{SL(4)}_{[010];s}
\end{equation}
To analyze the second term we  perform a Poisson  resummation on  half of  the integers  in the
 lattice  $\Gamma_{(4,4)}$   giving  the  representation   in  terms  of
 Kaluza-Klein momenta $p$ and windings $w$,
 \begin{equation}
   \Gamma_{(4,4)} =\left(\tau_2 r^{\frac45}\over V\right)^{2} 
   \,\sum_{(p,w)\in \Lambda_{(2,2)}}
   \,  e^{-\pi \tau_2\,  (V r^{-\frac45}  p^2 +  V^{-1} r^{\frac45}\,
     w^2)+2i\pi\,\tau_1\, p\cdot w}\,.
 \end{equation}
The  integral   over  $\tau_1$   projects  onto  the   subspace  $p\cdot
 w=0$ where $p^2=m^T\cdot g_4\cdot m$ and $w^2=n^T\cdot g_4^{-1}\cdot n$.  This  is solved by either $p=0$ or $w=0$.  So 
 the finite part  of the second term in~(\ref{e:Int1}) is  given by the contribution
 with $w=0$,
 \begin{eqnarray}
   I_s(\Lambda)^{(4,1)}|_{\Lambda^0}  &=&r^{\frac85}\, \int_0^\infty  dV \, V^{2s-3} \,
   \int_0^\infty dt\, \sum_{m\neq0\atop p\in\mathbb Z^4} 
   \, e^{-\pi V r^{16\over5} {m^2\over t}- \pi t\, {V p^2\over r^{\frac45}}
   }\\
\nn     &=&r^{6-{12s\over5}}     \,         \zeta(2s-3)\,
{\Gamma(s-\frac32)\Gamma(s-\frac12)\over \pi^{2s-2}}\, \bE^{SL(4)}_{[100];s-\frac12}\,.
 \end{eqnarray}
Thus, the  constant term for the parabolic $P(4,1)$ is
 \begin{equation}\label{e:A42}
 \int_{P(4,1)} \,  \bE^{SL(5)}_{[0100];s}=
     r^{8s\over 5} \bE^{SL(4)}_{[010];s}+r^{6-{12s\over5}} \,\pi^{3\over2}\, \zeta(2s-3)\, {\Gamma(s-\frac32)\over\Gamma(s)}\, \bE^{SL(4)}_{[100];s-\frac12}\,,
 \end{equation}
which  gives the  entry $A^{SL(5)}_s(4,2;r)$  of the  $ A^{SL(5)}_s$  matrix in
(\ref{tab:A4}).

For  the parabolic subgroup  $P(1,4)$ characterized  by the  metric in
(\ref{e:gP14}) the lattice sum takes the form
\begin{equation}
 \Gamma_{P(1,4)}= \sum_{(p,q)\in\mathbb Z^2} e^{-\pi V \, r^{-{16\over5}} \, {|p+q\tau|^2 \over \tau_2} }\,
   \sum_{(m,n)\in  \mathbb Z^8}  e^{-\pi V\,  r^{{4\over5}}  {(m -\tau
       n)^T\cdot g_4\cdot
       (m-\bar\tau n)\over \tau_2}}\,.
 \end{equation}
Performing  a  complete Poisson  resummation  on the  $\Gamma_{(1,1)}$
lattice and then using the same manipulations as before leads to the expression for the
constant term 
\begin{equation}\label{e:A12}
 \int_{P(1,4)}      \,      \bE^{SL(5)}_{[0100];s}=  \zeta(2s-1)  r^{12s\over5}
 \bE^{SL(4)}_{[001];s} +\pi
     r^{4-{8s\over 5}}\,{\Gamma(s-1)\over\Gamma(s)}\, \bE^{SL(4)}_{[010];s-\frac12}\,,
\end{equation}
which  gives the  entry $A^{SL(5)}_s(1,2;r)$  of the  $ A^{SL(5)}_s$  matrix in
(\ref{tab:A4}).

 \paragraph{Constant term of the series $\bE^{SL(5)}_{[0010];s}$}
 \label{sec:decomp-seri-ea40010}

This series is defined in section~\ref{e:Eanti}, as 
$\bE^{SL(5)}_{[0100];s}(g_5^{-1})$,   which is the same series as discussed in the previous paragraphs but evaluated with the inverse metric.  
Applying the previous results  it follows that the constant term on the parabolic subgroup
$P(1,4)$
is given by 
 \begin{equation}\label{e:A13}
 \int_{P(1,4)} \,  \bE^{SL(5)}_{[0010];s}=
     r^{8s\over 5} \bE^{SL(4)}_{[010];s}+r^{6-{12s\over5}} \, \pi^{3\over2}\, \zeta(2s-3)\, {\Gamma(s-\frac32)\over\Gamma(s)}\, \bE^{SL(4)}_{[001];s-\frac12}\,,
 \end{equation}
 which  gives the entry  $A^{SL(5)}_s(1,3;r)$ of  the $A^{SL(5)}_s$  matrix in
~(\ref{tab:A4}).

On the parabolic subgroup $P(4,1)$ the constant term is given by 
\begin{equation}\label{e:A43}
 \int_{P(4,1)}      \,      \bE^{SL(5)}_{[0010];s}= \zeta(2s-1)  r^{12s\over5}
 \bE^{SL(4)}_{[001];s} +\pi
     r^{4-{8s\over 5}}\,{\Gamma(s-1)\over\Gamma(s)}\, \bE^{SL(4)}_{[010];s-\frac12}\,,
\end{equation}
 which gives the entry $A^{SL(5)}_s(4,3;r)$ of the $A^{SL(5)}_s$ matrix in~(\ref{tab:A4}).
 \subsubsection{Parabolic subgroup $P(2,3)$ and $P(3,2)$}
 \label{subtwo}

The maximal parabolic subgroup $P_{\alpha_2}=P{(2,3)}$,  obtained by deleting the second  node, is characterized by the matrix
 \begin{equation}\label{e:gP23}
  g_5= \begin{pmatrix}
  r^{-{12\over 5}} g_2&0\\
0& r^{{8\over 5} } g_3
   \end{pmatrix}\,,
 \end{equation}
 where $g_3$ is square $3\times 3$ matrix
 and $g_2$ a  square $2\times 2$ matrix both  of unit determinant. The
 other parabolic $P_{\alpha_3}=P(3,2)$ is obtained
 by considering the matrix
  \begin{equation}\label{e:gP32}
  g_5= \begin{pmatrix}
  r^{-{8\over 5} } g_3&0\\
0& r^{{12\over 5}} g_2
   \end{pmatrix}\,.
 \end{equation}
For  these   parabolic  subgroups  the  Levi  subgroup   is  given  by
$GL(1)\times SL(2)\times SL(3)$.

 \paragraph{Constant term of the series $\bE^{SL(5)}_{[1000];s}$}
 \label{sec:decomp-seri-ea4100b}

For the parabolic $P(2,3)$ the metric takes the form given in~(\ref{e:gP23}),
leading to the integral representation
 \begin{eqnarray}
&&\int_{P(2,3)}                  \, \bE^{SL(5)}_{[1000];s}={\pi^s\over
  \Gamma(s)}\times\\
\nn &\times&\sum_{(m_1,\dots,m_3,n_1,n_2)\in\mathbb Z^5\backslash\{(0,\dots,0)\}}\int_0^\infty
   {dt\over   t^{1+s}}  \,  \exp\left(-{\pi\over   t}  [r^{{8\over5}}
     m\cdot g_3\cdot m^T
     +r^{-{12\over5}} \,
     n^T\cdot g_2 \cdot n]\right)\,.
 \end{eqnarray}
 Performing a Poisson resummation on the two integers $n_1$ and $n_2$ one gets one gets for the constant term for the
 parabolic $P(2,3)$
 \begin{equation}\label{e:A21}
 \int_{P(2,3)}\,      \bE^{SL(5)}_{[1000];s}=r^{12s\over 5} \bE^{SL(2)}_{[1];s}+ \pi\,{\Gamma(s-1)\over\Gamma(s)}\, r^{4-{8s\over5}}\, \bE^{SL(3)}_{[10];s-1}\,,
 \end{equation}
 which gives the entry $A^{SL(5)}_s(2,1;r)$ of the $ A^{SL(5)}_s$ matrix in~(\ref{tab:A4}).

 The parabolic $P(3,2)$ is obtained by using the metric~(\ref{e:gP32})
 and performing the Poisson resummation $(m_1,\dots,m_3)$ one gets
 gives the coefficient $A^{SL(5)}(4,3;r,s)$  of the $ A^{SL(5)}_s$ matrix in~(\ref{tab:A4}).
\begin{equation}\label{e:A31}
  \int_{P(3,2)}\,        \bE^{SL(5)}_{[1000];s}=               r^{8s\over              5}\,
      \bE^{SL(3)}_{[10];s}+\pi^{3\over2}\,{\Gamma(s-\frac32)\over\Gamma(s)}\,
      r^{6-{12s\over5}}\, \bE^{SL(2)}_{[1];s-\frac32}\,,
 \end{equation}
 which gives the element $A^{SL(5)}_s(3,1;r)$ of the $ A^{SL(5)}_s$ matrix in~(\ref{tab:A4}).

 \paragraph{Constant term of the series $\bE^{SL(5)}_{[0001];s}$}
 \label{sec:decomp-seri-ea4001b}

For the parabolic $P(2,3)$  the relevant metric is that in~(\ref{e:gP32})
and the integral representation for the constant term is given by
 \begin{eqnarray}
&&\int_{P(2,3)}                  \, \bE^{SL(5)}_{[0001];s}={\pi^s\over
  \Gamma(s)}\times\\
\nn &\times&\sum_{(m_1,\dots,m_3,n_1,n_2)\in\mathbb Z^5\backslash\{(0,\dots,0)\}}\int_0^\infty
   {dt\over   t^{1+s}}  \,  \exp\left(-{\pi\over   t}  [r^{-{8\over5}}
     m\cdot g_3^{-1}\cdot m^T
     +r^{{12\over5}} \,
     n^T\cdot g_2^{-1} \cdot n]\right)\,.
 \end{eqnarray}
 Performing a  Poisson resummation on the three  integers $m_1$, $m_2$
 and $m_3$ one gets the constant term for the
 parabolic $P(2,3)$,
 \begin{equation}\label{e:A24}
 \int_{P(2,3)}\,      \bE^{SL(5)}_{[0001];s}=r^{8s\over 5} \bE^{SL(3)}_{[01];s}+ \pi^{\frac32}\,{\Gamma(s-\frac32)\over\Gamma(s)}\, r^{6-{12s\over5}}\, \bE^{SL(2)}_{[1];s-\frac32}\,,
 \end{equation}
 which gives the entry $A^{SL(5)}_s(2,4;r)$ of the $ A^{SL(5)}_s$ matrix in~(\ref{tab:A4}).

In the case of  the  $P(3,2)$   parabolic  we perform  the  Poisson
 resummation on the two integers $n_1$ and $n_2$ one gets
\begin{equation}\label{e:A34}
  \int_{P(3,2)}\,        \bE^{SL(5)}_{[0001];s}=               r^{12s\over              5}\,
      \bE^{SL(2)}_{[1];s}+\pi\,{\Gamma(s-1)\over\Gamma(s)}\,
      r^{4-{8s\over5}}\, \bE^{SL(3)}_{[01];s-1}\,,
 \end{equation}
which gives the entry $A^{SL(5)}_s(3,4;r)$ of the $ A^{SL(5)}_s$ matrix in~(\ref{tab:A4}).
 \paragraph{Constant term of the series $\bE^{SL(5)}_{[0100];s}$}
 \label{sec:decomp-seri-ea40100b}
 
In the case of the parabolic $P(2,3)$ we decompose the lattice sum~(\ref{e:IntDef})   as
 $\Gamma_{P(2,3)}=\Gamma_{(2,2)}(r^{-{12\over5}}      g_2)           \,
 \Gamma_{(3,3)}(r^{{8\over5}}g_3)$.
Performing a Poisson resummation on the $\Gamma_{(2,2)}$ factor gives
\begin{equation}
  \Gamma_{(2,2)}= {r^{24\over5}\over V^2}\, \sum_{(m,n)\in\mathbb Z^4\bs\{0\}} e^{-\pi V^{-1}
    r^{12\over5}\, {(m-n\tau)^T\cdot g_2^{-1}\cdot (m-n\bar\tau)\over \tau_2}}\,,
\end{equation}
 and unfolding the lattice sum following the method described in appendix~\ref{sec:Rankin-Selberg} results in 
 \begin{eqnarray}
 \label{e:I23} I^{(2,3)}_s(\Lambda) &=& I^3_s(\Lambda)\\
\nn&+&r^{24\over5}\int_0^\Lambda dV\, V^{2s-3}\,
   \int_0^\infty{d\tau_2\over   \tau_2^2}  \,\sum_{
     n\in\mathbb    Z^2\backslash{(0,0)}}    \,   e^{-\pi  r^{{12\over5}}   {n^T\cdot
       g_2^{-1}\cdot n \over \tau_2\, V}\,
     }
   \,\int_{-\frac12}^{\frac12}d\tau_1\Gamma_{(3,3)}\\
 \nn&+&2r^{24\over5}\int_0^\Lambda dV\, V^{2s-3}\,\int_{\mathbb C^+}{d^2\tau\over \tau_2^2}\,
 \sum_{[M_{0,1}]}   e^{-\pi   \,    r^{{12\over5}}\,   {(1\,\bar\tau)
     M_{0,1}^Tg_2^{-1}M_{0,1} (1\,\bar\tau)^T\over V\tau_2}}\, \Gamma_{(3,3)}\,.
 \end{eqnarray}
We are interested in the finite part of this integral,
\begin{equation}
  I^{(2,3)}_s(\Lambda)|_{\Lambda^0}=  2 {\Gamma(2s-1)\over(2\pi)^{2s-1}}
  \int_{P(2,3)} \, \bE^{SL(5)}_{[0100];s}\,.
\end{equation}
The first term in the right-hand-side of~(\ref{e:I23}) leads to
\begin{equation}
  I^3_s(\Lambda)|_{\Lambda^0}= 2r^{8-{16s\over5}}\,{\Gamma(2s-3)\over(2\pi)^{2s-3}}\zeta(2s-3) \bE^{SL(3)}_{[01];s-1}\,.
\end{equation}
 The  second  term  is  treated   as  in  the  previous  section.  The
 integration over $\tau_1$ projects on  the sector $p\cdot w=0$ of the
 $\Gamma_{(3,3)}$ lattice  and the contribution  constant in $\Lambda$
 is given by the $p=0$ term
 \begin{eqnarray}
  \nn  (I^{(2,3)}_s(\Lambda)|_{\Lambda^0})_{2nd~line}&=&  r^{12\over5}
  \int_0^\infty {dV\over \,
  V^{\frac92-2s}}\,\int_0^\infty{d\tau_2\over \tau_2^{\frac12}}\,
 \sum_{ n\in\mathbb Z^2\backslash{(0,0)}\atop p\in\mathbb Z^3} e^{-\pi
   {r^{12\over5}\over V}\,
   { n^T\cdot g_2^{-1}\cdot n\over\tau_2}-\pi\tau_2 { w^2\over V r^{\frac85}}}\\
 &=&{1\over2}\,r^{2+{4s\over5}} \, \left(\Gamma(s-\frac12)\over \pi^{s-\frac12}\right)^2\, \bE^{SL(2)}_{[1];s-\frac12}\bE^{SL(3)}_{[10];s-\frac12}\,.
 \end{eqnarray}
 In  the last  line the  sum  is over  the representative  $[M_{0,1}]$
 defined in~(\ref{e:Rep})
 \begin{equation}
   M_{0,1}=
   \begin{pmatrix}
     m& j\\ 0 & n
   \end{pmatrix}
 \qquad 0\leq j< m, n\neq0\,.
 \end{equation}
 The finite contribution from the last line is given by
 \begin{eqnarray}
  \nn      (I^{(2,3)}_s(\Lambda)|_{\Lambda^0})_{3rd~line}&=&     2r^{24\over5}     \int_0^\infty     dV\,
  V^{2s-3}\,\int_{\mathbb C^+} {d^2\tau\over \tau_2^2}\,
 \sum_{[M_{0,1}]} e^{-\pi \, { r^{12\over5}\over V}\, {(1\,\bar\tau)
     M_{0,1}^Tg_2^{-1}M_{0,1}      (1\,\bar\tau)^T\over\tau_2}}\\
 &=&4r^{24s\over5} {\Gamma(2s-1)\over (2\pi)^{2s-1}} \zeta(2s)\zeta(2s-1)\,,
 \end{eqnarray}
where we have used the fact that this  contribution only arises from the sector with
$\Gamma_{(3,3)}\sim r^{24/5} V^{-3} $.

Collecting the various contributions, the constant term for the parabolic $P(2,3)$ reads
 \begin{equation}\label{e:A22}\begin{split}
 \int_{P(2,3)}   \,    \bE^{SL(5)}_{[0100];s}&=  2   r^{24s\over5}\,
 \zeta(2s)\zeta(2s-1)\cr
&+(2\pi)^2\,r^{8-{16s\over5}}
 {\Gamma(2s-3)\over\Gamma(2s-1)}\,\zeta(2s-3) \, \bE^{SL(3)}_{[01];s-1}\cr
&+{\sqrt\pi\over2}\,
 r^{2+{4s\over5}} \, {\Gamma(s-\frac12)\over\Gamma(s)}\, \bE^{SL(2)}_{[1];s-\frac12}\bE^{SL(3)}_{[10];s-\frac12}\,,
\end{split} \end{equation}
which gives the entry $A^{SL(5)}_s(2,2;r)$ of the $ A^{SL(5)}_s$ matrix in~(\ref{tab:A4}).

Similar manipulations apply to the analysis of the parabolic subgroup $P(3,2)$, leading to 
 \begin{equation}\label{e:A32}\begin{split}
\int_{P(3,2)}             \,              \bE^{SL(5)}_{[0100];s}&=           r^{16
   s\over5}\,\zeta(2s-1)\bE^{SL(3)}_{[01];s}\cr
&+\pi\,r^{4-{4s\over5}}\,{\Gamma(2s-2)\over\Gamma(2s-1)}\, \bE^{SL(2)}_{[1];s-1}
 \bE^{SL(3)}_{[10];s-\frac12} \cr
&+2(2\pi)^3\,r^{12-{24s\over5}} \, {\Gamma(2s-4)\over \Gamma(2s-1)}\, \zeta(2s-4)\zeta(2s-3)\,,
\end{split} \end{equation}
which gives the entry $A_s^{SL(5)}(3,2;r)$ of the $ A^{SL(5)}_s$ matrix in~(\ref{tab:A4}).

 \paragraph{Constant term of the series $\bE^{SL(5)}_{[0010];s}$}
 \label{sec:decomp-seri-ea40010b}
Applying the same manipulation as before one finds the constant term for the parabolic $P(2,3)$
 \begin{equation}\begin{split}
\label{e:A23} \int_{P(2,3)}             \,              \bE^{SL(5)}_{[0010];s}&=            r^{16
   s\over5}\,\zeta(2s-1)\,\bE^{SL(3)}_{[10];s}\cr
&+\pi\,r^{4-{4s\over5}}\,{\Gamma(2s-2)\over\Gamma(2s-1)}\, \bE^{SL(2)}_{[1];s-1}
 \bE^{SL(3)}_{[01];s-\frac12} \cr
&+2(2\pi)^3r^{12-{24s\over5}} \, {\Gamma(2s-4)\over \Gamma(2s-1)}\, \zeta(2s-3)\zeta(2s-4)\,,
\end{split} \end{equation}
which gives the entry $A^{SL(5)}_s(2,3;r)$ of the $ A^{SL(5)}_s$ matrix in~(\ref{tab:A4}).

Finally,   similar manipulations applied to the  parabolic subgroup $P(3,2)$ lead
to 
 \begin{equation}\begin{split}
 \int_{P(3,2)}  \,  \bE^{SL(5)}_{[0010];s}&=  2r^{{24s\over5}} \,
 \zeta(2s-1)\zeta(2s)\cr
&+{\sqrt\pi\over2}\,r^{2+{4s\over5}}\,{\Gamma(s-\frac12)\over\Gamma(s)}\, \bE^{SL(2)}_{[1];s-\frac12}
 \bE^{SL(3)}_{[01];s-\frac12} \cr
&+ (2\pi)^2         r^{8-{16
   s\over5}}\,{\Gamma(2s-3)\over\Gamma(2s-1)}\,\zeta(2s-3)\,\bE^{SL(3)}_{[10];s-1}\,,
\end{split}\label{e:A33} \end{equation}
which  gives the  entry $A^{SL(5)}_s(3,3;r)$  of the  $ A^{SL(5)}_s$  matrix in
(\ref{tab:A4}).

 \section{The $SO(d, d)$ Eisenstein series}\label{sec:Dnseries}

We will here consider Eisenstein series for $SO(d,d)$ groups defined with respect to the  Dynkin label $[1,0^{d-1}]$
(recall our convention for labelling the nodes in the case of  $SO(d,d)$ groups shown in figure~\ref{fig:dynkin})(ii).  These are analogous to the Epstein series discussed earlier in the case of $SL(d)$ groups.  In this case the series depend on the coset $SO(d)\times  SO(d)\bs SO(d,d)$.

In order  to define these  Eisenstein series we will  consider various
integrals involving the lattice sum $\Gamma_{(d,d)}$ 
  \begin{equation}\label{e:LatDefI}
   \Gamma_{(d,d)}=\sqrt{\det g}\, \sum_{(m^i,n^i)\in\ZZ^d\times
     \ZZ^d}  \,  \exp(-{\pi\over\tau_2}\, (g_{ij}+b_{ij}) (m_i -\tau  n_i)  (m_j
   -\bar\tau n_j) )\,,
 \end{equation}
which typically arises in compactifications of string or field theory loop integrals on $\calT^d$.  We will introduce the  volume  of the  $d$-torus, $V_{(d)}=\sqrt{\det
  g}$ and the rescaled metric, $\tilde g$,  defined by   $g_{ij}= V_{(d)}^{2\over d}\, \tilde  g_{ij}$, so that
$\det\tilde g=1$.
A sensible definition of the $SO(d,d)$ Eisenstein series of relevance to us is the manifestly invariant function
 \begin{equation}\label{e:DnDef1}
\bE^{SO(d,d)}_{[1,0^{d-1}];s}=   {\pi^{s}\over 2\zeta(2s+2-d)\Gamma(s)}\,\int_{\calF_{SL(2,\mathbb Z)}} {d^2\tau\over
     \tau_2^2} \bE_{s+1-{d\over2}}(\tau)\,(\Gamma_{(d,d)}-V_{(d)})\,.
\end{equation}
The  analysis  in  the  body   of  the  paper  and  in  the  following
demonstrates that,  for the  appropriate values of  $s$, this  has the
correct  behaviour   in  the  appropriate   limits.   Furthermore,  it
satisfies a  Laplace eigenvalue equation  of the appropriate  form, as
well as the correct functional equation.

 [The definition of the Eisenstein series in~(\ref{e:DnDef1}) differs from that given in~(3.10)
of~\cite{Pioline:Automorphic}     and
in~\cite{Kiritsis:1997em,Pioline:1997pu}.] 

 We are particularly interested in the series with $s=d/2-1$, which is  given
by 
\begin{equation}
  \label{e:DnDef2}
\bE^{SO(d,d)}_{[1,0^{d-1}];{d\over2}-1}=   {\pi^{{d\over2}-1}\over \Gamma({d\over2}-1)}\,\int_{\calF_{SL(2,\mathbb Z)}} {d^2\tau\over
     \tau_2^2}\,(\Gamma_{(d,d)}-V_{(d)})\,,
\end{equation}
where we have used $\bE_0(\tau)=-1$.
Instead of subtracting the volume factor we could have regularised the
series     by    analytically    continuing     in    $s$     as    in
appendix~\ref{sec:genusone}.

Using the differential equation for the lattice factor given in~\cite{Pioline:Automorphic}
\begin{equation}
 (\Delta_{SO(2)\bs SL(2)}- \Delta_{SO(d)\times SO(d)\bs SO(d,d)}-{d(d-2)\over 4})\Gamma_{(d,d)}=0
\end{equation}
we find that 
\begin{equation}\label{e:DiffDn}
    \Delta_{SO(d)\times SO(d)\bs SO(d,d)} \bE^{SO(d,d)}_{[1,0^{d-1}];s} = 2s(1-d+s) \bE^{SO(d,d)}_{[1,0^{d-1}];s}\,.
\end{equation}
These    equations    are  particular  cases
of~(\ref{e:LaplaEisenstein}) for the value of the weight vector  $\lambda$
specified by the Dynkin label $[s,0,\dots,0]$.

Using the method of
orbits~\cite{Dixon:1990pc,Kiritsis:1997em,Bachas:1997mc,Kiritsis:1997hf,Pioline:Automorphic,Pioline:2001jn}
reviewed in appendix~\ref{sec:Rankin-Selberg},
this Eisenstein series can be expanded in terms of $SL(d)$ series as
 \begin{eqnarray}
 && \bE^{SO(d,d)}_{[1,0^{d-1}];s}=V_{(d)}\,{\pi^{s} \over \Gamma(s)}\,\sum_{m_i\in\ZZ^d\bs\{0\}}\int_0^\infty{d\tau_2\over\tau_2} 
 \tau_2^{s-\frac{d}2} \, e^{-\pi {m_i g_{ij} m_j\over \tau_2}}\\
\nn &+&V_{(d)}\, {\zeta(2s+1-d)\over \zeta(2s+2-d)}{\pi^{s+\frac12}\over\Gamma(s)} {\Gamma(s+\frac{1-d}2)\over            \Gamma(s+1-\frac{d}2)}\,\sum_{m_i\in\ZZ^d\bs\{0\}}\int_0^\infty{d\tau_2\over\tau_2}
 \tau_2^{\frac{d}2-s-1} \, e^{-\pi {m_i g_{ij} m_j\over \tau_2}}\\
\nn &+&V_{(d)}{\pi^{s}\over \zeta(2s+1-d)\Gamma(s)}\,
 \sum_{(m_i,n_i)\in \ZZ^{2d}\bs\{0\}}\int_{\IC^+} {d^2\tau\over \tau_2^2} \, \bE_{s+1-\frac{d}2}(\tau)\,
 e^{-{\pi\over\tau_2} (g_{ij}+b_{ij}) (m_i-\tau n_i) (m_j- \bar\tau n_j)}\,,
\end{eqnarray}
leading to 
\begin{eqnarray}\label{e:DnToAn}
\bE^{SO(d,d)}_{[1,0^{d-1}];s}&=&V_{(d)}^{2s\over d}\,
  \bE^{SL(d)}_{[0^{d-2},1];s}\\
 \nn&+& V_{(d)}^{2-{2(s+1)\over d}}\,\pi^{d-1\over2}\,
{ \zeta(2s+1-d)\over \zeta(2s+2-d)}{\Gamma(s+\frac{1-d}2)\over\Gamma(s)}\,
\bE^{SL(d)}_{[1,0^{d-2}];s+1-{d\over 2}}+O(e^{-g_{ij}})
 \,,
 \end{eqnarray}
where we have made use functional equation~(\ref{e:EfundPois}) for the $SL(d)$ series. 
This expansion corresponds  to the constant term of  the series for the
parabolic subgroup obtained by  deleting the node $\alpha_d$ with Levi
subgroup $GL(1)\times SL(d)$.

For  the $d=3$ case comparison of  the expansion in \eqref{e:DnToAn} with
the   expansion  of    the  $SL(4)$ series,    $\bE^{SL(4)}_{[010];s}$
in~(\ref{e:SL4toSL3}) leads to 
\begin{equation}
  \label{e:D3eqA3}
  \bE^{SL(4)}_{[010];s}= \zeta(2s-1)\,\bE^{SO(3,3)}_{[100];s}\,.
\end{equation}
In the case of $s=d/2-1$  we get\footnote{ %
 This expansion matches the one of appendix~C
 of~\cite{Pioline:Automorphic} which uses $SL(d)$ series with non unit
 determinant. We would like to thank Boris Pioline for a clarification
 about this point.  }
\begin{equation}\label{e:DntoAn}
\int_{P_{\alpha_d}} \bE^{SO(d,d)}_{[1,0^{d-1}];{d\over2}-1}=V_{(d)}^{1-{2\over
    d}}\,\bE^{SL(d)}_{[0^{d-2},1];{d\over2}-1}
+{V_{(d)}\over                3}
{\pi^{{d\over2}}\over\Gamma({d\over2}-1)}\,,
\end{equation}
where we have used $\bE^{SL(d)}_{[1,0^{d-2}];0}=-1$.

\subsection{Constant term on the Parabolic subgroup $P_{\alpha_1}$ }
\label{sec:parab-subgr-dd}

The  constant term of the series  defined   in
(\ref{e:DnDef1})  on the parabolic  subgroup obtained  by removing
the first    node    of     the    Dynkin    diagram   in
figure~\ref{fig:dynkin}(ii) is expressed in terms of series for the  parabolic subgroup with
 Levi component $GL(1)\times SO(d-1,d-1)$.
This is analysed by splitting the metric of the $d$-torus in the form
\begin{equation}
  g_{IJ}=
  \begin{pmatrix}
    \tilde g_{ij}&0\\
0 & r^2
  \end{pmatrix}\,.
\end{equation}
so that  the lattice factor  $\Gamma_{(d,d)}= \Gamma_{(d-1,d-1)}\times
\Gamma_{(1,1)}$,giving
 \begin{equation}
\int_{P_{\alpha_1}} \, \bE^{SO(d,d)}_{[1,0^{d-1}];{d\over2}-1}= {\pi^{{d\over2}-1}\over \Gamma({d\over2}-1)}\,\int_{\calF_{SL(2,\mathbb Z)}} {d^2\tau\over
     \tau_2^2}             
  \left( \Gamma_{(d-1,d-1)}\Gamma_{(1,1)}-V_{(d)}\right)\,.
\end{equation}
Since $\Gamma_{(1,1)}$ is given by the sum
\begin{equation}
  \Gamma_{(1,1)}=r_d\,    \sum_{(m,n)\in    \mathbb    Z^2}
  \,e^{-\pi r_d^2{|m +n\tau|^2\over\tau_2}}\,,
\end{equation}
one can  evaluate this integral  by unfolding the $\Gamma_{(1,1)}$ factor as
in~\cite{Bachas:1997mc}, to get
\begin{eqnarray}
 \nn \int_{P_{\alpha_1}} \, 
\bE^{SO(d,d)}_{[1,0^{d-1}];{d\over2}-1}&=&{\pi^{{d\over2}-1}\over\Gamma({d\over2}-1)}\,r_d\,\Big(
 \int_{\calF_{SL(2,\mathbb Z)}} {d^2\tau\over
     \tau_2^2} (\Gamma_{(d-1,d-1)}-V_{(d-1)})\\
&+&\sum_{m\in\mathbb            Z\bs\{0\}}            \int_0^\infty
{d\tau_2\over\tau_2^2} e^{-\pi {r_d^2 m^2\over \tau_2}}\,\int_{-\frac12}^{\frac12}d\tau_1 \Gamma_{(d-1,d-1)}
\Big)
\end{eqnarray}
where  $V_{(d)}=r_d  \  V_{(d-1)}$.    Using  the  second
representation in~(\ref{e:LatDef})
for the lattice sum in the second line we find 
\begin{equation}
  \int_{P_{\alpha_1}}           \,    \bE^{SO(d,d)}_{[1,0^{d-1}];{d\over2}-1}=  
 2\zeta(d-2)\, r_d^{d-2}+
  r_d\, \sqrt\pi\,{\Gamma({d\over2}-{3\over2})\over \Gamma({d\over2}-1)}\,
 \bE_{[1,0^{d-2}];{d-1\over2}-1}^{SO(d-1,d-1)}\,.
\end{equation}
For the  $SO(5,5)$ case  used in the  main text we  have
\begin{equation}
  \label{e:D5toD4}
\int_{P_{\alpha_1}}\,   \bE^{SO(5,5)}_{[1,0^4];{3\over2}}
=2\zeta(3)\,r_5^3+ 2r_5\, \bE_{[1000];1}^{SO(4,4)}\,.
\end{equation}
%

 \section{Genus-one integrals in string theory}\label{sec:genusone}

 In this  appendix we  evaluate the one-loop  integrals arising  in the
 derivative  expansion of  the  genus-one four-graviton amplitude  in
 $10-d$  dimensions,  which  was  discussed  in~\cite{Green:2008uj}.

 First we will introduce some notation appropriate for the evaluation of the terms that contribute to the analytic part of the amplitude at any order in $\alpha' = \ell_s^2$ on a genus-$h$ world-sheet.  This expansion involves integration over the world-sheet moduli, $\calM$, with measure  $d\mu(\calM)$.  In principle, this leads to integrals of the form  
 \begin{equation}
   \label{e:I22Def}
 I^{(d)}_h(j_h^{(p,q)})=           \int_{\mathcal{M}_g}\,  d\mu(\calM)\,
 j_h^{(p,q)}(\calM)\, \Gamma_{(d,d)}\,,
 \end{equation}
 where $\Gamma_{(d,d)}$ is the genus-$h$ generalisation of the (even) Lattice sum defined in~(\ref{e:LatDef}) and 
 the function  $j_h^{(p,q)}(\calM)$ is a specific  modular function of
 the world-sheet complex structure. 
 This integral is invariant under $SO(d)\times SO(d)\bs SO(d,d)$.

For genus-$h \le 3$  the integration  over  the moduli  space  of  Riemann  surfaces can  be
 evaluated  directly by  integration over  the fundamental  domain for
 $Sp(h,\mathbb Z)$, which is evaluated   in
 appendix~\ref{sec:highergenus}. 
Beyond  that order  the dimension  of  the (complex)  moduli space  of
Riemann surfaces $3(h-1)$
is strictly smaller than the number of parameters in the period matrix
$h(h+1)/2$, which leads to technical difficulties in defining the integration over moduli for genus $h \ge 4$.  
 
Much more is known about the genus-one function $j^{(p,q)}_1$ than other values of
$h$\footnote{This  notation identifies $j_1^{(p,q)}$  with $j^{(p,q)}$
  introduced in the $h=1$ case in \cite{Green:2008uj}.}. 
 In the genus-one case ($h=1$) there is a single modulus so $\calM \to \tau$ and 
 $\int_{\mathcal{M}_1}  d\mu(\calM)  =  \int_{\mathcal{F}_{SL(2,\ZZ)}}
 d^2\tau/\tau_2^2$.   The functions $j_1^{(p,q)}(\tau)$  are invariant
 under $SL(2,\mathbb Z)$ transformations of $\tau$. 
 Although the  genus-one string amplitude is finite, when performing the derivative
expansion   the  separation  of the  analytic   contribution   from  the
non-analytic contribution may introduce divergences in each term separately, which cancel in the total amplitude.    In particular,~(\ref{e:I22Def}) diverges for  
large  $\tau_2$. Following
the  method of~\cite{Green:1999pv,Green:2008uj}  one can cut off
the fundamental  domain so that $\tau_2\leq L$. The  total string amplitude
is independent of $L$ and all dependence on $L$ cancels between 
$I^{(d)}_1(j_1^{(p,q)})$ and  the non-analytic  part  of the
amplitude.  This is a fairly simple procedure and in this appendix we will only quote the result for the $L$ independent contributions.

Determining the form of the functions $j_1^{(p,q)}$ was a major part of~\cite{Green:2008uj}.  
At  low orders  in the  expansion $j_1^{(p,q)}$  is simply  a  linear
combination of $SL(2)$ Eisenstein
 series   $\bE_s$  and one  can   apply  the   results  of
 appendix~\ref{sec:Dnseries}, giving
 s manifest $SO(d,d)$ invariance 
 \begin{equation}
 I^{(d)}_1(\bE_s)
    ={2\zeta(2s)\Gamma(s+{d\over2}-1)\over\pi^{s+{d\over2}-1}}\,\bE^{SO(d,d)}_{[1,0^{d-1}];s+{d\over2}-1}
+
 V_{(d)}\,\int_{\cF_{SL(2,\ZZ)}}{d^2\tau\over\tau_2^2}\, \bE_{s}(\tau)\,.
 \label{soddloop}
 \end{equation}
The last term is divergent for $\Re\textrm{e}(s)>1$ but can be
 regularised by  cutting off the  fundamental domain at $\tau_2  = L$,
 where $L\gg 1$, as in~\cite{Green:1999pv}. As mentioned above, 
 terms that diverge as positive powers of $L$ can be dropped since they cancel with contributions from nonanalytic terms in the amplitude, which we are not considering here.  The only real concern might have been  $\log L$ terms, which arise at poles in $s$ -- but these are regularised by subtracting them.     
For  $\Re\textrm{e}(s)\in ]0,1[$  the  integral of
$\bE_s$ converges, and since this function is an eigenfunction of the $SL(2)$
Laplacian in~(\ref{e:lalpaSl2}) we deduce that
\begin{equation}
    \int_{\cF_{SL(2,\ZZ)}}{d^2\tau\over\tau_2^2}\,   \bE_s(\tau)=   0,
    \qquad \textrm{for} \ \Re\textrm{e}(s)\in ]0,1[\, .
\end{equation}
By analytic continuation we set to zero the value of this integral for all
values of $s$ different from $s=0$ and $s=1$ so that 
\begin{equation}
  \label{e:genusOnedGen}
  I^{(d)}_1(\bE_s)
    ={2\zeta(2s)\Gamma(s+{d\over2}-1)\over\pi^{s+{d\over2}-1}}\,\bE^{SO(d,d)}_{[1,0^{d-1}];s+{d\over2}-1}\, .
\end{equation}
  Substituting      $s=0$     in      the
expansion of the $SO(d,d)$ series~(\ref{e:DnToAn}), and using the fact that
 $\bE^{SL(d)}_{[10\cdots 0];s=0}=\bE^{SL(d)}_{[0\cdots 01];s=0}=-1$
 and that the volume of the fundamental domain for $SL(2,\ZZ)$ is
 $\pi/3$   we find that
 \begin{equation}\label{e:genusOned}
  I^{(d)}_1(1)
= {\Gamma(\frac{d}2)\over\pi^{d\over2}}\, \bE^{SO(d,d)}_{[1,0^{d-1}];\frac{d}2-1}
={V_{(d)}\over\pi}\,\left(4\zeta(2)                      +V_d^{-{2\over
      d}}{\Gamma({d\over2}-1)\over\pi^{d\over2}}\,
  \bE^{SL(d)}_{[0^{d-2},1]; {d\over2}-1} + O(e^{-g_{ij}}) \right)\,.
\end{equation}
We will now consider the $d=2$ and the $d=3$ cases in more detail.

 \subsection{The genus-one amplitude on a two-torus}\label{sec:G1T2}
  For the special case with $d=2$ an application of 
  the method of  orbits of appendix~\ref{sec:Rankin-Selberg}, together
  with 
  the regularisation by analytic continuation described above, gives
  \begin{equation}
I^{(2)}_1(\bE_s)=   \int_{\mathcal{F}_{SL(2,\ZZ)}}\,{d^2\tau\over\tau_2^2}\,
 \bE_s(\tau)\,        \Gamma_{(2,2)}={\Gamma(s)\over \pi^s}\, \bE_{s}(T)\,\bE_{s}(U)\,,
   \label{e:EsLat}
  \end{equation}
 where $T$ and $U$ are  respectively the K{\"a}hler and complex structure
 of the $\cT^2$ of compactification.
 This   leads  to   the   following  expressions   for  the   one-loop
 contributions to  the higher-derivative interactions.  
 \begin{itemize}
 \item  The coefficient of the  $\cR^4$ interaction~\cite{Green:2008uj}  is  given by  the
   lowest order term in the  expansion of the genus-one diagram, which
   has $j_1^{(0,0)}=1$.  Setting $s=\epsilon$ and  considering the small
   $\epsilon$
 expansion of~(\ref{e:EsLat}) gives
  \begin{equation}
  \label{e:poleeq}
  I_1^{(2)}(\bE_\epsilon)=    \int_{\mathcal{F}_{SL(2,\ZZ)}}\,{d^2\tau\over\tau_2^2}\,\bE_\epsilon(\tau)
  \Gamma_{(2,2)}=        {1\over\epsilon}-{1\over       \pi}       \,
  (\hbE_1(T)+\hbE_1(U)+\log \mu)+o(\epsilon)
 \end{equation}
where the hat notation again  denotes the subtraction of the pole part
of $\bE_s$ and $\log\mu=\pi(\gamma_E-4\log(2)-3\log(\pi))$.
The $1/\epsilon$-pole corresponds to the ultraviolet divergence of the
one-loop  supergravity  amplitude.   This  pole  cancels  against  an
equivalent  
non-analytic        contribution        in        the        genus-one
amplitude~\cite{Green:2008uj}.  The same finite expression is obtained by decompactifying the analytic $D=7$
$\cR^4$ coefficient shown in~\eqref{e:finiteoneloop}.
Therefore, the analytic contribution is given by
 \bea
I^{(2)}_1(j_1^{(0,0)})
  =  {1\over\pi}\,(\hbE_1(T)+ \hbE_1(U)+\log\mu)\,,
 \label{e:R4oneloop}
 \eea
The  $\log  \mu$   term  is  interpreted as the  scale  of   the  massless  threshold
contribution, $\cR^4\,\log(-\ell_s^2\,s)$, to the nonanalytic part of the amplitude in eight dimensions.

 \item The $\partial^4\cR^4$ coefficient is determined by  the function $j_2^{(1,0)}=\bE_2(\tau)/(4\pi)^2$
~\cite{Green:1999pv,Green:2008uj}, which gives 
 \begin{equation}\label{e:D4R4oneloop}
   I^{(2)}_1(j_2^{(1,0)})  ={1\over 16\pi^4}\, \bE_2(T)\,\bE_2(U)\,.
 \end{equation}
 
 \item The genus-one contribution to the $\partial^6\cR^4$ coefficient~\cite{Green:2008uj} is determined  by the function $j_1^{(0,1)}=10\bE_3(\tau)/(4\pi)^3+\zeta(3)/32$, resulting in 
 \begin{equation}\label{e:D6R4oneloop}
    I^{(2)}_1(j_1^{(0,1)}) ={10\over 32\pi^6}
  \,\bE_3(T)\,\bE_3(U)+{\zeta(3)\over 32\pi}\, (\hbE_1(T)+\hbE_1(U)+\log\mu)\,.
 \end{equation}
 \end{itemize}
The $\log \mu$ term contributes to  the massless threshold contribution,
$\ell_s^6 s^3\cR^4\,\log(-\ell_s^2\,s)$, to the amplitude  in eight dimensions.

 \subsection{The genus-one amplitude on a three-torus}\label{sec:G1T3}
  In this section we evaluate the genus one contributions to the
 $\cR^4$, $\partial^4\cR^4$ and $\partial^6\cR^4$  interactions for the special case
 of a three-torus compactification $d=3$.

By     definition    of     the     $SO(d,d)$    Eisenstein     series
in section~\ref{sec:parab-subgr-dd}    the    one-loop    integral   of    the
three-dimensional torus gives
\begin{equation}
   I^{(3)}_1(\bE_s)                  ={2\zeta(2s)\Gamma(s+\frac12)\over
     \pi^{s+\frac12}}\,
     \bE^{SO(3,3)}_{[100];s+\frac12}\,.
 \end{equation}
For   $\Re\textrm{e}(s)$  large  this   integral  would   divergence  for
large-$\tau_2$ and it needs to  be regulated either by subtracting the
term proportional to the volume as in~(\ref{e:DnDef1}) or equivalently
by using the analytic continuation in $s$ as above.
 Applying~(\ref{e:genusOnedGen}) to the $d=3$ case
and  using  the relation~(\ref{e:D3eqA3})  between  the $SO(3,3)$  and
$SL(4)$ series, $I^{(3)}_1(\bE_s)$ can be expressed  in terms of $SL(4)$ series,
 \begin{equation}
   I^{(3)}_1(\bE_s) ={2\Gamma(s+\frac12)\over \pi^{s+\frac12}}\, \bE^{SL(4)}_{[010];s+\frac12}\,.
 \end{equation}

 \begin{itemize}
 \item  The  $\cR^4$ interaction~\cite{Green:1999pv,Green:2008uj}  is  given by  the
   lowest order term in the  expansion of the genus-one diagram, which
   has $j_1^{(0,0)}=1$.   Applying the result  in~(\ref{e:genusOned}) to
   the case $d=3$ and comparing to the expansion of the $SL(4)$ series
   into $SL(3)$ series given in~(\ref{e:SLtoSL}) gives

\begin{equation}\label{e:R4oneloopT3}
  I^{(3)}_1(1) = \bE^{SO(3,3)}_{[100];\frac12}=2\bE^{SL(4)}_{[010];\frac12}= {2\over\pi}\, \bE^{SL(4)}_{[100];1}
\end{equation}
  where    we  have   made    use     of     the    relation
$\pi\,\bE^{SL(4)}_{[010];\frac12}=\bE^{SL(4)}_{[100];1}$ derived in appendix~\ref{sec:Slnseries}.

 \item For the $\partial^4\cR^4$ interaction~\cite{Green:1999pv,Green:2008uj}
   the function $j_2^{(1,0)}=\bE_2(\tau)/(4\pi)^2$ which gives
   \begin{equation}\label{e:D4R4oneloopT3}
     I^{(3)}_1(j_1^{(1,0)})={\zeta(4)\over 960}\bE^{SO(3,3)}_{[100];\frac52}={1\over 960}\, \bE^{SL(4)}_{[010];\frac52}
   \end{equation}
 
 \item For the $\partial^6\cR^4$ interaction~\cite{Green:2008uj}  the contribution to
   the analytic part of the interaction is given by the function $j_1^{(0,1)}=10\bE_3(\tau)/(4\pi)^3+\zeta(3)/32$, resulting in 
 \begin{equation}\label{e:D6R4oneloopT3}
   \begin{split}
      I^{(3)}_1(j_1^{(0,1)}) &={25\zeta(6)\over 8!}
  \,\bE^{SO(3,3)}_{[100];\frac72}+{\zeta(3)\over 32}\, \bE^{SO(3,3)}_{[100];\frac12}\cr
&={25\over 8!}
  \,\bE^{SL(4)}_{[010];\frac72}+{\zeta(3)\over 16\pi}\, \bE^{SL(4)}_{[100];1}\,.
   \end{split}
    \end{equation}
 \end{itemize}

Upon decompactification, $r_3\to\infty$, the results of the previous
section must be recovered. This is the limit corresponding to   the constant term of the
$SO(3,3)$ Eisenstein series on the parabolic  subgroup obtained by deleting the node
$\alpha_1$ in Dynkin diagram represented in figure~\ref{fig:dynkin}(ii),
\begin{equation}\begin{split} 
  \int_{P_{\alpha_1}} \, I^{(3)}_1(\bE_s) &=r_3\,
  I^{(2)}_1(\bE_s)+4  r_3^{1+2 s} {\zeta (2 s) \zeta
   (2 s+1) \Gamma \left(s+\frac{1}{2}\right)\over\pi ^{s+\frac{1}{2}}} \cr
&+ r_3^{3-2s} \frac{ \zeta (2 s-2) \zeta (2 s-1) \Gamma
   (2 s-2)}{2^{2s-5} \pi ^{s-2} \Gamma (s)}\,.
\end{split} \label{e:D3toD2}\end{equation}
Equivalently, using the $SL(4)$  representation, this expression  corresponds to the 
 parabolic   $P(2,2)$ obtained  by  deleting  the  node
$\alpha_2$. 
The constant term of the $SL(4)$ series $\bE^{SL(4)}_{[100];s}$ on the
parabolic subgroup $P(2,2)$ is given by 
\begin{equation}\begin{split} 
  \int_{P(2,2)} \,\bE^{SL(4)}_{[100];s}&=
  r_3^{s}  \,   \bE_s(T)+  r_3^{2-s}\,  \pi^{2s-2}\,{\Gamma(2-s)\over
    \Gamma(s)} \bE_{2-s}(U)
\end{split} \label{e:A3toD2}\end{equation}
The $SL(4)$ representation makes explicit the factorized dependence on
the K{\"a}hler  modulus $T$ and  the complex structure modulus  $U$. The
equivalence   of  the   two  formula   is   due  to   the  fact   that
$SO(2,2)=SL(2)\times SL(2)$.

For the case of the $\cR^4$ interaction in~(\ref{e:R4oneloopT3}) we have
\begin{equation}
  \begin{split}
\int_{P_{\alpha_1}}\,   I^{(3)}_1(\bE_{\epsilon})&= r_3\,( I^{(2)}_1(\bE_\epsilon)  -{1\over\epsilon}
+2\log(r_3)-\log(\pi)-\gamma_E)+O(\epsilon)\cr
&=   r_3\,(   \bE_{1+\epsilon}(T)+\bE_{1-\epsilon}(U))+   \epsilon\,
  \log(r_3)\, (\bE_{1+\epsilon}(T)-\bE_{1-\epsilon}(U))\cr
&+2\epsilon\,
  r_3\, 
(\gamma_E+\log(\pi))\, \bE_{1-\epsilon}(U)+O(\epsilon)\cr
&=r_3 \,\left(\hbE_1(T)+\hbE_1(U)+2 \log (r_3/\pi)-2 \gamma_E\right)+O(\epsilon)
  \end{split}\label{e:finiteoneloop}
\end{equation}
leading to a finite answer in the decompactification limit (apart from
the $\log r_3$ term which is needed to build the correct eight-dimensional thresholds~\cite{Green:2008uj}).  The
explicit  $1/\epsilon$ pole  in  the first  line  cancels against  the
$1/\epsilon$  pole  of   $I^{(2)}_1(\bE_\epsilon)$  evaluated  in  the
previous section.

 \section{Genus-two string integrals}\label{sec:genus2}

In this section we consider the genus-two partition function   arising  from the  compactification  of  string amplitudes  on
$d$-torus $\calT^d$.  The leading term in the $s,t,u \to 0$ limit is
 \begin{equation}\label{e:g2lat}
  I^{(d)}_2(1) =\int_{\cF_{Sp(2,\mathbb Z)}}\,{|d^3\tau|^2\over (\det\Im\textrm{m}\tau)^3}\, \Gamma_{(d,d)}\,.
 \end{equation}
This integral~\cite{D'Hoker:2002gw,D'Hoker:2005jc} is over the  Siegel upper
 half-plane  for $Sp(2,\mathbb  Z)$.  The resulting  expression is  an
 automorphic  form  invariant under  the  $T$-duality group,  $SO(d,d;
 \mathbb Z)$.   The  lattice factor for  a compactification on a  two-torus is
 given by a theta series summed over the even-lattice,
 \begin{equation}
   \Gamma_{(d,d)}= (V^{(d)})^2\, \sum_{(m^i_{a},n^{ia})\in\ZZ^{2d}\times\ZZ^{2d}} \exp\left(-\pi
     (g_{ij}+b_{ij}) (m^i_{a}-\tau_{ab} n^{ib}) (\Im\textrm{m}\tau^{-1})^{ac}(m^j_{c}-\tau_{cd} n^{jd})\right)\,.
 \end{equation}
 It  was  remarked   in~\cite{Pioline:Automorphic}  that  the  lattice  factor
 satisfies  the differential  equation\footnote{Our  normalisations for
   the $SO(d,d)$ laplacian differ by a factor of 2 compared to~\cite{Pioline:Automorphic}.}
 \begin{equation}
   \Big(\Delta_{SO(d)\times SO(d)\bs SO(d,d)} - \Delta_{Sp(2)}+ d(d-3)\Big)\, \Gamma_{(d,d)}=0\,,
 \end{equation}
 so that the integral in~(\ref{e:g2lat}) satisfies the differential equation
 \begin{equation}\label{e:diffI2}
    \Big(\Delta_{SO(d)\times  SO(d)\bs SO(d,d)} +d(d-3)\Big)\, I^{(d)}_2(1)=0\,.
 \end{equation}

\medskip\noindent$\bullet$ For  $d=2$ the $SO(2,2)$ Laplace operator  is a sum
of the $SL(2)$ Laplace operators acting on the $T$ and the $U$ moduli
and~(\ref{e:diffI2}) gives
\begin{equation}
  (\Delta_T+\Delta_U -2)\, I^{(2)}_2(1)=0\,,
\end{equation}
which is solved by 
\begin{equation}\label{e:Genus2Lat2D}
  I^{(2)}_2(1)={1\over6\pi}( \bE_2(T)+ \bE_2(U))\,.
\end{equation}
The normalisation has been determined from the large-volume limit 
 The normalisation is determined by the  large volume limit the integral~(\ref{e:g2lat}) behaves as 
 \begin{equation}
 \lim_{T_2\to\infty} I_2^{(2)}(1)= {\zeta(4)\over3\pi}\, T_2^2+O(T_2)\,,
 \end{equation}
 where we have used  the value of the fundamental  domain for $Sp(2,\mathbb
 Z)$ given in~\cite{Siegel}
 \begin{equation}
 \int_{\mathcal{F}_{Sp(2,\ZZ)}}\,{|d^3\tau|^2\over(\det\Im\textrm{m}\tau)^3}=
 {\zeta(4)\over 3\pi}\, . 
 \end{equation}

\medskip\noindent$\bullet$      For      $d=3$     the      eigenvalue
in~(\ref{e:diffI2})   vanishes  as   expected  since   there  two-loop
supergravity amplitude has an ultraviolet divergence in $D=7$. In this
case  the integral in~(\ref{e:g2lat})  needs to  be regulated  and the
finite part is given by
\begin{equation}
  \begin{split}
I^{(3)}_2&={1\over 6\pi}\,\left(\hbE^{SO(3,3)}_{[010];2}+\hbE^{SO(3,3)}_{[001];2}\right)\cr
&={1\over 6\pi}\,\left(\hbE^{SL(4)}_{[100];2}+\hbE^{SL(4)}_{[001];2}\right)  \,.  
  \end{split}
 \label{e:Genus2Lat3d}
\end{equation}
The normalisation has been fixed using the large-volume limit and the expansion~(\ref{e:SLtoSL}).

\medskip\noindent$\bullet$ For $d\geq  4$ the differential equation is
not sufficient to determine the solution. 
The    Eisenstein     series    $\bE^{SO(d,d)}_{[0^{d-1},1];s}$,
$\bE^{SO(d,d)}_{[0^{d-2},1,0];s}$    associated with  the    nodes
$\alpha_{d-1}$ and $\alpha_d$ of the $D_d$ Dynkin diagram of figure~\ref{fig:dynkin}(ii) satisfy 
(\ref{e:LaplaEisenstein})
\begin{eqnarray}
  \Delta_{SO(d)\times SO(d)\bs SO(d,d)} \bE^{SO(d,d)}_{[0^{d-1},1];s}&=& {d s(1-d+s) \over2} \bE^{SO(d,d)}_{[0^{d-1},1];s}\, ,\\
  \Delta_{SO(d)\times SO(d)\bs SO(d,d)}  \bE^{SO(d,d)}_{[0^{d-2},1,0];s}&=& {ds(1-d+s)\over2}\,
  \bE^{SO(d,d)}_{[0^{d-2},1,0];s}
\end{eqnarray}
The  series associated  with the  other nodes  $\alpha_u$  with $1\leq
u\leq d-2$ satisfy the differential equation
\begin{equation}
\Delta_{SO(d)\times SO(d)\bs SO(d,d)} \bE^{SO(d,d)}_{[0^{u-1},1,0^{d-u}];s}= u \,s(2s-2d+u+1)\, \bE^{SO(d,d)}_{[0^{d-1},1,0,0];s}\,.
\end{equation}
Therefore, (\ref{e:diffI2}) is satisfied by 
$\bE^{SO(d,d)}_{[0^{d-1},1];2}$,     $\bE^{SO(d,d)}_{[0^{d-2},1,0];2}$,
$\bE^{SO(d,d)}_{[0^{d-3},1,0,0];1}$,
$\bE^{SO(d,d)}_{[0^{d-3},1,0,0];d/2}$  for  all  values of  $d$. 
 With other solutions for each value of $d$. 

It  would be interesting to confirm the conjecture  in~\cite{Pioline:Automorphic}    the
only solution is  the sum of $\bE^{SO(d,d)}_{[0^{d-1},1];2}$,
$\bE^{SO(d,d)}_{[0^{d-2},1,0];2}$.   

 \section{Integrals over Siegel fundamental domains}\label{sec:highergenus}
 
 For genus $h \ge 4$ the parametrisation of the moduli space
$\mathcal{M}_h$ of genus $h$
curves  is  given by  period  matrices  supplemented  by the  Schottky
relations~\cite{D'Hoker:1988ta}, and the integration is not over
  the  Siegel fundamental  domains  for  $Sp(h,\mathbb
Z)$.  The quantities protected by supersymmetry, such as
the $\cR^4$, $\partial^4\,\cR^4$ and $\partial^6\cR^4$ interactions evaluated in
the main  text receive  perturbative contributions up  to genus-three
and  are given  by integrals  over the  Siegel fundamental  domain for
$Sp(h,\mathbb Z)$. 

For the case of the two-torus we consider the integral 
\begin{equation}
I^{(2)}_h =  \int_{\cF_{Sp(h,\mathbb        Z)}}        {|d^{h(h+1)\over2}\tau|^2\over       (\det
      \Im\textrm{m}\tau)^{h+1}}\, \Gamma_{(2,2)}\,.
\end{equation}
This  integral is an  automorphic function  invariant under  the T-duality
group  $SO(2,2)$.  By  applying  the  $SO(2,2)$  Laplace  operator  we
obtain~\cite{Pioline:Automorphic}
\begin{equation}\label{e:DiffTT}
  (\Delta_T+\Delta_U ) \, I^{(2)}_h= h(h-1) \, I_h^{(2)}\,,
\end{equation}
where $\Delta_{SO(2)\times SO(2)\bs SO(2,2)}=\Delta_T+\Delta_U$.
The large-volume limit of $I^{(2)}_h$ is given by
\begin{equation}
  \lim_{T_2\to\infty} I^{(2)}_h= \textrm{vol}(\cF_{Sp(h,\mathbb Z)})\, T_2^h\,,
\end{equation}
where   $\textrm{vol}(\cF_{Sp(h,\mathbb  Z)})$   is   the  volume   of
$\cF_{Sp(h,\mathbb Z)}$ computed in~\cite{Siegel}
\begin{equation}
    \textrm{vol}(\cF_{Sp(h,\mathbb         Z)})=        2\prod_{k=1}^h
    {\zeta(2k)\Gamma(k)\over \pi^k }\, .
\end{equation}
With  this  boundary  condition   the  solution  to (\ref{e:DiffTT}) is given by
\begin{equation}\label{e:genusgT2}
I^{(2)}_h=          {\textrm{vol}(\cF_{Sp(h,\mathbb          Z)})\over
  2\zeta(2h)}\,\left(\bE_h(T)+\bE_h(U)\right)\ .
\end{equation}
Now consider the case of the three-torus compactification,
\begin{equation}
  I^{(3)}_h= 
 \int_{\cF_{Sp(h,\mathbb        Z)}}        {|d^{h(h+1)\over2}\tau|^2\over       (\det
      \Im\textrm{m}\tau)^{h+1}}\, \Gamma_{(3,3)}\ .
\end{equation}
This is a $SO(3,3)$  automorphic function, which satisfies the differential equation derived in~\cite{Pioline:Automorphic}\,,
\begin{equation}
 \Delta_{SO(3)\times SO(3)\bs SO(3,3)} \, I^{(3)}_h= {3\over2}\, h\,(h-2)\, I^{(3)}_h\,,
\end{equation}
which is satisfied by $I^{(3)}_h = a\, \bE^{SO(3,3)}_{[010];h} + b\, \bE^{SO(3,3)}_{[001];h}$ for any $a$ and $b$.  The large-volume limit
\begin{equation}
  \lim_{V_3\to\infty} I^{(3)}_h=  \textrm{vol}(\cF_{Sp(h,\mathbb Z)})\, V_3^h\,,
\end{equation}
determines the solution to be
\begin{eqnarray}\label{e:genus3T3}
  I^{(3)}_h&=&  {\textrm{vol}(\cF_{Sp(h,\mathbb          Z)})\over
  2\zeta(2h)}\,
  \left(\bE^{SO(3,3)}_{[010];h}+\bE^{SO(3,3)}_{[001];h}\right)\\
\nn &=& {\textrm{vol}(\cF_{Sp(h,\mathbb Z)})\over
  2\zeta(2h)}\, \left(\bE^{SL(4)}_{[100];h}+\bE^{SL(4)}_{[001];h}\right)\,.
\end{eqnarray} 
 %
 \section{Supergravity loop amplitudes}
 \label{sec:sugra}
 \subsection{One-loop amplitudes in $D=11$ and the Epstein series}\label{sec:11d}

 In this appendix the expressions for the scalar box function and the scalar
 triangle function reduced on a $d+1$-dimensional torus $\cT^{d+1}$ will be evaluated.
The scalar box function arises as the coefficient of $\cR^4$ in the four-graviton one-loop amplitude in eleven-dimensional supergravity~\cite{Green:1997as}. This diagram has a one-loop divergence that is subtracted by a $\cR^4$ counterterm.  The  scalar triangle function arises from the contribution of this counterterm as a vertex in a 
one-loop four-graviton amplitude, which cancels the sub-divergences of the two-loop eleven-dimensional supergravity amplitude.    
and multiplies   $\partial^4\cR^4$~\cite{Green:1999pu}.  These results generalize the $d=1$ discussion given in~\cite{Green:2008bf}  to higher values of $d$.  

The  expression   for the  scalar box  function is, 
 \begin{equation}
   I^{(D-d-1)}_4(S,T)   = \frac{\pi^{\frac{D - d-1}{2}}}{\cV_{d+1}}
   \int_{\Lambda^{- 2}}^{\infty} \frac{dt}{t}\, t^{\frac{d-D+9}{2}} \,
    \int_{\cT_{ST}} \prod_{r = 1}^3 d \omega_r \sum_{m_I \in
   \mathbb Z^{d+1}} e^{- \pi \,t\, g^{IJ} m_I m_J + \pi
  \, t\, Q_4(S,T)}\ ,
 \end{equation}
 where $D=11+2\epsilon$, 
  $\cT_{ST}=  \{0\leq  \omega_1\leq
 \omega_2\leq \omega_3\leq 1\}$, and the function $Q_4(S,T)$ is defined by~\cite{Green:1999pu}
\begin{equation}
Q_4(S,T)=-S \omega_1(\omega_3 - \omega_2) - T (\omega_2 - \omega_1)(1 - \omega_3)\,,
\end{equation}
 with an equivalent definitions for the $(S,U)$ and $(T,U)$ regions.
The scalar triangle function is given by
 \begin{equation}
   I^{(D-d-1)}_3(S)   = \frac{\pi^{\frac{D - d-1}{2}}}{ \cV_{d+1}}
   \int_{\Lambda^{- 2}}^{\infty} \frac{dt}{t}\, t^{\frac{d-D+7}{2}} \,
    \int_{0\leq \omega_1\leq \omega_2\leq 1} \prod_{r = 1}^2 d \omega_r \sum_{m_I \in
   \mathbb Z^{d+1}} e^{- \pi \,t\, g^{IJ} m_I m_J + \pi
  \, t\, Q_3(S)}\ ,
 \end{equation}
where the function $Q_3(S,T)$ is defined by~\cite{Green:1999pu}
\begin{equation}
Q_3(S,T)=-S \omega_1\omega_2\,.
\end{equation}
The masses of the Kaluza--Klein states running in the loop are denoted
$g^{IJ}m_{I}m_{J}$ and the volume of the $d+1$-torus is $\cV_{d+1}$.

 We will first  analyze the momentum  expansion of the scalar  box function.
 This expression contains a non-analytic contribution from the
  massless  supergravity states in  ($10-d$) dimensions  together with
  analytic terms,
  \begin{equation}
 I^{(D-d-1)}_4(S,T)= I^{(D-d-1)}_{4,\textrm{nonan}}(S,T)+ \hat I^{(D-d-1)}_4(S,T)\,.
  \end{equation}
 The non-analytic part is the usual field theory contribution,
  \begin{equation}
 I^{(D-d-1)}_{4,\textrm{nonan}}(S,T)\sim \int_{\cT_{ST}} \prod_{r = 1}^3 d \omega_r \, (Q_4(S,T))^{d-D+9\over2}\ .
 \end{equation}
 For $d=-1$ this  is the eleven-dimensional supergravity contribution,
 $M_{4;1}\sim (-\ell_{11}^2 S)^{3/2}$; for $d=0$ it is the ten-dimensional supergravity
 contribution $M_{4;1}^{(10)}\sim S\log(-\ell_{11}^2\,S)$; for $d=1$ it is the nine-dimensional contribution
 $M_{4;1}^{(9)}\sim (-\ell_{11}^2\,S)^{-1/2}$, with an extra power of $S^{-1/2}$  for each extra compact dimension.  A detailed discussion of the relation between these various expressions obtained by decompactifying successively from $d=1$ to $d=0$ and $d=-1$ is given 
 in~\cite{Green:1997as, Green:1999pu,Green:2006gt}.

 It is convenient to separate the zero-momentum part of the analytic part of the amplitude
 \begin{equation}
 I^{(D-d-1)}_4(S,T)= I^{(D-d-1)}_4(0,0)+\tilde I^{(D-d-1)}_4(S,T) \,.
\end{equation}
 In order to isolate the divergences one must perform a Poisson resummation over the
 Kaluza--Klein               modes               $m_I$              in
 $I^{(D-d-1)}_4(0,0)$~\cite{Green:1997as,Green:1999pu}.
Evaluating  this  integral with  $D=11$  and  a
momentum cut-off $\Lambda$  gives
 \begin{equation}
   \begin{split}
     I^{(10-d)}_{4}(0,0) & = \pi^{\frac{10 - d}{2}} \int_0^{\Lambda^2} d \hat{t}
    \, \hat{t}^{\frac{1}{2}} \hspace{0.75em} \sum_{\{ \hat{m} \}
     \in \mathbb{Z}^{d+1}} e^{- \pi \hat{t}\, g_{IJ} \hat{m}^I
     \hat{m}^J}\cr
     & =\pi^{10-d\over2}\, \Lambda^3 +{\pi^{10-d\over2}\over 2\pi\cV_{d+1}^{3\over d+1}}\,\bE^{SL(d+1)}_{[1,0^{d-1}];\frac{3}2}\, ,
     \label{e:EisI}
   \end{split}
 \end{equation}
 where  $g_{IJ}=\cV_d^{2/d}\,  \tilde g_{IJ}$  is  the  metric of  the
 $d$-torus   and  $\det\tilde   g_{IJ}=1$.  
 The ultra-violet divergence is now localised in the zero winding sector $\hat{m}_I
 = 0$.  The finite part is  the contribution from the  non zero winding, which
 is  invariant under  large diffeomorphisms, described by the action of  $SL(d+1,\ZZ)$ on  the $d+1$
-dimensional  torus  and  is   proportional  to  the  Epstein  series,
 $\bE^{SL(d+1)}_{[1,0^{d-1}];(D-8)/2}$.
 The   same   integral  evaluated   in   dimension
$D=11+2\epsilon$ gives
\begin{equation}
   \begin{split}
     I^{(10-d+2\epsilon)}_{4}(0,0) & = \pi^{\frac{10 - d}{2}+\epsilon} \int_0^\infty d \hat{t}
    \, \hat{t}^{\frac{1}{2}+\epsilon} \hspace{0.75em} \sum_{\{ \hat{m} \}
     \in \mathbb{Z}^{d+1}} e^{- \pi \hat{t}\, g_{IJ} \hat{m}^I
     \hat{m}^J}\cr
     & = \pi^{10-d\over2}\,{1\over \cV_{d+1}^{3+2\epsilon\over d+1}}\,{\Gamma(\frac32+\epsilon)\over\pi^{\frac32}} \,\bE^{SL(d+1)}_{[1,0^{d-1}];\frac{3}2+\epsilon}\,.
     \label{e:EisIDimReg}
   \end{split}
 \end{equation}
The higher-order terms in the expansion in powers of the external momenta give
  \begin{equation}\label{e:HDbox}
 \tilde    I^{(D-d-1)}_{4}(S,T)    =2\sum_{n\geq1}    
 (\cV_{d+1}^{2\over d+1})^{n-{D-d-1\over2}}\,
 {\mathcal{G}_{ST}^n\over n!}\,
 {\Gamma\left({d-D+9\over2}+n\right) \over   \pi^{d+5+n-D}}\,\bE^{SL(d+1)}_{[0^{d-1},1];{d-D+9\over2}+n}\,,
 \end{equation}
 where
 \begin{equation}
 \mathcal{G}_{ST}^n\equiv \int_{\cT_{ST}}\, \prod_{r=1}^3d\omega_{r}\,(Q_{4})^n\ .
 \end{equation}

Similarly,  the triangle diagram will be written as the sum of analytic and non-analytic terms,
\begin{equation}
  I^{(D-d-1)}_3(S) = I^{(D-d-1)}_{3,\textrm{nonan}}(S)+\hat I^{(D-d-1)}_3(S)\,,
\end{equation}
where
\begin{equation}
  I^{(D-d-1)}_{3,\textrm{nonan}}(S)                                 \sim
  \int_{0\leq\omega_1\leq\omega_2\leq1}        \prod_{r=1}^2d\omega_r\,
  (Q_3(S))^{d-D+7\over2}\,,
\end{equation}
and  the analytic part will be separated into a zero-momentum part and a
momentum-dependent part,
\begin{equation}
  \hat I^{(11-d)}_3(S)=\hat I^{(11-d)}_3(0)+\tilde I^{(11-d)}_3(S)\, .
\end{equation}
The zero-momentum part is given by
 \begin{equation}
   \begin{split}
     I^{(10-d)}_{3}(0) & = \pi^{\frac{10- d}{2}} \int_0^{\Lambda^2} d \hat{t}
    \, \hat{t}^{\frac{3}{2}} \hspace{0.75em} \sum_{\{ \hat{m} \}
     \in \mathbb{Z}^{d+1}} e^{- \pi \hat{t}\, G_{IJ} \hat{m}^I
     \hat{m}^J}\\
     & =\pi^{\frac{10- d}{2}}\, \Lambda^{5} +\pi^{\frac{10- d}{2}}\,{1\over \cV^{5\over d+1}_{d+1}}{3\over(2\pi)^2}\,\bE^{SL(d+1)}_{[1,0^{d-1}];{5\over2}}\, .
     \label{e:EisII}
   \end{split}
 \end{equation}
 The momentum expansion of $\tilde I^{(D-d+1)}_3(S)$ is given by
 \begin{equation}\label{e:HDtri}
 \tilde    I^{(D-d+1)}_{3}(S)   =   2    \sum_{n\geq1}   
 (\cV_{d+1}^{2\over d+1})^{n-{D-6\over2}}\,
 {\mathcal{S}^n\over n!}\,
{ \Gamma\left({d-D+7\over2}+n\right)\over \pi^{3+d+n-D}}\,\bE^{SL(d+1)}_{[0^{d-1},1];{d-D+7\over2}+n}\,,
 \end{equation}
 where
 \begin{equation}
 \mathcal{S}^n\equiv \int_{0\leq\omega_1\leq\omega_2\leq1}\, \prod_{r=1}^2d\omega_{r}\,(Q_{3})^n\ .
 \end{equation}

 \subsection{Two-loop amplitudes in $D=11$ and Eisenstein series}\label{sec:Twoloop11d}

The  finite  part of  the $L=2$ four-graviton amplitude in  eleven-dimensional supergravity compactified on $\cT^{d}$ will be evaluated in this appendix.  The leading term in the low-energy limit has the form~\cite{Bern:1998ug}
$(s^2+t^2+u^2)\, I_{L=2}$.
 Following~\cite{Green:1999pu}  $I_{L=2}$ can be rewritten in the form of a genus-one string theory amplitude, which has the low energy limit 
 \begin{equation}
   \label{e:I22}
 I^{(11-d)}_2= \int_0^{\Lambda^2} dV \, V^3\, \int_{\mathcal{F}_{\Lambda}}\,{d^2\tau\over\tau_2^2}\, \sum_{(m^i,n^i)\in\ZZ^d\times
     \ZZ^d} \, e^{-V\,\calV_d^{2\over d}\,{\pi\over\tau_2}\, g_{ij} (m^i -\tau n^i) (m^j
   -\bar\tau n^j) }\,,
 \end{equation}
 where   $\mathcal{F}_\Lambda=   \{   \tau=\tau_1+i\tau_2  |   -1/2\leq
 \tau_1\leq 1/2,  |\tau_1|^2+|\tau_2|^2\geq \Lambda^2\}$. Using the
 method    of    orbits    this    integral   has    three    kinds    of
 pieces~\cite{Green:1999pu}
 \begin{equation}
   \label{e:Il2}
  I^{(11-d)}_2= \Lambda^8 I^{(0)} + \Lambda^5 \, I^{(1)} + I^{fin}\,.
 \end{equation}
 We are interested in the finite part of the integral, which can be evaluated  by the method  of orbits as
 detailed in appendix~\ref{sec:Rankin-Selberg} and is given by
 \begin{eqnarray}
 \nn    I^{fin}&=&     2\int_0^\infty    dV    \,    V^3\,\int_{\mathbb
   C^+}\,{d^2\tau\over\tau_2^2}\,     \sum_{1\leq    k\leq    d-1\atop
   [M_{0,k}]} \, e^{-V\, {\cal V}_d^{2\over d}\,{\pi\over\tau_2}\, g_{ij} (m^i -\tau n^i) (m^j
   -\bar\tau n^j) }\\
 \nn &=&{2\over {\cal V}_d^{2\over d}}\sum_{1\leq k\leq d-1\atop[M_{0,k}]} \,{1\over \sqrt{\det \mathcal{M}}}\,\int_0^\infty dV \,
 V^2\, e^{-2\pi V{\cal V}_d^{2\over d} \sqrt{\det \mathcal{M}}}\\
 &=&{1\over 2\pi^3\, {\cal V}_d^{8\over d}}\, \sum_{1\leq k\leq d-1\atop [M_{0,k}]} \,{1\over
   (\det \mathcal{M})^2}= {1\over 2\pi^3\, {\cal V}_d^{8\over d}}\, \bE^{SL(d)}_{[0,1,0^{d-3}];2}\,,
   \label{e:I22Result}\end{eqnarray}
 where   the   sum   is   over   the   representatives   $M_{0,k}$   in
 \eqref{e:Rep} and the matrix ${\cal M}$ is defined in~(\ref{e:calM}).

 \section{Laplacians on $K\bs G$ manifolds}
\label{sec:laplacian}

In the next subsection we will  discuss the Laplace operator on some of the cosets of explicit relevance to the discussions in the text.   In the subsequent subsection we will use an iterative method to relate the Laplace operator and its eigenvalues for different values of $D$, which leads to equations~\eqref{laplaceeigenone}-\eqref{laplaceeigenthree}.

\subsection{Explicit examples for $D=8,9,10$}

These cosets are parameterised by scalar (moduli) fields.  These scalars enter in the supergravity in the form of a sigma model with action
\begin{equation}
     S_{scalar}=  {1\over \ell_D^{D-2}}\int               d^Dx              \sqrt{-G^{(D)}}\,
     h_{ij}(\sigma)\, \partial_\mu\sigma^i\partial^\mu \sigma_j\,,
   \end{equation}
and the associated Laplace operator is given by
 \begin{equation}
   \Delta_\sigma=                                                {1\over
     \sqrt{h(\sigma)}}\,\partial_{\sigma^i}\big(\sqrt{h(\sigma)}\,
   h^{ij}\partial_{\sigma^j}\big)\ .
   \label{laplacedef}
 \end{equation}
The  explicit expressions for these Laplacians  in terms of
our choice of fields in the Einstein frame in various dimensions is as
follows.

\noindent $\bullet$ The scalar field action of $D=10$ type~IIB is
 \begin{equation}
 S^{scalar}_{10d}= {1\over 2\ell_{10}^{8}} \int  d^{10}x\, \sqrt{-G^{(10)}}\, 
    {\partial_\mu \Omega\partial^\mu\Omega\over\Omega_2^2}\,.
 \end{equation}
 The $SL(2,\IR)$  symmetry acts on the  complexified coupling constant
 $\Omega$, and 
the $SO(2)\backslash SL(2)$ Laplacian is defined as
 \begin{equation}
   \Delta_\Omega                                                 \equiv
   4\Omega_2^2\,\partial_\Omega\bar\partial_{\bar\Omega} =\Omega_2^2\,
   (\partial^2_{\Omega_1}+ \partial^2_{\Omega_2})\,.
 \end{equation}
 Note that our normalisation conventions are such that the Eisenstein series $\bE_s(\Omega)$ has
 eigenvalue $s(s-1)$.

\noindent $\bullet$  The nine-dimensional  scalar field action with $GL(2,\IR)= SL(2,\IR)\times \IR^+$ invariance is 
 \begin{equation}
 S^{scalar}_{D=9}={1\over \ell_9^7}\int   d^{9}x\,  \sqrt{-G^{(9)}}\,  \left({2\over  7}\,
   \,\partial_\mu \log\nu_1\partial^\mu \log\nu_1- {1\over2}\,
   {\partial_\mu  \Omega\partial^\mu\bar\Omega\over\Omega_2^2}\right)\,.
 \end{equation}
 Here the  $SL(2,\IR)$ symmetry acts on  $\Omega$  and
 $\IR^+$ acts as a shift  on $\log\nu_1\to \log\nu_1+\lambda$.
 The  Laplace  operator acting on scalars in $D=9$ is 
 \begin{equation}\label{e:9DLapla}
 \Delta^{(9)} \equiv \Delta_\Omega
  +{7\over 4} \nu_1 \partial_{\nu_1}(\nu_1 \partial_{\nu_1}) +
 {1\over 2} \nu_1 \partial_{\nu_1 }\ .
 \end{equation}
\noindent $\bullet$  In   eight   dimensions  the   U-duality   group  is   $SO(3)\backslash
 SL(3,\IR)\times SO(2)\backslash SL(2,\IR)$
 where   $SL(3,\IR)$  acts  on  $\Omega$,  the eight-dimensional volume $\nu_2$, and
 the combination of Ramond--Ramond and Neveu--Schwarz---Neveu--Schwarz  $B$-fields, $B=B_{\rm RR}+\Omega B_{\rm NS}$. The
 $SL(2,\IR)$ group acts on the complex structure $U$.
  The $SO(3)\backslash SL(3)$ laplacian is given by~\cite{Kiritsis:1997em}
 \begin{equation}
 \Delta_{SO(3)\bs SL(3)}   =  4\Omega_2^2\partial_\Omega\bar\partial_{\bar\Omega}+
 {|\partial_{B_{\rm NS}}-\Omega\partial_{B_{\rm RR}}|^2\over\nu_2\Omega_2}+
 3\partial_{\nu_2}(\nu_2^2\partial_{\nu_2})\, .
 \end{equation}
 The full Laplacian for the  eight-dimensional theory is the sum of the $SO(3)\backslash SL(3)$ and the
 $SO(2)\backslash SL(2)$ Laplacians,
 \begin{equation}
   \label{e:Delta8d}
   \Delta^{(8)}\equiv \Delta_{SO(3)\bs SL(3)}+\Delta_{SO(2)\bs SL(2)}= 4U_2^2\partial_U\bar\partial_{\bar U}+4\Omega_2^2\partial_\Omega\bar\partial_{\bar\Omega}+
 {|\partial_{B_{\rm NS}}-\Omega\partial_{B_{\rm RR}}|^2\over\nu_2\Omega_2}+
 3\partial_{\nu_2}(\nu_2^2\partial_{\nu_2})\ .
 \end{equation}

\subsection{Connections between Laplace equations in different dimensions.}
 \label{sec:eigenD} 
  
   In this appendix we will  give a derivation of~(\ref{laplaceeigenone})-(\ref{laplaceeigenthree}).
We  will take the dimensionless radius  of the $(d+1)$'th
dimension on the string theory torus to be large, i.e., large 
$r_{d+1}/\ell_s$.   This
corresponds  to  deleting the  last  node  of  the Dynkin  diagram  in
fig.~\ref{fig:dynkin}(i) for the group
$G_d=E_{d+1(d+1)}$, which reduces its rank.  In this limit the Laplace
operator,      $\Delta^{(D)}      \equiv     \Delta^{G_d}$    decomposes as  (where $d=10-D$)
 \begin{equation}
  \Delta^{(D)}\to \Delta^{(D+1)}-  a_D (r_d\partial_{r_d})^2 - b_D  (r_d\partial_{r_d})\,,
\label{eigD}
\end{equation}
where $a_D$ and $b_D$ are numerical coefficients whose determination is discussed below.
In the decompactification limit
\be
\ell_{D+1}^{D-1} = \ell_D^{D-2} r_d\ .
\ee

We will now determine the Laplace equations, ~(\ref{laplaceeigenone})-(\ref{laplaceeigenthree}), by a recursive method, as follows.
Given a modular function $\cE^{(D)}_{(p,q)}$ in
dimension $D$, the modular function   $\cE^{(D+1)}_{(p,q)}$ in $D+1$ dimensions can be obtained via the relation
\begin{equation}
  \ell_D^{8+2k-D}\, \int d^Dx \sqrt{-G^{(D)}}\, \cE^{(D)}_{(p,q)}\, \partial^{2k}\cR^4  =   \ell_{D+1}^{7+2k-D}\, \int d^Dx  \sqrt{-G^{(D)}}\, (r_d\cE^{(D+1)}_{(p,q)} +\cdots)\p^{2k}\cR^4\, 
  \,,
\label{Isu} 
\end{equation}
where $k=2p+3q$ and  the ellipsis ``\dots'' stands for  the terms that
either   grow   faster   than   $r_d$   or   vanish   in   the   limit
$r_d\to\infty$. As  we have seen  in the examples  in the body  of the
paper the  divergent terms contribute to the  threshold behaviour, and
not to the analytic part  of the $D+1$ dimensional amplitude. They can
therefore be ignored.
Therefore, the $r_d$ dependence in (\ref{Isu}) is completely determined by the requirement that the term decompactifies to $D+1$,
\be\label{e:RelE}
\cE^{(D)}_{(p,q)}         =         \left(         {r_d\over\ell_{D+1}
  }\right)^{4p+6(q+1)\over D-2} (\cE^{(D+1)}_{(p,q)}+\cdots)\, .
\ee
The  formula~(\ref{eigD}) then  establishes  a recursive  relationship
between  the  eigenvalues   $\lambda^{(D)}_{(p,q)}$:   knowing  the
eigenvalues in ten dimensions, one  can  derive the eigenvalues in
all lower dimensions.

The direct determination of the numerical coefficients $a_D$ and $b_D$ in low dimensions is complicated, due to the complicated structure of the Laplace operator. However, a simple way to find them is by using as input the eigenvalues for the $\cR^4$ and $\partial^4 \cR^4$ interactions
in $D$ and $D+1$ dimensions where they are known. Then the eigenvalue for the $\partial^6 \cR^4$ interaction is a prediction.
It is actually sufficient to determine $a_D$ and $b_D$ in $7\leq D\leq
9$. We find
\be
(a_7,b_7)=(-{5\over 12},{5\over 2})\ ,\qquad   (a_8,b_8)=(-{3\over 7},{9\over 7})\ ,\qquad (a_9,b_9) =(-{7\over 16},{1\over 4})\ .
\ee
With this information one can consider the ansatz
\be
\lambda^{(D)}_{(p,q)}=        {A_{(p,q)}(B_{(p,q)}-D)(D-C_{(p,q)})\over
  D-2}\, .
\ee
The   ($D$-independent)   coefficients   $A_{(p,q)},\   B_{(p,q)},   \
C_{(p,q)}$ are determined 
by substituting  the relation~(\ref{e:RelE}) between  the coefficients
into the Laplace equation satisfied by 
$\cE^{(D)}_{(p,q)}$.  It follows
that, for $7\leq D\leq 9$,
\be
\lambda^{(D)}_{(p,q)} = \lambda^{(D+1)}_{(p,q)} -a_D\, {(4p+6(q+1))^2 \over (D-2)^2} - b_D\, {4p+6(q+1) \over D-2}\,.
\label{juno}
\ee
For the three cases under consideration 
\bea
\lambda^{(D)}_{(0,0)} &=& {3(11-D)(D-8)\over D-2}\,,
\label{kuno}\\
\lambda^{(D)}_{(1,0)} &=& {5(12-D)(D-7)\over D-2}\,,
\label{kdos}\\
\lambda^{(D)}_{(0,1)} &=& {6(14-D)(D-6)\over D-2}\,.
\label{ktres}
\eea
Assuming  that~(\ref{juno})  holds for  $(p,q)=(0,0)$ and
$(p,q)=(1,0)$ in any generic dimension $3\leq D\leq 10$,
one can determine $a_D$ and $b_D$
\be
a_D= -{D-2\over 2(D-1)}\ ,\qquad b_D=- {D^2-3D-58\over 2(D-1)}\,.
\label{solge}
\ee 
As  a check  that this  extrapolation to  arbitrary  dimensions $3\leq
D\leq   10$  makes   sense,  one   verifies   that~(\ref{solge})  also
solves~(\ref{juno}) for $(p,q)=(0,1)$.
Another check is that~(\ref{kuno})--(\ref{ktres}) (or, equivalently,
(\ref{juno}) with (\ref{solge}) and $0\leq2p+3q\leq3$)
give the correct eigenvalues in six dimensions. Since the information about the $D=6$ eigenvalues was not used at all, this is  a non-trivial check.

Summarizing,  the  basic  rule  behind  the above  derivation  is  the
requirement   that  a   modular  function   in  $D$   dimensions  
decompactifies to a
finite term  in $D+1$ dimensions. This determines the $r_d$ dependence, and hence the shift in the eigenvalues.
Since this rule applies equally to the $3\leq D<6$ modular functions, we expect that in these dimensions the
modular functions for the interactions $\cR^4$, $\partial^4 \cR^4$ and $\partial^6 \cR^4$ satisfy the differential equations~(\ref{laplaceeigenone})--(\ref{laplaceeigenthree}).
It should  be noted that the  source term in~(\ref{laplaceeigenthree})
is   also  determined  by   the  decompactification   procedure  since
$\cE^{(D)}_{(0,0)}$ decompactifies appropriately.
 
  \section{Determination of   $\cE^{(8)}_{(0,1)}$.}
 \label{sec:solutionD6R4}
  
 We will here solve the inhomogeneous Laplace equations that define the coefficients of the $\partial^6\cR^4$ interactions in  $D=8$ dimensions.  In each case we will find a unique solution satisfying certain boundary conditions  obtained from string perturbation theory.
 
We wish to solve~(\ref{e:DiffD6R4Bis}),
  \be\label{e:DiffD6R4}
 \Delta^{(8)}\,\cE^{(8)}_{(0,1)}
 =12\,\cE^{(8)}_{(0,1)}  
 -\,(\cE^{(8)}_{(0,0)})^2\ .
 \ee
 The general form of the solution is the sum of a particular solution and a solution of the homogeneous equation.
 The  homogeneous  equation  $(\Delta^{(8)}-12)\mathcal{F}=0$ has  one solution that is compatible with string perturbation theory, 
  \begin{eqnarray}
   f_{-\frac32,3}&=&\bE^{SL(3)}_{[10] ;-{3\over2}}\, \bE_3(U)\,.
 \label{homosol}
\end{eqnarray}
 There are other solutions, such as  $\bE^{SL(3)}_{[10] ;s}$  with
 $s_\pm=3/4(1\pm                    \sqrt{17})$               and $\bE_4(U)$.  However, none of these solutions is compatible with string perturbation theory.
Therefore
 \begin{equation}\label{e:Pf}
  \cE_{(0,1)}^{(8)} = \alpha_{-\frac32,3}\,\bE^{SL(3)}_{[10];- {3\over2}}\, \bE_3(U)+ \mathcal{P}
 \end{equation}
 where  the particular solution $\mathcal{P}$ can be expressed by separation of variables as
 \begin{equation}
   \mathcal{P}= A^{SL(3)}+ B^{SL(2)}(U) + C^{SL(3)}\, D^{SL(2)}(U)
 \end{equation}
where  $A^{SL(3)}$ and $C^{SL(3)}$  are $SL(3,\mathbb  Z)$ automorphic
functions and $B^{SL(2)}(U)$ and $D^{SL(2)}(U)$ are $SL(2,\ZZ)$-invariant  functions of $U$.
 By  expanding the  source term, each  piece is found to satisfy the
 following equations
 \begin{eqnarray}\label{e:Alapla}
     (\Delta_{SO(3)\bs SL(3)} -12) A^{SL(3)}& =& -\, (\hbE^{SL(3)}_{[10];\frac32})^2\,,\\
 \label{e:BU} (\Delta_{SO(2)\bs SL(2)} -12) B^{SL(2)}(U)& =& -4\, (\hbE_{1}(U))^2\,,\\
 \label{e:CD} (\Delta_{SO(3)\bs SL(3)\times SO(2)\bs SL(2)} -12) C^{SL(3)}D^{SL(2)}(U)& =& -4\, \hbE^{SL(3)}_{[10]; {3\over 2}}\,\hbE_{1}(U)\,.
 \end{eqnarray}
 The $SL(3,\mathbb Z)$ functions can be expanded using the variables
 $(\nu_2,\Omega)$ with an explicit $SL(2,\mathbb Z)$ invariance acting on $\Omega$ or
 using  the variables  $(y_8,T)$  with an  explicit  $SL(2,\mathbb Z)$ symmetry  acting on
 $T$.  The T-duality group  in eight dimensions is $SO(2,2)= SL(2)\times
 SL(2)$, where the  $SL(2)$ factors act on $T$ and $U$ respectively. This ensures that the perturbative answer is symmetric under
 exchange $T\leftrightarrow U$. 

The first differential equation in~\eqref{e:Alapla} defines an  $SL(3,\mathbb Z)$ invariant function
\be
  A^{SL(3)}\equiv {\cal E}^{SL(3)}_{(0,1)}\ .
 \ee
The $SL(3)$ functions will be written as functions of  the $(y_8,T)$ variables, in terms of which the $SO(3)\bs SL(3)$ Laplacian takes the form
 \begin{equation}
   \Delta_{SO(3)\bs SL(3)}\sim                             T_2^2(\partial_{T_1}^2+\partial_{T_2}^2)+
   3 \partial_{y_8}(y_8^2\partial_{y_8})\,.
 \end{equation}
 Using the expansion given in~\eqref{e:E32reg} for $\hbE^{SL(3)}_{[10];3/2}$, one
 can determine the perturbative expansion of $\cE_{(0,1)}^{SL(3)}$. The ansatz
  \begin{equation}\begin{split}
 \int_{-\frac12}^{\frac12}d\Omega_1dB_{\rm  RR} \, \cE_{(0,1)}^{SL(3)} &={a_0\over
   y_8^2} + {1\over y_8}\, (A_1(T)+a_1 \log (y_8) )+ A_2(T,y_8)\cr
&+ \sum_{n\geq 3} A_n(T) y_8^{n-2} \, ,
\end{split} \end{equation}
leads to
\be
a_0 =  {2\zeta(3)^2\over 3}\ ,\qquad a_1= {2\pi\over 9}\zeta(3)\ , 
\ee
and the set of equations
 \begin{eqnarray}
 \label{e:A1}  (\Delta_T - 12) \, A_1(T)&=&- 8\zeta(3) \, \hbE_1(T)+{2\pi\over 3}\zeta(3)\,, \\
 \label{e:A2} (\Delta_T+ 3 \partial_{y_8}(y_8^2\partial_{y_8})-12) \, A_2(T,y_8)&=& - (2\hbE_1(T) +{2\pi\over 3}\log (y_8) )^2\,,\\
 \label{e:A3} (\Delta_T-6) \, A_3(T)&=&0\,,\\
 \label{e:An} (\Delta_T- 3(2+3n-n^2)) A_n(T)&=&0; \qquad n\geq 4\,,
 \end{eqnarray}
with $\Delta_T=T^2_2(\partial^2_{T_1}+\partial^2_{T_2})$.

\begin{itemize}
 \item Equation~(\ref{e:A1}) gives the genus-one  contribution.
 Because the source term  is linear~\eqref{e:A1} is solved
 by
 \begin{equation}
   A_1(T) = a_1'\,\bE_4(T) +{2\over 3}\zeta(3)\, \hbE_1(T)\,.
 \end{equation}
  An   explicit   evaluation   of   the  genus-one   contribution   in
 \eqref{e:D6R4oneloop} shows that $a_1'=0$. 
 \item
 Equation~(\ref{e:A2}) is  solved by
\bea
A_2(T,y_8) &=& a_2'\,\bE_4(T)+f(T) + {7\pi^2\over 216} +{\pi\over 18}\, \hbE_1(T)
\nn\\
&+&\left({\pi^2\over 27} +  {2\pi\over 9} \, \hbE_1(T)\right)\, \log(y_8)+ {\pi^2\over 27}\, \log (y_8)^2\ ,
\eea
 where $f(T)$ is the particular solution to
 \begin{equation}
   (\Delta_T-12) \, f(T)=-4\, \hat \bE_1^2(T)\,.
 \end{equation}
This is  the same  as  the equation  for
 $B^{SL(2)}(U)$ in~\eqref{e:BU} as is required by T-duality at genus-two.  
  The structure of this equation is similar to that of  $\cE_{(0,0)}^{(10)}$.  This is complicated to solve explicitly, but it is straightforward to determine the power-behaved terms in the  large-$T_2$ expansion, as given in~\cite{Basu:2007ck},
\begin{equation}
   \begin{split}
      f(T)&=     {\zeta(2) \over 180}\,    \left(65-20\pi    T_2+48\pi^2
 T_2^2\right)+ {\zeta(3)\zeta(5)\over 6\pi T_2^3}\cr
& -{\zeta(2)\over3} \log T_2\, \left(4\pi T_2-6\log
 T_2+1\right)+ O(e^{-T_2})\ .
   \end{split}\label{fexpand} 
  \end{equation} 
 Since there cannot be a $T_2^4$ contribution to the genus-two $\partial^6\cR^4$ we conclude that $a_2'=0$.
 \item Equation~(\ref{e:A3}) is solved by $A_3(T)= \alpha_3\, \bE_3(T)$.

 \item  Equation~(\ref{e:An})  has  solutions  $ A(T)=b\, \bE_s(T)$
   where  $s$ is  not real.   Therefore they  do not  fit  with string
   perturbation  theory, so we must  set $b=0$, which  is compatible
   with  the   absence of contributions  beyond  genus-three.

 \end{itemize}
 The perturbative expansion for $\cE^{SL(3)}_{(0,1)}$ therefore has the form
 \begin{equation}\begin{split}
 \int_{-\frac12}^{\frac12}d\Omega_1dB_{\rm  RR} \, \cE^{SL(3)}_{(0,1)} &=  {2\over
   3}{\zeta(3)^2\over   y_8^2}  +{2\zeta(3)\over  9}   {1\over  y_8}\,
 \big(3\hat \bE_1(T)+\pi \log (y_8) \big)\cr
&+ A_2(T,y_8) +
   \alpha_{3}\, y_8\, \bE_3(T)\,.
\end{split} \end{equation}
 By considering the powers  of $y_8$ in~(\ref{e:BU}) and~(\ref{e:CD}) we
 see  that~(\ref{e:CD}) has  genus-one  and genus-two
 contributions, 
 \begin{equation}\label{e:CDexp}
 \int_{-\frac12}^{\frac12}d\Omega_1dB_{\rm RR}  \, C^{SL(3)} D^{SL(2)}(U) = {h_1(T,U)\over y_8}+ h_2(T,U,y_8)\,,
 \end{equation}
 where
 \begin{eqnarray}
   (\Delta_T+\Delta_U-12) \, h_1(T,U)&=- 8\zeta(3)\, \hbE_1(U)\,,
 \label{h1def}
 \eea
 \begin{equation}
 (\Delta_T+\Delta_U+ 3 \partial_{y_8}(y_8^2\partial_{y_8})-12) \, h_2(T,U,y_8) = -8\, \hbE_1(T)\hbE_1(U)-{8\pi\over 3}\hbE_1(T)\log (y_8)  \, .
 \label{h2def}
 \end{equation}
 These equations are solved by 
 \begin{eqnarray}
    h_1(T,U)&=& \hat h_1(T,U)+{2\over3}\zeta(3)\hat \bE_1(U)  +{\pi\over 18}\, \zeta(3)\,,\\
    h_2(T,U,y_8)&=& \hat h_2(T,U)+ {2\over3} \hbE_1(T)\hbE_1(U)+ {\pi\over 9}\hbE_1(U)+ {\pi\over 18}\hbE_1(T)
\nn\\\
&+& {2\pi\over 9} \hbE_1(U)\log (y_8)+{\pi^2\over 54}\log (y_8)+{\pi^2\over 54}\,,
\end{eqnarray}
 where 
 \begin{eqnarray}
   (\Delta_T+\Delta_U-12) \, \hat h_i(T,U)=0, \qquad i=1,2\,.
 \end{eqnarray}
The only solution  to this  equation which is symmetric under $T\leftrightarrow U$, and that can {\it a priori} be compatible with 
 the decompactification limit
has the form
 \be
 \nn   \hat   h_i(T,U)=  
\beta_i   \, \bE_3(T)\bE_3(U) \, .
 \ee
 General solutions with eigenvalue equal to 12 of the form $\bE_{s_1}(U)\bE_{s_2}(T) + \bE_{s_2}(U)\bE_{s_1}(T)$ would have non-rational values
 of $s_1,\ s_2$ and thus would lead to non-rational powers of $r_2$ in the decompactification limit.
On the other hand, a possible solution proportional to $\bE_4(U)+\bE_4(T)$ is ruled out for the reasons explained above.
 
 Finally, the perturbative contributions from  the homogeneous solution (\ref{homosol}) are
 \be
 \int_{-\frac12}^{\frac12}d\Omega_1dB_{\rm RR}\, \bE^{SL(3)}_{[10] ;-{3\over2}}\, \bE_3(U) = {3\over 2\pi^5} \Big( y_8^{-1} \, \bE_3(T) + \pi \zeta(4) y_8 \Big)\bE_3(U).
  \ee
This expression contains genus-one and genus-three terms.
  
 Collecting     the        perturbative       contributions        to
 $\cE_{(0,1)}^{(8)}$ we have
 \begin{equation}
 \int_{-\frac12}^{\frac12}d\Omega_1dB_{\rm RR}   \cE_{(0,1)}^{(8)}=    {f_0\over y_8^2} +
  {f_1\over y_8}+ f_2+ y_8\,f_3\,,
 \end{equation}
with
 \begin{eqnarray}
   f_0&=&{2\over3}\zeta(3)^2\,,\\
 f_1&=&  (\alpha_{-\frac32,3}   {3\over  2\pi^5}+\beta_1)\  \bE_3(T)\,
 \bE_3(U)   +   {2\over3}\zeta(3)\,
 (\hbE_1(T)+\hbE_1(U))\\
\nn&+&{2\pi\zeta(3)\over    9}\log(y_8)+{\pi\over 18}\, \zeta(3) \,,\\
 f_2&=&{2\over3}\hbE_1(T)\,\hbE_1(U)+f(T)+f(U) +\beta_2 \bE_3(T)\, \bE_3(U)\\
  \nn &+& {11\pi^2\over 216} +{\pi\over 9}\, (\hbE_1(T)+ \hbE_1(U))\\
\nn&+&{\pi\over 18}\left(\pi +  4  \hbE_1(T)+ 4\hbE_1(U)\right)\, \log(y_8)+ {\pi^2\over 27}\, \log (y_8)^2\,,\\
 f_3&=& {\alpha_{-\frac32,3}\over 60} \, \bE_3(U)+\alpha_3 \, \bE_3(T)\,.
 \end{eqnarray}
 Strikingly, after combining the different log contributions the final result containing log parts 
is symmetric under the exchange of $U$-$T$ variables.

 Symmetry  under  the  $T\leftrightarrow U$  also determines the relation
 \begin{equation}
   \alpha_{-\frac32,3}= 60\alpha_3\,.
 \end{equation}
 Decompactification to ten dimensions and the value of the genus-three coefficient
 found in~\cite{Green:2005ba} fixes
 \begin{equation}
   \alpha_3={2\over27}\,.
 \end{equation}
 Comparison  with the  genus-one computation  in~(\ref{e:D6R4oneloop})
 fixes
 \begin{equation}
   \alpha_{-\frac32,3}+{2\pi^5\over3}\beta_1= {40\over9}\ ,
 \end{equation}
 which implies that $\beta_1=0$. The large volume limit of the genus-two contribution fixes $\beta_2=0$.
 Thus we find
 \begin{eqnarray}
   f_0&=&{2\over3}\zeta(3)^2\,,\\
 f_1&=&  {20\over 3\pi^5}\ \bE_3(T)\, \bE_3(U)+ {2\over3}\zeta(3)\, (\hbE_1(T)+\hbE_1(U)) +{2\pi\zeta(3)\over 9}\log(y_8)+{\pi\over 18}\, \zeta(3)  
\,,\\
 f_2&=&{2\over3}\hbE_1(T)\,\hbE_1(U)+f(T)+f(U)+{\pi\over 9}\, (\hbE_1(T)+ \hbE_1(U)) \\
\nn &+& 
{\pi\over 18}\left(\pi + 4 \hbE_1(T)+ 4\hbE_1(U)\right)\, \log(y_8)
+ {2\zeta(2)\over 9}\, \log (y_8)^2+{11\zeta(2)\over 36} \ ,
\nn\\
 f_3&=& {2\over 27}\left( \bE_3(U)+ \bE_3(T)\right) \,.
 \end{eqnarray}
 Finally, the $SL(3,\ZZ)\times SL(2,\ZZ)$  invariant expression for $C^{SL(3)}D^{SL(2)}(U)$ that solves
 (\ref{e:CD}) and has the above perturbative expansion is given by
 \begin{eqnarray}
  \nn C^{SL(3)}D^{SL(2)}(U)= {1\over 3} \hbE^{SL(3)}_{[10]; {3\over 2}} \hbE_1(U) +{\pi\over 36} \hbE^{SL(3)}_{[10]; {3\over 2}}+
{\pi\over 9}  \hbE_1(U) +{\zeta(2)\over 9}\,.
 \end{eqnarray}
 This  has terms  that were  not present  in  \cite{Basu:2007ck}, that
 originate from the regularisation of the source term.
 
The complete form of the solution is given by 
 \begin{equation}\begin{split}
 \mathcal{E}^{(8)}_{(0,1)}&= \mathcal{E}_{(0,1)}^{SL(3)}
 +{40\over9}           \,           \bE^{SL(3)}_{[10]; -{3\over2}}\,\bE_3(U)+{1\over3}
 \hbE^{SL(3)}_{[10]; {3\over2}} \,\hbE_1(U) +f(U)\cr
&+{\pi\over36}\, (\hbE^{SL(3)}_{[10];\frac32}+4\hbE_1(U))+ {\zeta(2)\over9}
\,.
\end{split}
 \end{equation}

\end{document}